%% file: Fog_survey_root.tex
\pdfoutput=1
\documentclass[10pt,journal]{IEEEtran}

\usepackage{multirow}
\usepackage{graphicx}
\usepackage{comment}
\usepackage{color}
\usepackage{url}
\urldef\myurl\url{https://portal.etsi.org/portals/0/tbpages/mec/docs/mobile-edge_computing_-_introductory_technical_white_paper_v1%2018-09-14.pdf}
\usepackage{amsmath}
\usepackage{amssymb}
\usepackage{cite}
\usepackage{array}
\newcolumntype{L}{>{\centering\arraybackslash}m{3cm}}

\fontfamily{pcr}\selectfont

\newcommand{\etal}{\textit{et al.}}

\begin{document}

\title{A Survey of Fog Computing and Communication: Current Researches and Future Directions}

\author{
Shubha Brata Nath, Harshit Gupta, Sandip Chakraborty, Soumya K Ghosh
\thanks{S. B. Nath is with the Department of Computer Science and Engineering, Indian Institute of Technology Kharagpur, INDIA 721302 (E-mail: nath.shubha@iitkgp.ac.in)}%
\thanks{H. Gupta is with the Georgia Institute of Technology, Atlanta, GA 30332 (E-mail: harshitgupta1337@gmail.com)}%
\thanks{S. Chakraborty is with the Department of Computer Science and Engineering, Indian Institute of Technology Kharagpur, INDIA 721302 (E-mail: sandipc@cse.iitkgp.ernet.in)}%
\thanks{S. K. Ghosh is with the Department of Computer Science and Engineering, Indian Institute of Technology Kharagpur, INDIA 721302 (E-mail: skg@cse.iitkgp.ernet.in)}
}

\maketitle

\input{0_abstract}

\input{1_introduction}

\input{2_evolution_of_distributed_computing}

\input{3_classification_of_existing_work}

\input{4_systems_of_fog_computing}

\input{5_qos_parameters}

\input{6_applications}

\input{9_fog_computing_associations}

\input{10_overhead_of_using_fog_computing.tex}

\input{7_limitations_and_future_scope}

\input{8_conclusion}

\bibliographystyle{IEEEtran}
\bibliography{fog_survey}

\end{document}

%% file: 0_abstract.tex
\begin{abstract}

The computing world has seen a paradigm shift from the traditional personal computing to present day client-server computing with the advancements in computer networking. The client-server computing has completely evolved into cloud computing, which provides flexibility, low cost deployment, fault tolerance and high availability, to build virtualized services. 
Currently with the proliferation of the Internet of Things (IoT) devices, the computing needs latency-sensitive support, which a cloud cannot provide. In the year of $2012$, a group of researchers from Cisco has presented a new computing paradigm, called fog computing, where the IoT devices can be given effective and enhanced support by bringing back a part of the computation from the cloud to the edge or near edge devices. Fog computing is a computing paradigm where some of the computations take place in the edge devices, and these fog devices interplay with the cloud server to provide better quality of service (QoS) to the end users. It can be noted that fog computing is not an offloading solution, rather it is a continuum; it is a range that goes from the cloud to the ground-level, where the computers and the end devices are located. 
In this survey, we discuss the evolution of distributed computing from the utility computing to the fog computing, various research challenges for the development of fog computing environments, the current status on fog computing research along with a taxonomy of various existing works in this direction. Then, we focus on the architectures of fog computing systems, technologies for enabling fog, fog computing features, security and privacy of fog, the QoS parameters, applications of fog, and give critical insights of various works done on this domain. Lastly, we briefly discuss about different fog computing associations that closely work on the development of fog based platforms and services, and give a summary of various types of overheads associated with fog computing platforms. Finally, we provide a thorough discussion on the future scopes and open research areas in fog computing as an enabler for the next generation computing paradigm.  

\end{abstract}


%% file: 1_introduction.tex
\section{Introduction}

\IEEEPARstart{I}{n} the present era of computing, device ubiquity is of prime importance for extending computing services over multiple end devices situated in different places and being used by various end users. 
 The anywhere, anytime presence of mobile devices is making our daily activities easier with proper utility based services and monitoring. The proliferation of {\em Internet of Things} (IoT) devices has created large networks with sensors and actuators, which provide delay-sensitive response to users. However, these large number of geographically distributed IoT devices have privacy and security concerns. Also, these sensors and actuators have minimal power with which it can provide storage and services. As the sensors need to continuously monitor the users, they cannot be recharged frequently. So, low power usage is of great importance for the sensor nodes. In order to save power, IoT sensors communicate with low-power protocols to a gateway (typically a computer), which then sends the data to the server.  
In most of the cases with current computation platforms, these servers are cloud based servers that can provide many instances of its virtualized services in order to provide the benefits like scalability, ease-of-setting-up, device ubiquity, seamless computation, hardware independence etc. According to the National Institute of Standards and Technology (NIST) definition of cloud computing~\cite{mell2010nist}, a shared pool of computing resources (e.g. networks, storage etc.) can be provided to the customers for computation and analysis of data collected from various sources. 

Computing resources over a cloud based system can be easily provided and released with minimal management interaction. This way the cloud infrastructure build up a two layer platform, where basic data collection tasks are done in the edge devices, and then the analytics related activities are performed over the cloud. 
Cloud computing has many advantages including on-demand self-service, infinite scaling, storing of large amount of data etc. 
That is why, the IoT devices communicate with the remote cloud server for executing the respective services. Cloud provides different service models, namely {\em Software as a Service} (SaaS), {\em Platform as a Service} (PaaS) or {\em Infrastructure as a Service} (IaaS) etc. to the requested users in a pay-as-you-go basis. Several recent works, such as \cite{zhou2013cloudthings,truong2015principles,kovatsch2014californium,muhammad2017smart,nastic2016middleware,barcelo2016iot} and the references therein, talk about platforms which integrate the IoT applications with cloud computing frameworks. The essence of these papers is that the proposed architectures accommodate IaaS, PaaS and/or SaaS for developing, deploying, running as well as composing various IoT applications.

Nevertheless, the computation over a cloud also has its own issues while providing services in the context of IoT, such as large response time for transferring the raw data to the cloud and then processing it there, disruption in the underlying communication network, issues related to data security and privacy, and so on. 
A typical IoT based platform works in three phases -- {\em sense-process-action}, where (i) sensors at the edge sense environmental parameters and send sensed data to the cloud, (ii) the cloud performs analytics on sensor data, and (iii) the processed information is forwarded back to the edge for actuators to perform necessary actions. As an example, in case of a typical smart-home environment, once the temperature sensor senses the room temperature above a threshold, it switches on the air condition subjected to the information from the window sensor that the window is closed. In this example, both the sensing and action are taken place at the temperature control module, however in a cloud based processing environment, the processing task can be offloaded to the cloud. 
In such a scenario, the response time can get affected due to unavailability of sufficient communication resources as well as high communication latency at the backbone network. 
Further, for third party cloud based solutions, there are privacy and security concerns of storing personal data in the cloud. 

\begin{figure}[!t]
	\centering
	\includegraphics[width=0.5\textwidth]{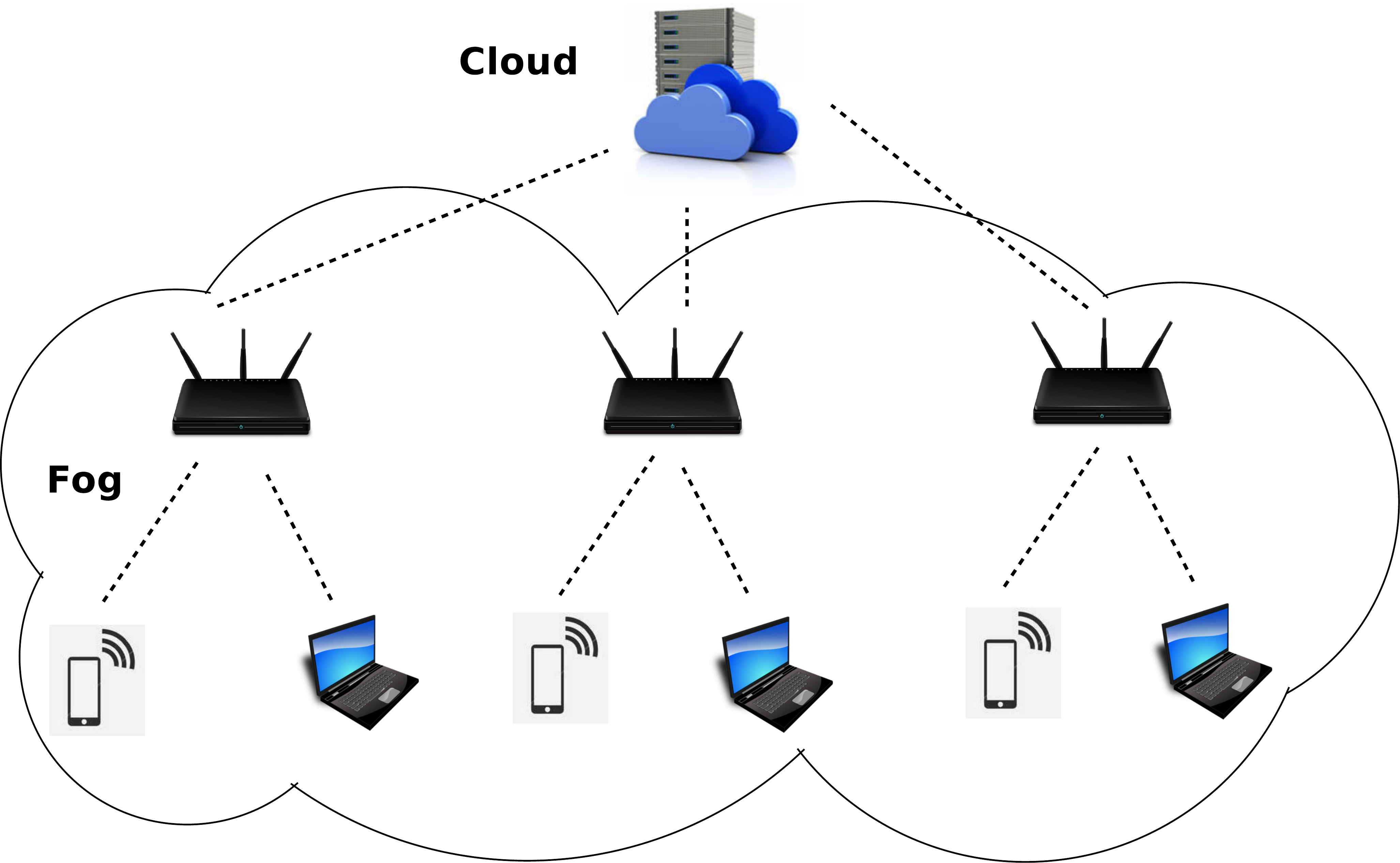}
	\caption{Typical Components of a Fog Architecture}
	\label{fig:fogarchi}
\end{figure}

In order to overcome these issues, the concept of {\em fog computing}~\cite{bonomi2012fog} has been emerged recently, which talks about doing the inter-play between the edge devices and the cloud servers. 
Fog computing is a distributed computing platform where the edge devices, i.e. routers, gateways, as well as even the sensor nodes etc., interplay with the cloud servers in order to give services. Bonomi \etal ~\cite{bonomi2012fog} defined fog computing as follows -- \textit{``Fog Computing is a highly virtualized platform that provides compute, storage, and networking services between end devices and traditional Cloud Computing Data Centers, typically, but not exclusively located at the edge of network"}. 
A fog network consists of different edge nodes with limited computing capability -- these are often termed as {\em fog nodes}. These fog nodes have storage as well as some limited computation facility. Sometimes in the fog network, we have different servers associated with the edge, known as {\em cloudlets}~\cite{whaiduzzaman2016mobicore,chen2016packetcloud}, which participate in the distributed computing environment within the edge network. By utilizing the fog devices, the users may get real-time reply for latency sensitive applications. 
Fig.~\ref{fig:fogarchi} shows the typical architecture of a fog computing platform. The end devices are connected to the routers as well as gateways. The routers as well as gateways are in turn connected to the remote cloud server. The fog is a multi-tier architecture i.e. it is spread from the edge devices to the cloud servers. 

Apart from the delay improvements, fog computing also have the potential of providing additional services~\cite{bonomi2012fog}, as follows. 
\begin{enumerate}
	\item[(i)] {\em Location awareness}: The fog device of a particular location can better know its context information.
	\item[(ii)] {\em Wide-spread geographical distribution}: The fog nodes are distributed around a large geography.
	\item[(iii)] {\em Mobility based services:} Mobile devices can move with uninterrupted fog enabled services.
	\item[(iv)] {\em Supporting very large number of nodes:} Large number of end devices can be served in the fog architecture.
	\item[(v)] {\em Omnipotent role of wireless access:} Wireless network has provided the advantage of accessing the fog services.
	\item[(vi)] {\em Device heterogeneity:} Different heterogeneous devices can reside and participate in the fog computation with minimal effort.
\end{enumerate}


This survey gives a comprehensive overview of the exiting works done over fog computing platforms along with its architectural impacts, and discusses about several open directions for fog computing research. Although there exists a few preliminary surveys on fog computing, they lack to produce an overall analysis of fog environment from different critical aspects. Mandlekar \etal ~\cite{mandlekarsurvey} have presented a survey, which talks about how fog computing can be used in order to eliminate data theft attacks. The works of \cite{saharan2015fog} have focused on some of the application areas of fog computing in comparison with cloud computing. In \cite{dastjerdi2016fog}, some applications and possible research directions of fog computing are discussed. Mahmud \etal ~\cite{mahmud2016fog} have discussed about challenges and key features of fog computing. Impact of fog computing over 5G networks has been discussed in a recent survey~\cite{kitanov20165g}. The security and privacy issues of fog computing have been discussed in~\cite{yi2015security}. In~\cite{yi2015survey}, application scenarios and issues of fog computing are discussed. However, these existing surveys lack several insights to bring out the overall researches and scenarios in the context of fog computing and its interplay with other computing platforms.  
In this paper, we provide a comprehensive survey of fog computing from different directions by developing a clear taxonomy of the existing works. We also extract the possible future directions of research areas over fog based platforms, both as a shortcoming or limitations of the existing works, as well as the open research problems. 

The rest of the survey is organized as follows. First, we discuss the evolution of the distributed computing framework in Section~\ref{evolution}, that talks about how the distributed computing technologies gradually evolved from utility computing in early $1960$'s to today's fog computing paradigm. Section~\ref{taxonomy} presents various research challenges and current status of the fog computing research environments, and based on that, we develop a taxonomy of the existing works on fog computing. Section~\ref{archi} discusses different architectures and frameworks to facilitate the fog computing framework. We primarily discuss about the service oriented architecture (SOA), and other application specific architectures in this context. Section~\ref{technology} highlights the different technologies for enabling fog computing. Section~\ref{features} presents the fog computing features for developing services. Section~\ref{securityprivacy} focuses on the security and privacy in fog computing. Section~\ref{qossection} discusses about the quality of service (QoS) parameters. Section~\ref{application} is about exploring the application areas of fog computing. Section~\ref{associations} discusses about the fog computing associations. Section~\ref{overheadfog} analyzes the fog computing overhead. The future scopes and open research areas have been discussed in section~\ref{futurescope}. Finally, section~\ref{conclusion} concludes the paper.

%% file: 2_evolution_of_distributed_computing.tex
\section{Evolution of Distributed Computing}
\label{evolution}

\begin{figure}[!t]
	\centering
	\includegraphics[width=0.5\textwidth]{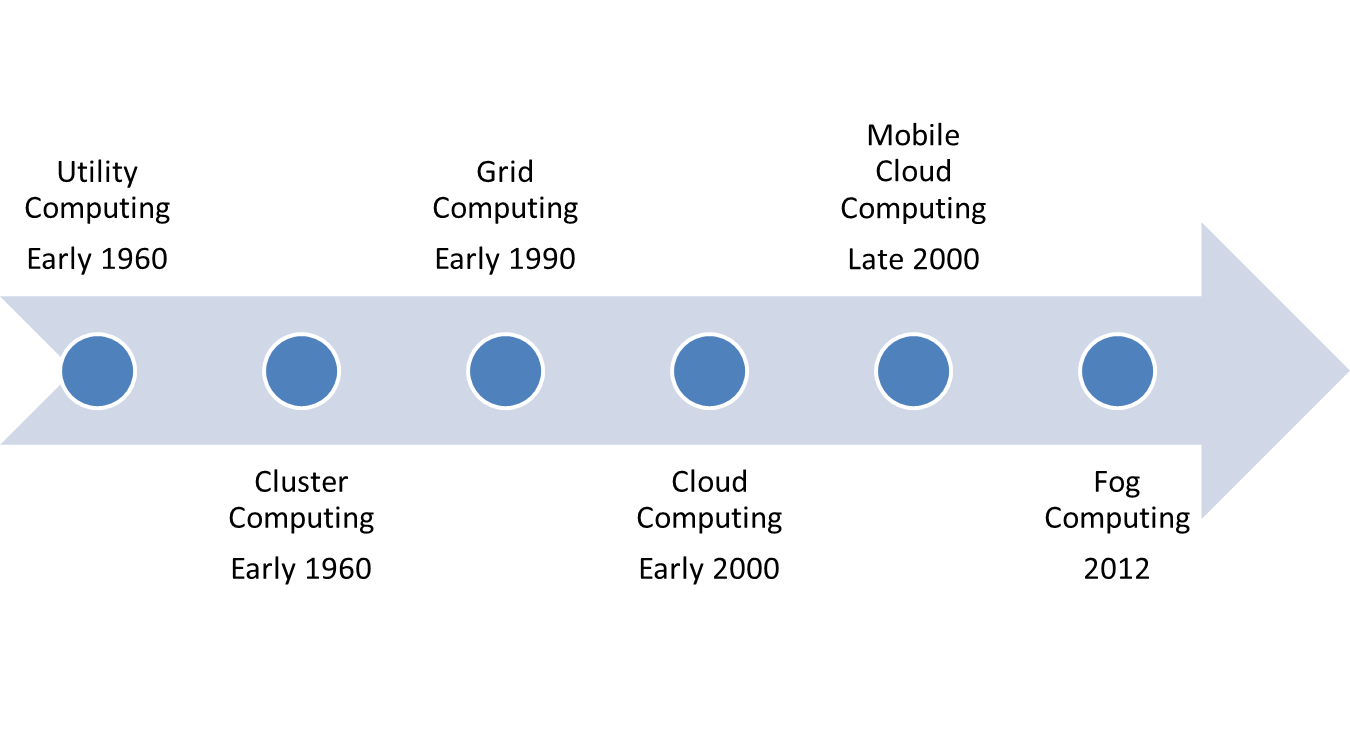}
	\caption{Evolution of Distributed Computing}
	\label{fig:timeline-fog}
\end{figure}

Distributed computing~\cite{leopold2001parallel,bermond1995distributed,huebscher2008survey,bal1989programming} refers to the study of decentralized models of systems where the computation task is divided between several network devices, and these devices communicate between themselves through a message passing interface. These resources, i.e. computers, are connected over the network. 
The concurrent processes communicated by message passing was studied in the early $1960$s~\cite{gropp1996high}. The first known distributed system was present in way back in $1970$s which is known as local area network (LAN)~\cite{clark1978introduction,limb1983distributed,tsao1984local} that interconnects multiple computers for making applications to communicate with each other for developing a collective solution. The distributed computing has gone through several new computation paradigms after that. Starting from the concept of utility computing over a distributed framework, the computing domain has gradually moved towards the concept of cloud computing, and very recently, mainframe computing has evolved into the concept of fog computing. 
In this section, we give a brief description of the distributed computing timeline to highlight the idea about how the computing paradigm has gradually evolved to today's fog computing concepts, based on the requirements from the end users. 
Fig.~\ref{fig:timeline-fog} shows the timeline view of the evolution of distributed computing. In the early $1960$s, utility and cluster computing~\cite{broberg2008market,huebscher2008survey,buyya1999high} were conceptualized. Grid computing~\cite{krauter2002taxonomy,yu2005taxonomy,ernemann2002advantages} has emerged as the computing paradigm in the early $1990$s, where a set of computers, connected together over a grid, takes the computing decisions collectively. Cloud computing~\cite{sharkh2013resource,agrawal2011big} has become popular in the early $2000$s. The concept of cloud computing has gradually extended over the mobile devices, and mobile cloud computing~\cite{othman2014survey} has come into picture in the late $2000$s. 
Fog computing is the new computing paradigm in the pervasive computing scenario, where involves computation over the end devices such as mobiles, sensor boards, control systems etc., and it has come into picture in $2012$ as shown in fig.~\ref{fig:timeline-fog}.

\subsection{Utility Computing}

\begin{figure}[!t]
	\centering
	\includegraphics[width=0.5\textwidth]{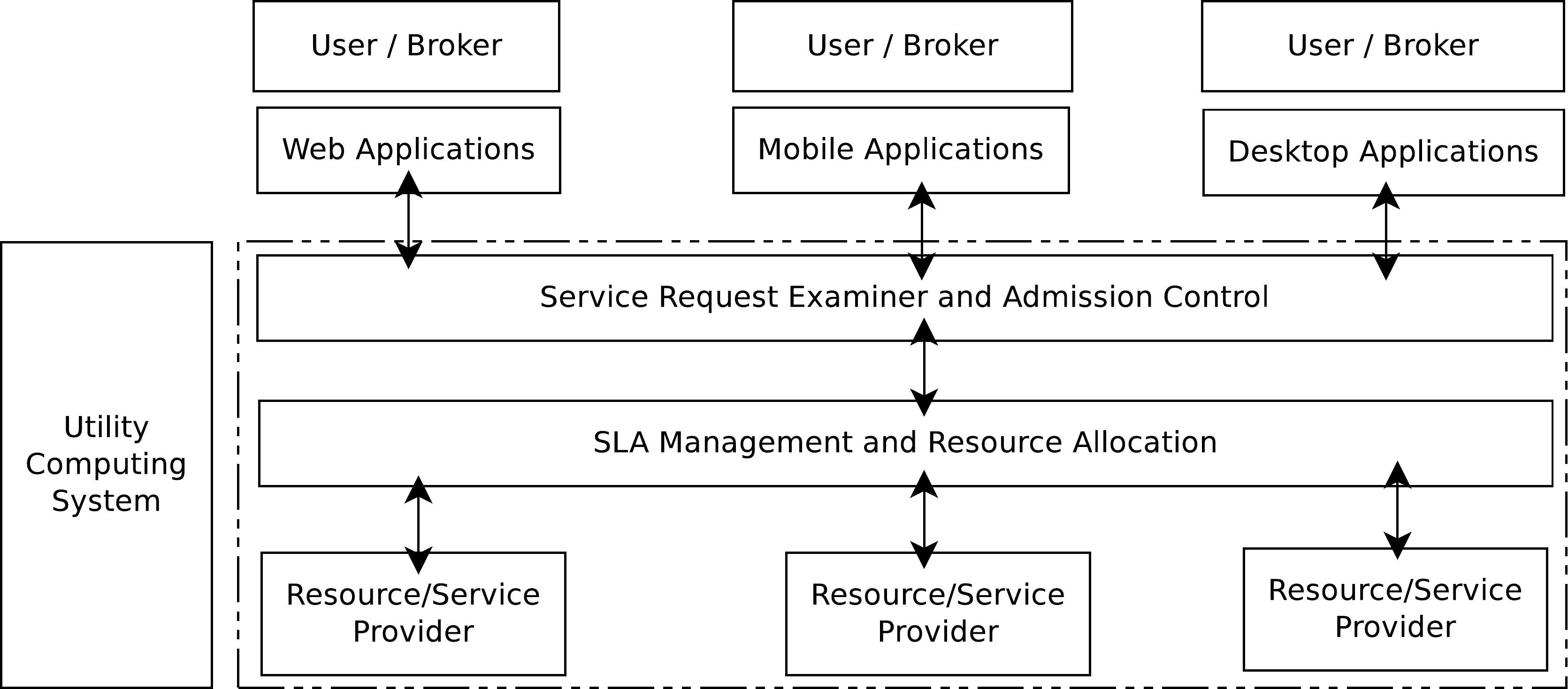}
	\caption{A Generic Framework for Utility Computing Architecture}
	\label{fig:utilitycomputing}
\end{figure}

A major requirement from the end-users is that they need to get computing and storage services within a short period of time. This strict requirement of deadline driven services has created the demand for getting services without the botheration about deployment and operationalization of custom hardware. The traditional mainframe based systems lack these features of providing deadline driven services, as the users need to purchase hardware, customize them, and install tools and software to make them operational. That is why, the end users have shifted towards getting real-time services from the vendors. These services are provided to the end users as an utility. The users need not worry about the underlying hardware infrastructure. The services are provided to the end users whenever they require that. 

The provision of providing computations and services to the end users based on their need has created the distributed computation model named as \textit{utility computing}. In this early computing model over distributed systems, users have to pay for the services whenever they use it. The concept was first presented by John McCarthy in $1961$~\cite{garfinkel1999architects}. Though the utility computing was not very popular in those times, it was again introduced in late $90$'s as the cost for computation hardware gradually dropped and miniaturization of servers become practical. The excessive demands for services have generated the need for this utility based service provisioning by the service providers. Previously, there was no proper access to resources in several systems. Further, there was no provision of supporting a fixed and predefined deadline for response time over these systems. However, the utility computing has given the user a proper valuation of their services. Utility computing systems can be considered as a marketplace where the users compete for getting their service by the service providers. The advantages of such utility computing systems are much more in comparison with the single time-sharing system. Utility computing supports the users by giving higher throughput than a single time-sharing system as multiple servers are placed in utility computing~\cite{padala2007adaptive,nurmi2008eucalyptus,ross2004preparing,buco2004utility}.  The different components of utility computing have been shown in fig.~\ref{fig:utilitycomputing}. Under the utility computing framework, web, mobile and desktop applications communicate with the service request examiner and admission control module, which determine the specific computation requirements for the user tasks. Service Level Agreement (SLA) management and resource allocation module communicate with the service request examiner and admission control module to allocate the resources for executing user tasks. Also, it communicates with the resource or service provider for building up a bridge between the service providers and the end users. 


\subsection{Cluster Computing}

\begin{figure}[!t]
	\centering
	\includegraphics[width=0.5\textwidth]{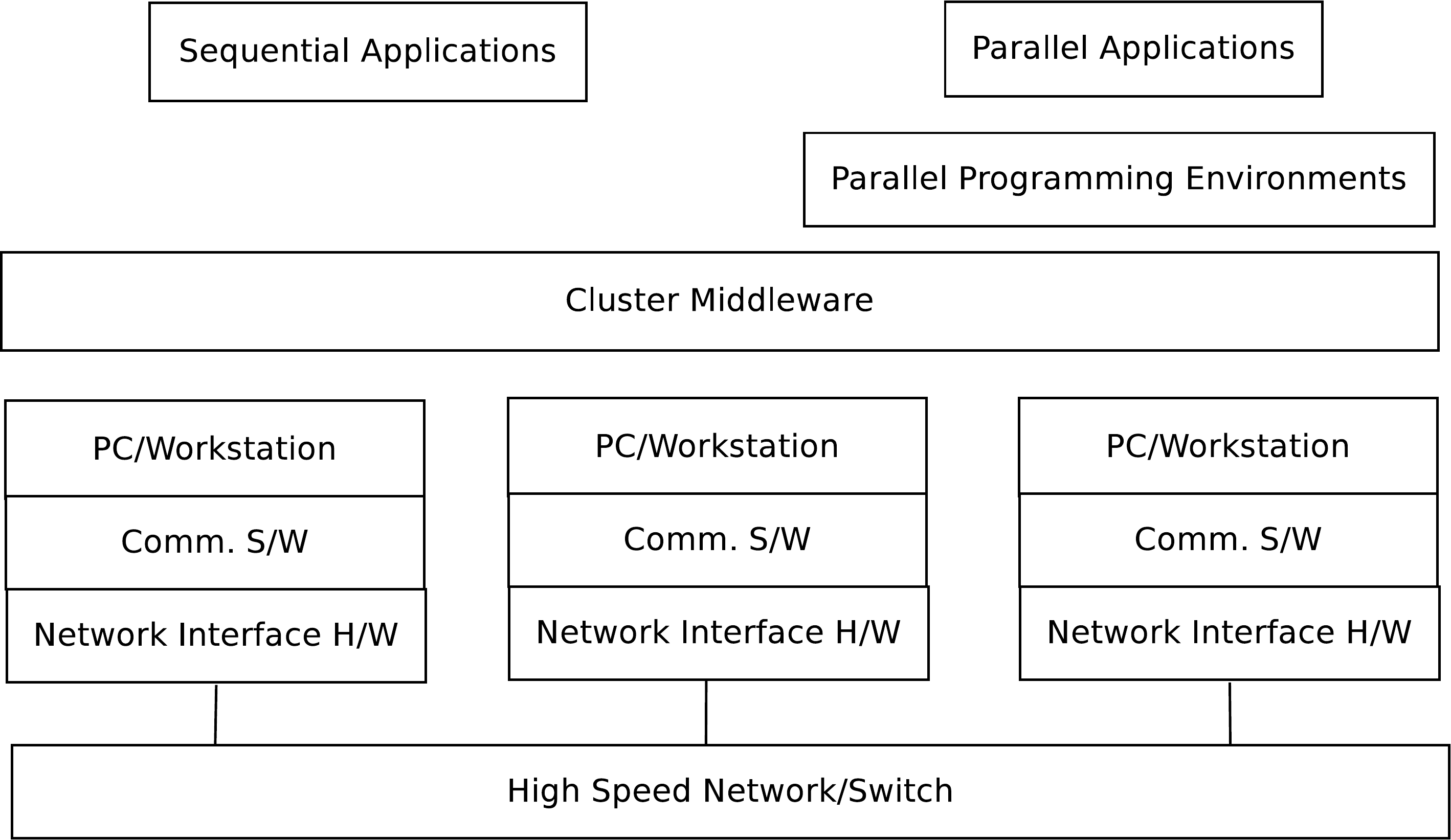}
	\caption{Cluster Computing Architecture}
	\label{fig:clustercomp}
\end{figure}

In cluster computing~\cite{valentini2013overview,buyya1999high}, many connected computers work together in order to behave like a single system. These computers do the same work, which is controlled by a scheduling software. In the advent of low cost microprocessors, the cluster computers have emerged as a new computing platform. The different components of cluster computing have been shown in fig.~\ref{fig:clustercomp}. The different personal computers along with communication software and network interfaces are connected to the high speed network or switch. The cluster middleware is the software environment that interconnects the cluster computing nodes with different applications.

The advantages of the cluster computing system are the reliability and the availability. The end users can be given more computing power and storage facilities by allowing several cluster computers to work. The system failure rate is decreased in case of cluster computing as we have redundancy within the clusters. The dedicated and high speed network connects these cluster nodes in order to provide more reliability in case of system failure. These are the driving forces behind the development of cluster computing concepts. There are several types of applications of cluster computing, such as load balancing~\cite{werstein2006load,bohn2002load,maguluri2012stochastic}, high availability clusters~\cite{agbaria1999starfish,fu2010failure} etc. In load balancing, a single task can be divided between several cluster nodes in order to provide the particular service. The high availability clusters provide the users with the required service in case of system failure. 
The cluster computing systems have data redundancy among the cluster nodes. That is why, cluster computing systems have greater reliability than utility computing systems. 


\subsection{Grid Computing}

\begin{figure}[!t]
	\centering
	\includegraphics[width=0.5\textwidth]{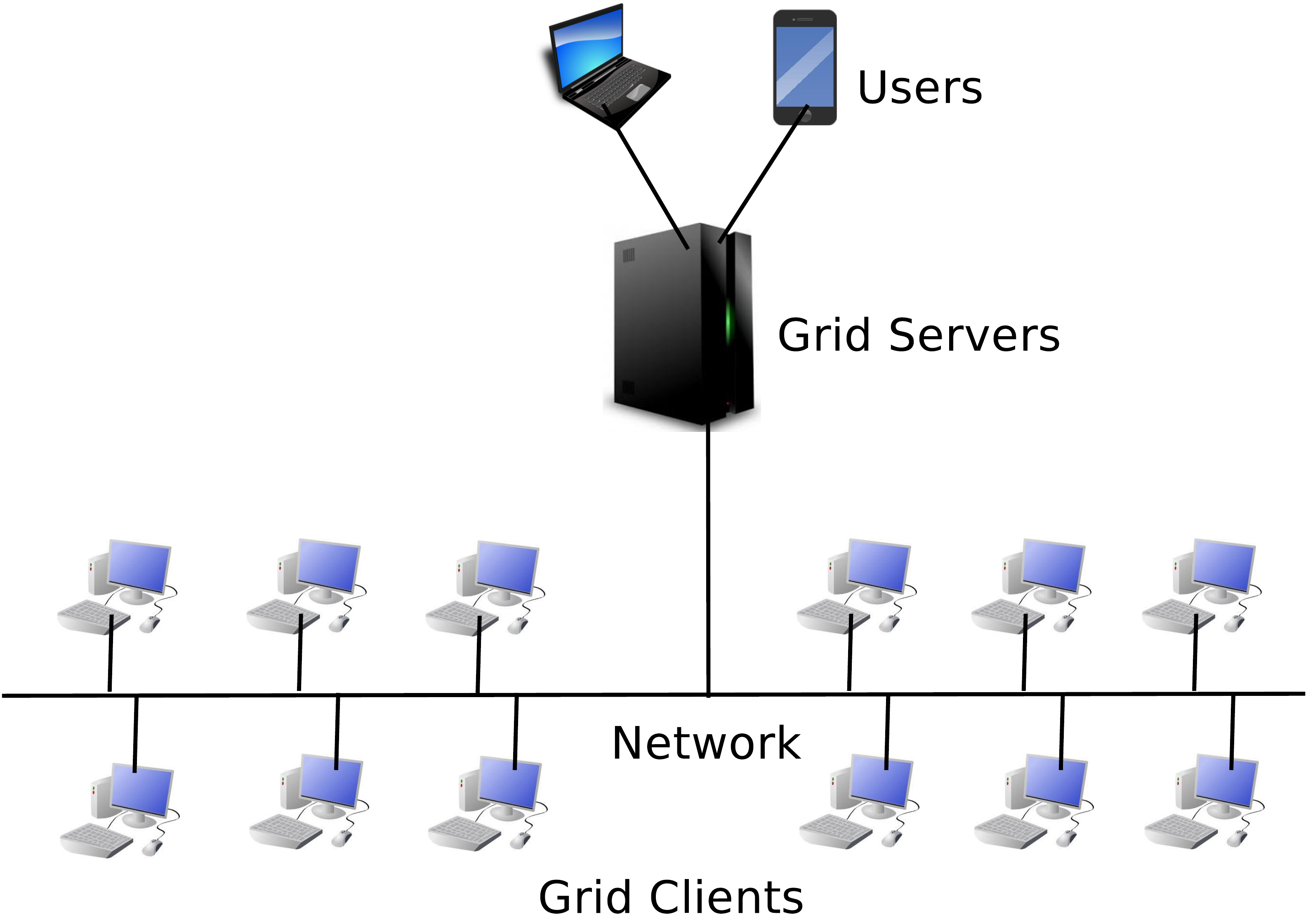}
	\caption{A Framework for Grid Computing Architecture}
	\label{fig:gridcomp}
\end{figure}

Grid computing~\cite{krauter2002taxonomy,yu2005taxonomy,berman2003grid,chervenak2000data} was originated in early $1990$s as an effect of extending the distributed computation models beyond the cluster computing framework. Cluster computing environment faced several limitations during its implementation. One of the major problems faced over the cluster architecture is that, there can be node failures in clusters; however as the cluster size increases, the complexity of finding the location of failures also increases. However, the grid computing systems are much modular and have very less number of points of failure. The grid software is responsible for policy management. In fig.~\ref{fig:gridcomp}, we have shown a generic architecture of the grid computing framework. The system consists of several grid clients that are connected by an underlying computer network to the grid server, and the grid servers are in turn connected to the end users. Recent works, such as~\cite{berman2003grid,yu2005taxonomy,buyya2002economic,yu2005taxonomy} and the references therein, talk about the utilities of grid computing architecture. In grid computing, several computers, termed as {\em computing grids}, work together to provide a high performance distributed environment. The computation is divided and distributed among several nodes. The computing nodes are loosely coupled in the sense that they make use of little or no knowledge of the definitions of other separate nodes. On the contrary, in tightly coupled systems, the computing nodes are not only linked together but also dependent upon each other. The disadvantages of tightly coupled system is that the entire system becomes down in case of even a single node failure. 
However, the resources are generally heterogeneous and largely distributed. Also, there is an overhead of liscencing of many servers in grid computing systems. Now a days, grid computing is mainly used in commercial organizations for its advantage of workload distribution. Grid computing are great in the sense that they provide fault tolerance which helps to provide better quality of service (QoS) requirements. 

\subsection{Cloud Computing}

 Cloud computing offers higher storage capacity with lower computation cost. Instead of buying the computer hardware, the end users only need an Internet connection in order to get the most of the applications. The cloud refers to the concept of remotely providing computing resources to the end users. Also, cloud means an important aspect which is to provide virtualized service as an abstraction of service. By abstraction, we mean that the services are provided to the users without giving the details of the location of the services, infrastructures involved as well as the physical devices responsible for the services. Virtualized services are provided by the cloud from the available poll of resources. The cloud framework is a utility based distributed computing paradigm where the end users are provided with computing resources and services whenever they need it from anywhere. The ubiquitous access is the main feature of the cloud. These computing resources reside in remote locations with the provision of providing the virtualized services. Cloud computing accelerated the industry with its scalability features dynamically. Cloud has the potential of providing the resources on demand to the end users. These pay-as-you-go model of computing has services like software-as-a-service (SaaS)~\cite{cusumano2010cloud}, platform-as-a-service (PaaS)~\cite{dikaiakos2009cloud} and infrastructure-as-a-service (IaaS)~\cite{nurmi2009eucalyptus}. There are many cloud software available such as Amazon Elastic Compute Cloud\footnote{\url{https://aws.amazon.com/documentation/ec2/} (last accessed: July 2017)}, Microsoft Azure\footnote{\url{https://azure.microsoft.com} (last accessed: July 2017)} etc. Cloud computing paradigm uses the concept of service oriented architecture (SOA)~\cite{namjoshi2009service} in order to break a user's problem into different services and work towards these services in order to solve the problem.
 
 The emergence of cloud computing systems~\cite{mell2011nist,armbrust2010view,buyya2009cloud,buyya2008market,zhang2010cloud,dillon2010cloud} has eliminated many limitations of the grid computing framework as experienced by the users, and has moved the computing framework towards a new horizon. In grid computing systems, licensing across many servers may limit some applications to be deployed. Again, the data sharing among several individuals or organizations (also known as administrative domains) in grid computing is another issue. Cloud computing is very much able to eliminate these issues of grid computing.
 
\begin{figure}[!t]
	\centering
	\includegraphics[width=0.5\textwidth]{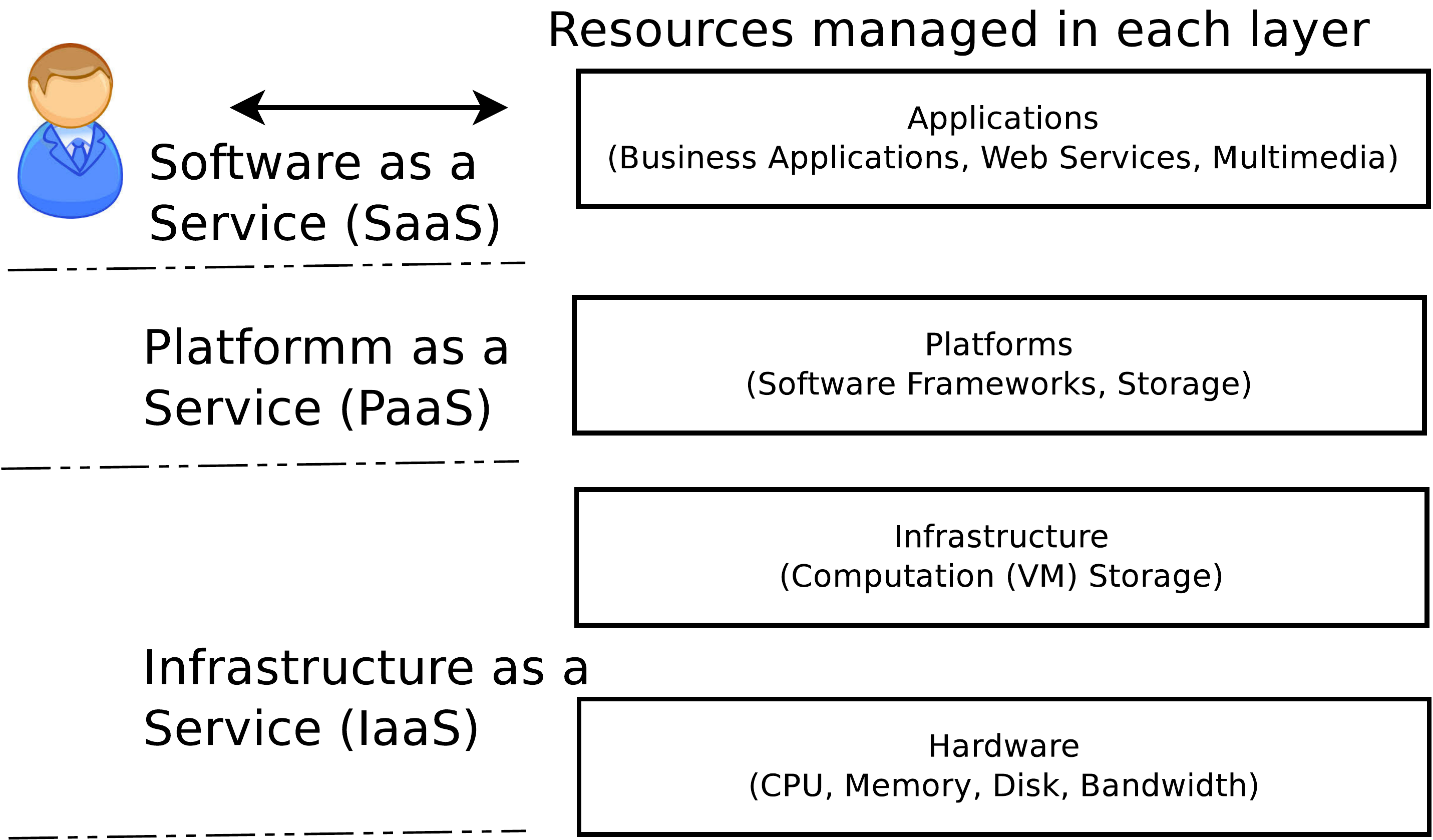}
	\caption{A Generic Framework for Cloud Computing}
	\label{fig:cloudcomp}
\end{figure}

 Clouds support multi-tenancy~\cite{bezemer2010enabling,wang2012application} that refers to a software architecture where a single instance of software runs on a server, and it serves multiple tenants. In a multi-tenant architecture, a software application is designed to provide each tenant a dedicated share of the instance. 
 Cloud also supports ubiquitous service provisioning along with multi-tenancy. We have shown a generic cloud computing architecture in fig.~\ref{fig:cloudcomp}. Cloud computing architecture consists of software-as-a-service (SaaS), platform-as-a-service (PaaS) and infrastructure-as-a-service (IaaS) layer. In the SaaS layer, we have different business applications, web services, and multimedia. In the PaaS layer, different software frameworks and storage facilities are provided by the cloud. The IaaS layer provides different hardware i.e. CPU, memory, disk and bandwidth.

However, over the years cloud computing has faced a few limitations. The most important issue is that it requires a constant Internet connection. Also, the cloud computing services do not work well with low-speed connections, such as dial-up services etc. 

\subsection{Mobile Cloud Computing}

The concept of cloud computing has gradually evolved and spreaded over the mobile computing framework. In mobile cloud computing systems~\cite{huang2011mobile,dinh2013survey,fernando2013mobile,kumar2010cloud,khan2013towards}, we have different smartphones and tablets as the end devices. These end devices generate various types of data which are sent to the servers via cellular networks or Wi-Fi access points. In an intermediate framework between the cloud computing and mobile cloud computing, people used to deploy small computing servers with limited computation capability, called cloudlets~\cite{whaiduzzaman2016mobicore,chen2016packetcloud,ibrahim2009cloudlet,chen2015computation,zhang2015offloading,shires2012cloudlet}. The cloudlet servers need some computing as well as storage which are provided by the cloud computing servers.

\begin{figure}[!t]
	\centering
	\includegraphics[width=0.5\textwidth]{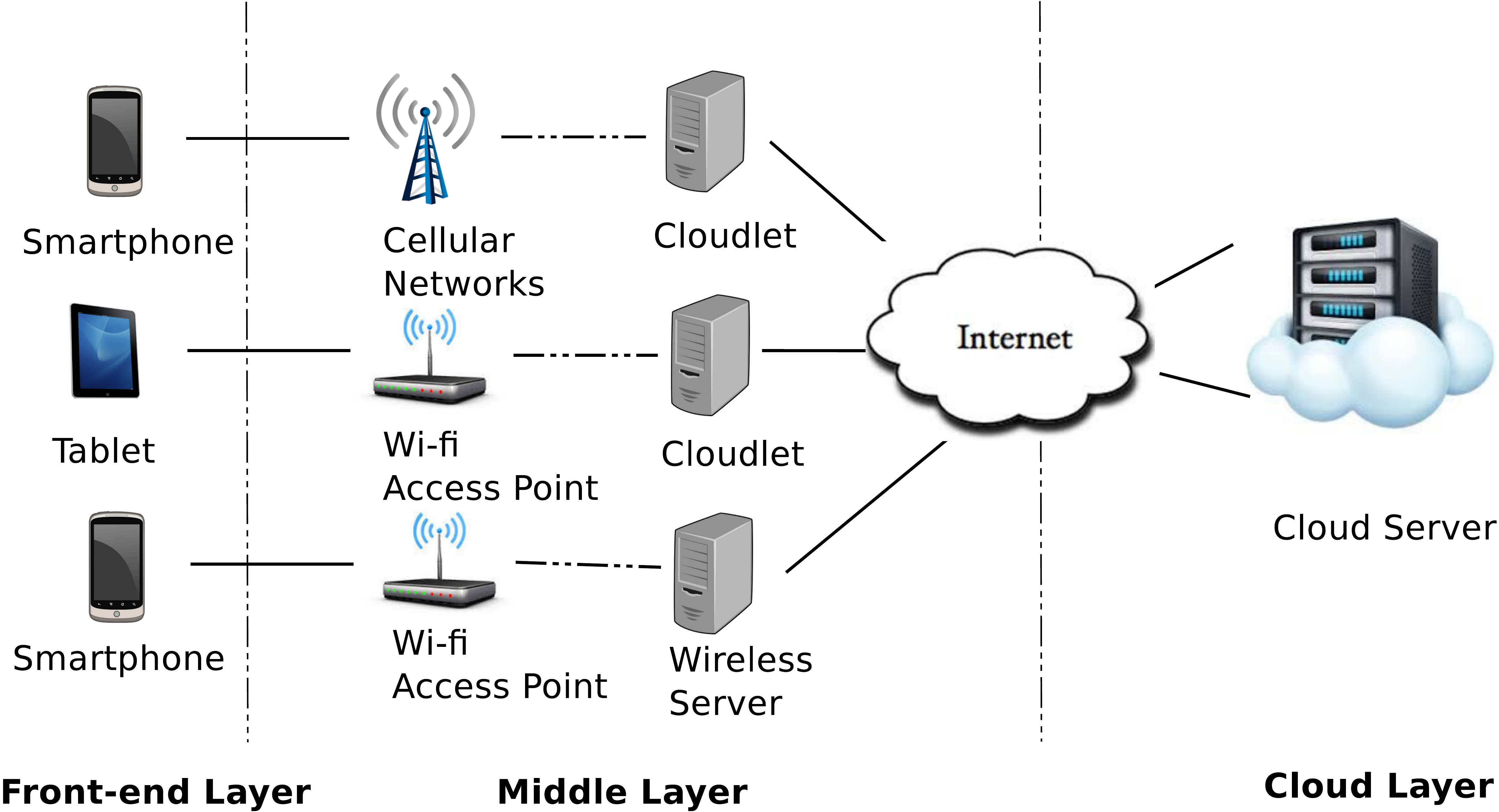}
	\caption{Mobile Cloud Computing Framework}
	\label{fig:mccomp}
\end{figure}

Fig.~\ref{fig:mccomp} depicts the mobile cloud computing architecture. Mobile cloud computing is a combination of mobile computing, cloud computing and mobile Internet. In mobile cloud computing, the data processing and storage is moved from mobile devices to the centralized cloud servers. These platforms can be accessed through wireless connections via web browsers on the mobile devices. The mobile devices can be smartphones, laptops or personal digital assistants (PDA). The mobile cloud computing framework has brought about a concept where a part of the application runs on user's smartphone, and another part is executed over the cloud. Though the mobile users are getting services whenever they are on the go, there are also few limitations of mobile cloud computing systems. The most important issue of the mobile devices is the resource constraints.
These mobile devices are sometimes not suitable for the deployment of complex applications which require more storage as well as more energy consumption. To reduce this problem, there will be a need to reduce the data exchange between the mobile devices and the cloud end. 

\subsection{Fog Computing} 
The concept of fog computing~\cite{stojmenovic2014fog,stolfo2012fog,vaquero2014finding,aazam2014fog,luan2015fog} emerged from the concept that part of the computing can be brought back near the edge devices. The term fog computing has been proposed by Cisco \cite{bonomi2012fog} in $2012$. The fog computing architecture has been shown in fig.~\ref{fig:fogarchi}. It refers to extend the cloud computing paradigm to the edge of the network. The edge devices (i.e. routers, gateways etc.) can be used as the computing nodes along with the existing cloud data centers. Fog computing has been envisioned to provide computation from the network edge, through the network core and to the cloud data centers. The different services are hosted in the fog nodes, which are using its resources through the hypervisor, the management software for virtualizing the computing environment. Fog computing does the proper interplay of the services with the cloud. The applications which require real time response and context aware computing rely on the fog computing framework. Further, there are situations where there are need for supporting huge amount of data generated from the IoT devices. Cloud computing alone is not sufficient in these situations as there is a requirement of real-time service provisioning. The typical applications of fog computing paradigm can be in real time health-care monitoring systems, smart cities, smart grids, vehicular ad-hoc network (VANET) etc. Being loosely coupled and highly distributed, QoS management and dynamic adaptability are the key challenges faced by the fog computing domain which need to be solved. 

\subsection{What is the Driving Force behind the Evolution of Distributed Computing?}

In Table~\ref{table:tablefog}, we provide a comparative study of various features and utilities of different distributed computing techniques as discussed above.
We can see that the distributed computing framework has gradually moved from user based computing to computing over a remotely located server, and with the incorporation of fog computing framework, researchers are interested to bring back a part of the computing over the devices at the edge. The technologies have evolved based on the requirements from the end-users as well as driven by the innovations at the hardware developments. For example, the cloud computing model suggested to use complex computation hardware at the user's end. However as people started to deploy devices with some capability of computing, the concept of edge and fog computing emerged as a result of bringing the simplified computation at the edge. The objective is to reduce data transfer overhead to the cloud, provide flexibility to the user application and support security and privacy for the user's data.   


One important takeaway from the above discussion on the evolution of distributed computing is that there is an oscillation between centralized and distributed computing, centralized mainframe to distributed personal computer (PC) to centralized cloud to distributed fog. However, there is more than just the oscillation. In each paradigm shift, we have learned from the previous one. For example, we learn that a user was provided the abstraction of a PC from personal computing era and then the user applied it to cloud computing via virtual machines. Similarly with fog computing, we learn the scalability and pay-as-you-go from the cloud, and apply the concepts over the edge or near-to-edge devices.

\begin{table*}[h!]
	\centering
	\caption{Architectures and utilities of different distributed computing techniques}
	\begin{tabular}{|p{2.5cm} ||p{1.6cm} |p{2cm} ||p{1.6cm} |p{1.6cm} |p{1.6cm} |p{2.8cm} |} 
		\hline
		& \multicolumn{2}{|c||}{\textbf{Architectural Aspects}}  & \multicolumn{4}{|c|}{\textbf{Utility Aspects}} \\ [0.5ex] 
		\hline
		\hline
		& \textbf{Connectivity} & \textbf{Device Coupling} & \textbf{Virtualization Support} & \textbf{Computation Mode} & \textbf{Response Time} & \textbf{Resource Availability}\\ \hline
		
		\textbf{Utility Computing} & Distributed & Tightly Coupled & No & Distributed & Moderate & Partially Distributed \\ \hline
		\textbf{Cluster Computing} & Centralized & Tightly Coupled & Yes & Centralized & Moderate to low & Partially Distributed \\ \hline
		\textbf{Grid Computing} & Distributed & Loosely Coupled & Yes & Distributed & High & Distributed \\ \hline
		\textbf{Cloud Computing} & Distributed & Tightly Coupled & Yes & Distributed & High & Distributed\\ \hline
		\textbf{Mobile Cloud Computing} & Distributed & Loosely Coupled & Yes & Distributed & High & Distributed\\ \hline
		\textbf{Fog Computing} & Fully Distributed & Loosely Coupled & Yes & Distributed & High to Moderate & Distributed\\ \hline
	\end{tabular}
	\label{table:tablefog}
\end{table*}

%% file: 3_classification_of_existing_work.tex
\section{Fog Computing: Research Challenges, Current Status and Taxonomy of Existing Works}
\label{taxonomy}

Fog computing is a domain emerged from the success of cloud computing framework as a commercial and commodity solution for providing computing resource to the end users. However, with the development of low cost computing hardware and devices like IoT sensors and smartphones, the research community realized that a part of the computation can be brought back to the devices near the edge, which can reduce the cost for data offloading at the cloud, as well as can provide privacy and security solution to the user data. However, computation at the edge also has its own challenges that the researchers are currently exploring for the end-to-end development of the fog computing framework. Consequently, a number of research outcomes have come out recently. In this section, we briefly discuss about the general challenges for the development of a fog computing solution, and accordingly classify the existing works into different groups for further discussion. 

\subsection{Challenges in Fog Computing Research}
As mentioned earlier, fog computing is a distributed computing architecture that involves network related challenges, computing related research directions, security related challenges, as well as management related challenges. Being highly distributed, it makes the system more vulnerable towards computation correctness. Here, we discuss about these issues of fog computing system.

\subsubsection{Network and Device Related Challenges}
The various network and device related challenges that the fog computing framework faces are as follows.

\textbf{Distributed architecture:} The distributed architecture makes the fog computing more prone towards having a redundant system. The same code is replicated in several locations in the edge devices of the network~\cite{tang2015hierarchical,chiang2017clarifying}. Therefore, the computing framework should have sufficient sophistication to reduce the redundancy over the distributed environment.

\textbf{Network resource distribution:} The networking resources are scattered in the edge or near-to-edge devices in the fog architecture. This makes the system more complex in terms of the network connectivity aspects. A common network middleware is required to be developed, which can manage the common pool of resources over the edge or near-to-edge devices, and accordingly should be able to allocate resources to the application workloads.  

\textbf{Heterogeneity of devices:} The fog environment has several end devices that are heterogeneous in nature. This heterogeneity of the devices has made the system more diverse~\cite{bonomi2014fog}. The computing platform should consider this device and network heterogeneity while developing the fog applications. 


\subsubsection{Computing Challenges}

The computation over a fog environment is itself challenging because of the following reasons. 

\textbf{Computation hierarchy development:} 
The fog computing system always communicates with the distant cloud servers. There is a trade-off between the response time and computation power in the fog computing system. The fog computing devices, that is the edge as well as the end devices, perform computations and should response to the users within a time guarantee. At the same time, some computations are also offloaded to the cloud, and these computations at the cloud may take higher time compared to the time required to execute the computation at the edge devices but with a less computation cost. Therefore, it is always a challenge to identify what parts of the computations have to be offloaded to the cloud, and what fractions of the computations have to be performed at the fog devices. This also has the trade-off in accuracy-interoperability that need to be addressed by the application developer.  

\textbf{Computation resource distribution:} Computation of different applications need proper resources. The edge devices may not always have all the resources deployed in them. Some of the resources have to be used from other fog nodes. This requirement has generated the need to distribute the computation resources among the edge devices. 
Therefore, there is a requirement for developing a converged framework to integrate the computation, memory as well as networking resources for building up the common pool. Applications can reserve resource from this common pool. The current researches in this direction are exploring whether the container technology~\cite{celesti2016exploring,bellavista2017feasibility,yannuzzi2017new} can be used to develop a common pool of resources over the edge devices for computation. 

\textbf{Distributed computation:} The distributed computation in the fog has created the need to verify the computation correctness. Fog applications need to be designed and developed in such a way that there are few inconsistencies in computation, and also such inconsistencies should be verifiable~\cite{sekar2011verifiable,aazam2015dynamic,chen2013towards,yannuzzi2017new,li2017coding}. 

\textbf{Mobility:} With the advent of mobile and hand-held devices, the current computing framework demands for the computation over anywhere, anytime and anything connectivity, and therefore a pervasive computing framework needs to be emerged over the fog computing framework. The edge nodes may be mobile in the fog computing environment. This mobility is another barrier for computing in the fog domain. Therefore, the researchers need to develop integrated, pervasive and ubiquitous solutions for handling mobility over the fog computing framework~\cite{vaquero2014finding,stojmenovic2014fog,truong2015software,orsini2015computing,ottenwalder2014mcep,makinen2015streaming,kai2016fog}.

\subsubsection{Security Related Challenges} 

The fog computing system, being distributed with different heterogeneous devices, is vulnerable towards various security attacks. The existing literature discusses man-in-the-middle attack in fog computing domain~\cite{stojmenovic2015overview}. Data and network security are the main issues in fog. Further as the fog computing framework also depends on the services from the cloud servers, the computation framework becomes vulnerable for trust and authentication issues. Privacy of the data is another concern in this highly distributed fog computing architecture \cite{yi2015security,stojmenovic2014fog,stojmenovic2015overview}.

Another security vulnerability is that, fog devices are not deployed in highly secure data centers, but in locations that may be easy to have physical access for attackers~\cite{greenberg2008cost,bhardwaj2016fast}. Hence, the system software itself may not be trusted. Therefore, there is a requirement to securely execute the edge functionalities over the fog. 


\subsubsection{Management Challenges}

Fog computing framework, being a distributed system architecture, poses several challenges related to system management. 

\textbf{Service oriented computing:} In the fog computing framework, a user service is divided into multiple micro-level services and these micro-services are distributed accross the edge devices and the cloud. This particular distribution of services over the fog devices is a mode of service oriented computation over edge devices. However, executing micro-services over the fog nodes has their own challenges. The proper management of the architecture in order to get the services is one of the prime challenges in fog computing domain. There are several challenges in micro-service management. These are service placement, service combination, tracking of execution steps etc. We need a proper orchestration system so that the services are provided to the end users within very less amount of time over the fog framework~\cite{truong2015software}.

\textbf{Resource management:} 
The different networking as well as computation resources are distributed in fog computing domain \cite{aazam2015fog}. Fog computing has to be flexible and adaptive (like cloud) to respond to issues like transient failures or resource shortages. The failure of fog nodes make the whole system down as the resource would not be available from that fog node. Again, the resources are virtualized in the fog network. The virtualization of resources creates many challenges. These challenges are the latency, initiation, placement, migration of virtual network devices in fog network etc.  
In these cases, we need to properly manage the resources so that the downtime can be avoided ensuring the high availability. This is primarily because a fog computing system deals with the latency-sensitive applications, such as smart home, smart health-care monitoring system etc.~\cite{aazam2014fog,stantchev2015smart,gia2015fog}. The modern technologies like software defined networking can be utilized for resource management~\cite{truong2015software} over the fog nodes which poses several research directions.

\textbf{Orchestration between fog nodes and the cloud:} Another challenge is the end-to-end orchestration of the fog-cloud resource ecosystem so as to provide QoS guarantees for various user level services~\cite{munoz2016cttc,vilalta2016end}. The fog computing system consists of the edge network as well as the cloud infrastructure. The integration of heterogeneous edge devices needs to be taken care of in the fog environment. Also, the cloud infrastructure should be properly handled in order to perform distributed computation ansd storage. Thus, there is a requirement of end-to-end orchestration of cloud servers and heterogeneous fog devices so that the resources can be allocated dynamically.

Based on these diverse challenges to develop an end-to-end fog framework, the researchers and industrial developers have targeted to solve various aspects of the fog computing framework. Accordingly, we classify the existing literature on fog computing, as discussed next.


\subsection{Taxonomy of Existing Works}

\begin{figure}[!t]
	\centering
	\includegraphics[width=0.5\textwidth]{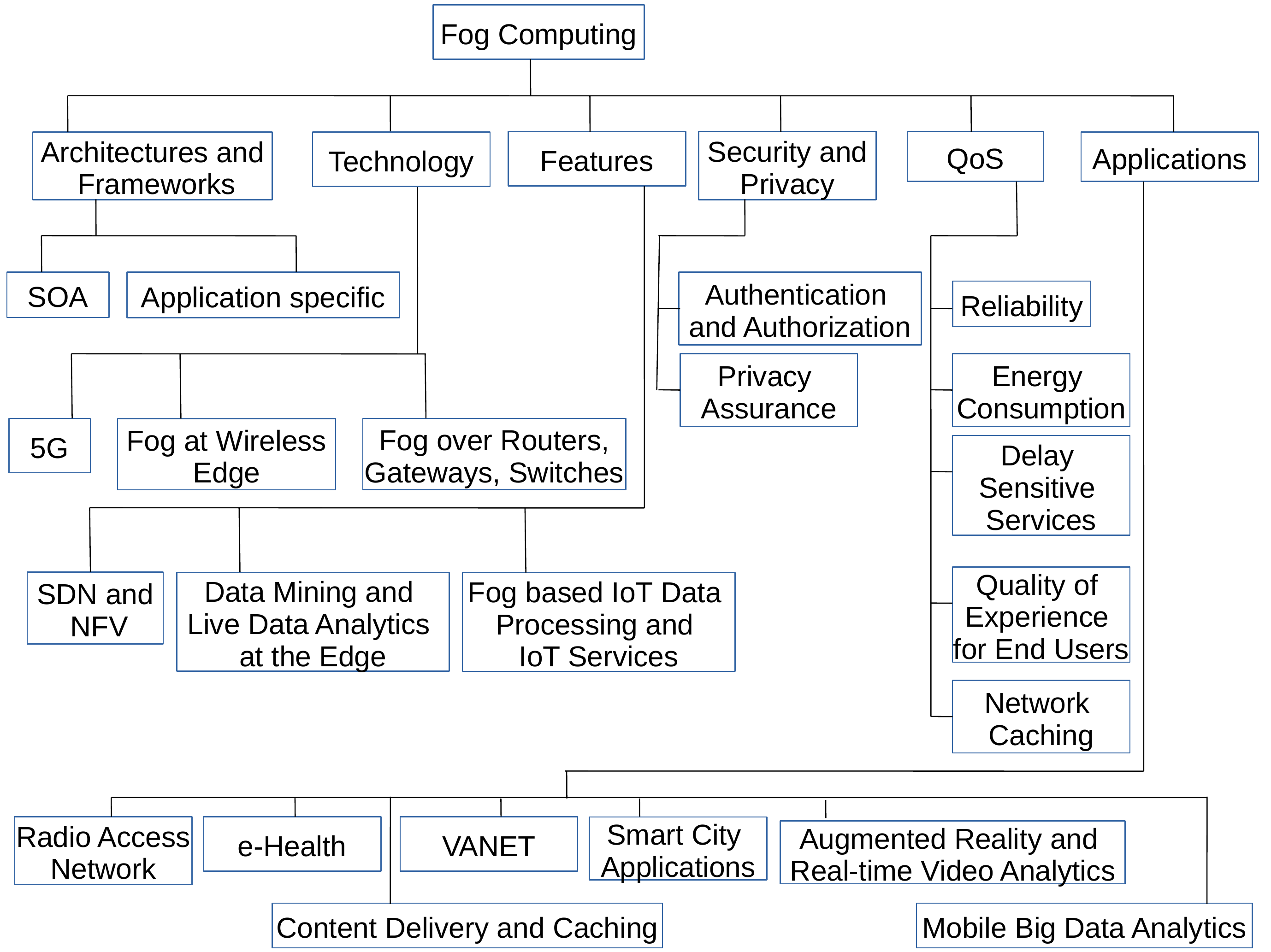}
	\caption{A Taxonomy of the Research Activities over the Fog Computing Framework}
	\label{fig:taxonomy}
\end{figure}

The discussion on various works on fog computing has been classified in this paper based on the thorough analysis of the existing works and their major contributions in this field. Fig.~\ref{fig:taxonomy} broadly shows the classification and taxonomy of fog computing based researches as per existing literature. We have classified the contributions as follows. First, we talked about various system level architectures and frameworks of fog computing exploiting the needs from the end users. Next, we talked about the technology aspects of fog computing. Then, we briefly touched upon various features, security and privacy, QoS and application domains developed over the fog domain. 


An extensive survey has been done in these aspects in order to have an understanding of the contributions of the existing literature. This thorough analysis gives us an in-depth insight of the existing developments of fog computing researches, which further helps us to extract the open gaps and limitations in these existing literature to put forward several open innovative research areas.  
  
%% file: 4_systems_of_fog_computing.tex

\section{Fog Computing Architectures and Frameworks}
\label{archi}
In this section, we broadly discuss about the ongoing research activities, which are being carried out to develop a computational architecture over fog. It can be noted that fog based systems are mainly inspired by the applications 
that drive a framework development based on their specific needs. Accordingly, we can observe two different architectural models which have been evolved over time. These involve -- (a) {\em service oriented architecture}, where the computing infrastructure is optimized based on the modularity of the end-to-end services, and (b) different {\em application specific architectures} as evolved from various IoT based platforms and their computation needs. Next, we discuss the research activities on these two directions in details. 

\subsection{Service Oriented Architectures}

\begin{figure}[!t]
	\centering
	\includegraphics[width=0.5\textwidth]{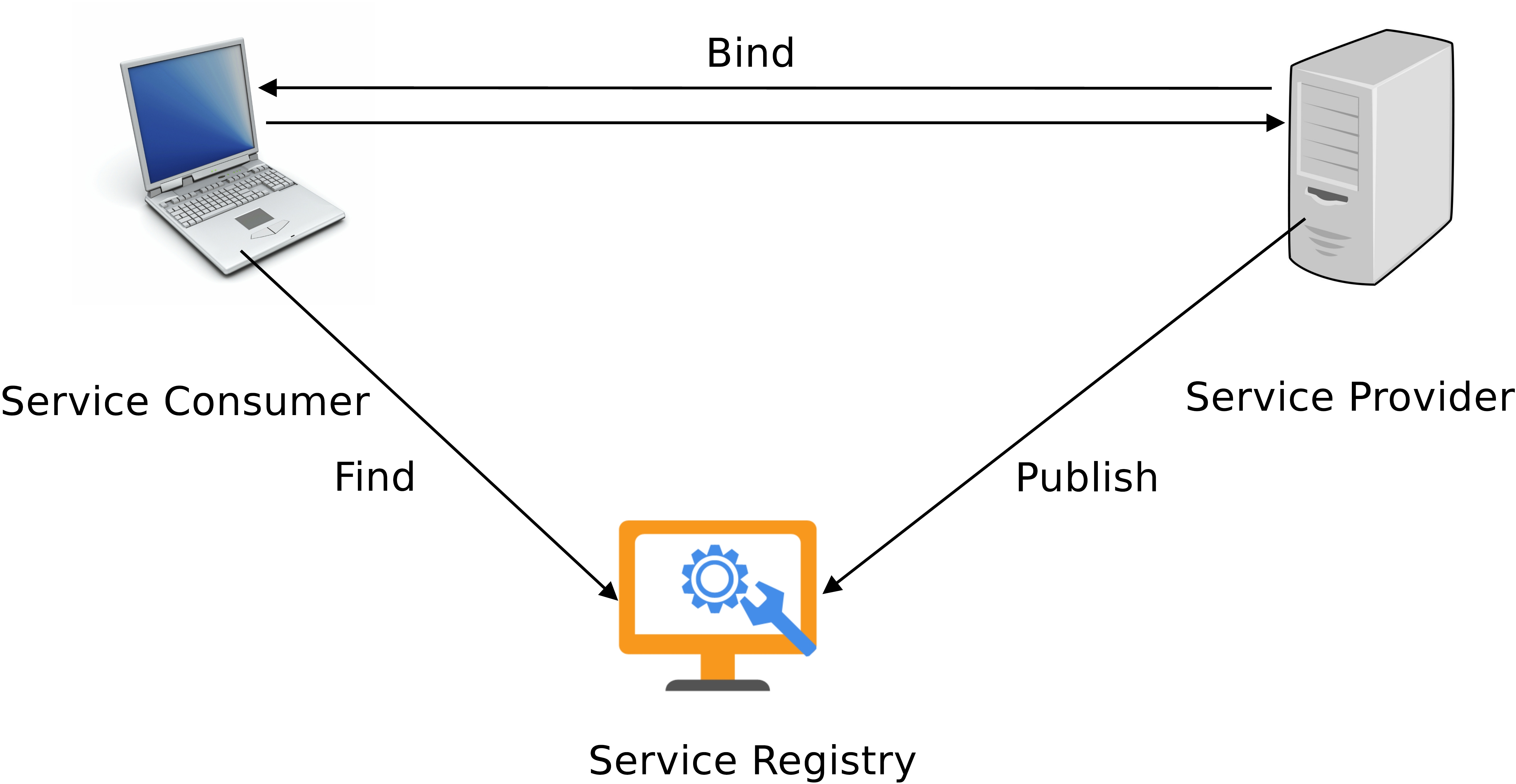}
	\caption{A Generic View of Service Oriented Architecture}
	\label{fig:soadiagram}
\end{figure}

Service orientation~\cite{perrey2003service,zhang2007service,erl2005service,krafzig2005enterprise} basically divides the application into a set of services or tasks. A sequential (or sometime parallel) execution of these tasks or services lead to the output of the application. In the era of cloud computing, the concept of service orientation became popular, where the computations to get the output of a user application was subdivided into multiple tasks, and these tasks were performed over the cloud~\cite{tsai2010service,wei2010service,tao2011cloud,duan2012survey,cellary2009government,chen2010robot,chauhan2011migrating,mohamed2016rsla,moghaddam2016policy}. With the widespread development of fog computing concepts and infrastructure, the philosophy of service oriented computation further got revolutionized. In the fog computing architecture, the services are hosted into different fog nodes as well as at the cloud nodes. The services are provided whenever the users need it. Service orientation over the fog environment is basically done by a middleware software which is capable of perfectly break a user's problem into several micro level services~\cite{issarny2011service}, \cite{teixeira2011service}. 
A fog based SOA provides the necessary business processing whenever there is a request from the end devices. SOAs over fog have three components: {\em consumers}, {\em producers} and {\em registry}. Any new services are registered in the registry by obtaining the service from the producer. The consumers are provided with the services from the available services registered in the registry. 
Fig.~\ref{fig:soadiagram} depicts a general framework for SOA over the fog nodes. In the diagram, we have service consumer, service provider and service registry. The service provider publishes the services in the service registry, and the service consumer finds the required services from the service registry. Service consumer and service provider communicate between themselves over the Internet. There are many applications of service oriented architecture over the fog computing framework~\cite{chen2017intelligent,al2016energy,liyanage2016mepaas,leitao2016industrial,zamfir2016towards,zhang2016infrastructure,zhang2016home,wu2016fog,kart2007distributed,prazeres2016soft,begum2016towards,butzin2016microservices}. For example, the service orientation can be used to facilitate the end to end quality of service in the healthcare systems. The authors in \cite{kart2007distributed} have discussed about a healthcare system based on service orientation. They have a clinic module and a pharmacy model. The division of the whole application work into several services has helped in designing the application tasks in a modular way, while ensuring interoperability between different platforms.

\subsubsection{How fog can help in SOA development} Fog computing helps to host the micro level services (also called micro-services) of a service oriented system over the edge or near-to-edge devices, which has its own benefits. Distributing the micro-level services over the fog nodes helps in providing security services, health monitoring, fitness monitoring, person tracking and so on. As the number of sensors mounted over the edge devices are increasing in a rapid pace, they generate large amount of heterogeneous data which is typically known as {\em Big data}. This big data demands proper data processing by mining and analytics in order to get to know about future insights~\cite{bellazzi2014big}. For example, in case of a smart hospital where the edge devices are the sensors mounted over the body of the patients, they provide the patients their required suggestions based on health conditions, while making a remote connection with the medical practitioners. In such a scenario, the micro-services, like heart beat monitoring, temperature monitoring, blood pressure monitoring, cholesterol monitoring etc. can be done at the edge devices. It can be noted that in this example, the overall service (that is monitoring the patient's condition) can be divided into such micro-services, which have their own computation models, and therefore fog based computing can be a good use-case in this scenario, where the output from the micro-services are only offloaded to the cloud rather than offloading the raw sensor data from each of the sensors, and thus reducing the amount of data transfer, response time and processing overhead at the cloud. Further, data privacy is also protected here because the private data from sensors is not offloaded to the third party sensors which have their own security implications as discussed on several existing literatures~\cite{li2010data,yi2016privacy,nissenbaum2016biosensing,zeadally2016security}. 
In summary, the advantages of SOA are as follows:- (i) Modularity, (ii) Distribution, (iii) Parallelism, (iv) Security, (v) Efficient resource usage.
A number of research works~\cite{chen2017intelligent,al2016energy,liyanage2016mepaas,leitao2016industrial,zamfir2016towards,zhang2016infrastructure,zhang2016home,wu2016fog,kart2007distributed,prazeres2016soft,begum2016towards,butzin2016microservices,dubey2015fog,dubey2015echowear,stantchev2015smart,aazam2015hamc,alazawi2011intelligent,cao2015fast} have explored fog based architecture for service oriented computing in various directions. Next, we discuss those in details.

\subsubsection{Service oriented fog architecture for pervasive health-care}
\begin{figure}[!t]
	\centering
	\includegraphics[width=0.5\textwidth]{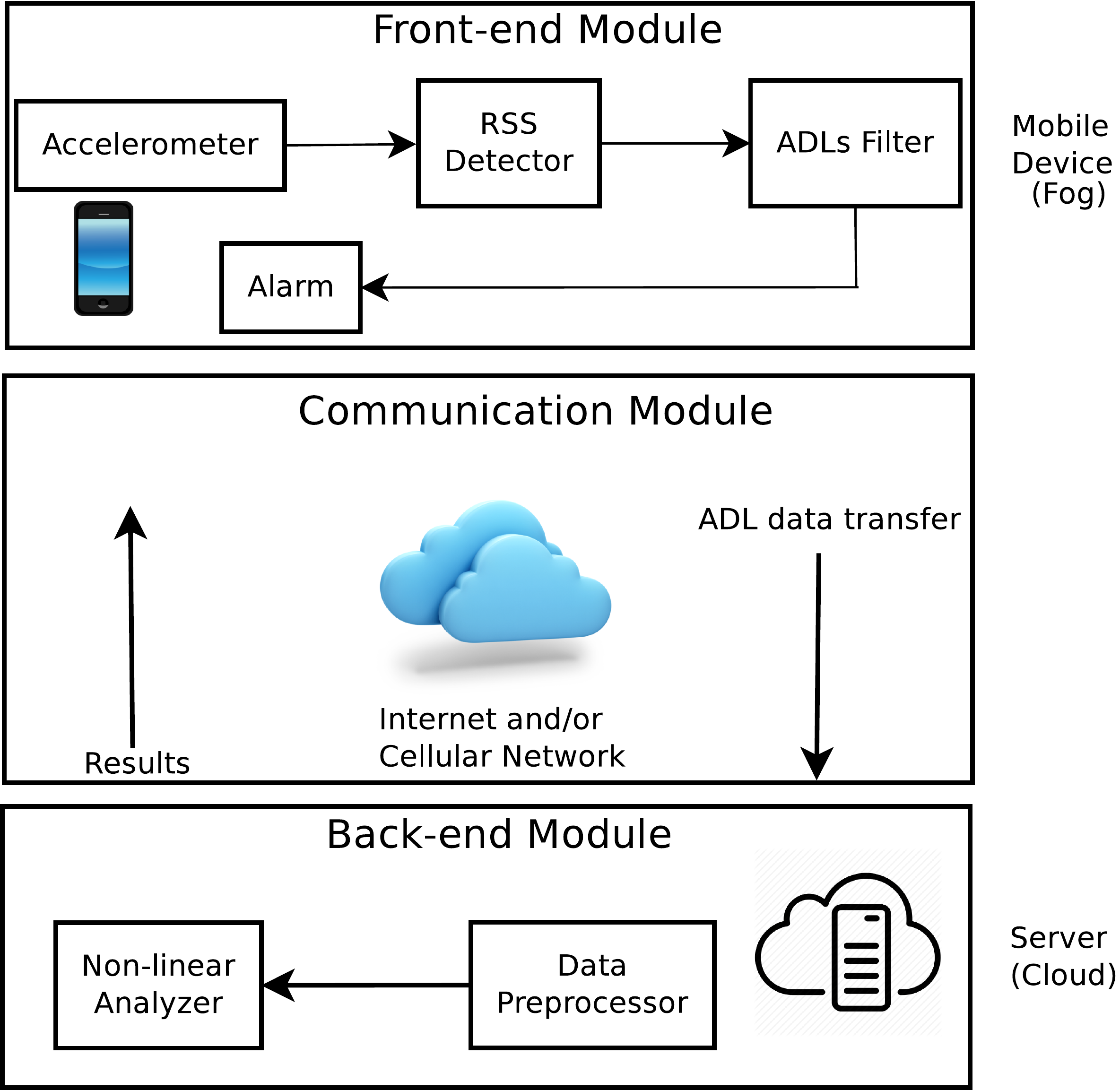}
	\caption{U-Fall: A Fall Detection System based on Fog Computing~\cite{cao2015distributed}}
	\label{fig:ufall}
\end{figure}

Fog computing based service oriented architecture has been able to efficiently process medical data in order to provide ubiquitous health-care to the patients. The service oriented fog architecture divides the large end service into many micro-services in order to have a modular structure so that some of the computations can be solved at fog level and some computations can be offloaded to remote cloud servers. In the following discussions, we present the works leveraging the service oriented fog architecture.
In \cite{cao2015distributed}, the main objective of the work is to develop a smart-device based, real-time system for use by stroke patients based on service oriented fog computing framework. Fig.~\ref{fig:ufall} shows the proposed architecture, where the front end is an app running on the edge devices such as smartphones, and the back end is the cloud server. Apart from that, the system has a communication module which provides channels for communication between the front end and the back end. The fog application running at the smartphones collect the accelerometer data and use the root-sum-of-square (RSS) of acceleration magnitude followed by an activities of daily living (ADL) filter to generate an alarm for fall like events. This alarm information is forwarded to the cloud via the communication channel, where the cloud executes a classification based learning mechanism to detect the actual fall from the fall like events. Here the fall detection is divided into four micro-services, like (i) RSS computation, (ii) ADL filtering for fall like event detection, (iii) data preprocessing and (iv) classification module. The first two services have been executed at the fog devices, whereas the final two services, which are indeed resource consuming, have been executed over the cloud to get the result for the U-Fall application. A similar work on fog based fall detection mechanism has been discussed in~\cite{cao2015fast}. There are multiple other works that use a service oriented fog architecture for pervasive health-care applications, such as~\cite{zamfir2016towards,gia2015fog,zao2014pervasive,cao2015distributed,aazam2015hamc,ahmad2016health} and the references therein. 

%

\subsubsection{Fog architecture for clinical data processing based on service orientation} The clinical data processing based on service orientation has been performed in the context of fog. The edge devices act as smart clinical data processing unit in order to improve the response time. Here, we discuss about how fog can be implemented for service oriented clinical data processing. In \cite{dubey2015fog}, the authors have proposed a SOA for fog computing over body sensor networks (BSN). They term this architecture as {\em Fog Data}, which has three components: the BSN, a fog gateway and a cloud server. It aims to reduce the data storage and data transfer  costs, as well as tries to minimize the overall power consumption with high efficiency. {\em Fog Data} architecture has the advantage of carrying out analytics at the edge devices mounted over the BSN, so that less amount of data is stored and transferred to the cloud server. The fog gateway works as a smart gateway, which does the following works.
\begin{enumerate}
	\item[(i)] It connects the sensor nodes with the Internet to transfer the sensed data to the cloud and also does the proper actuation.
	\item[(ii)] It processes the incoming data and sends the relevant data to the cloud for further analysis.
	\item[(iii)] It creates a local database containing the patient's features.
	\item[(iv)] It also allows the users to incorporate the security layer for data protection.
\end{enumerate}
While the BSN is used for data gathering from the patient's body, the fog gateway does the initial data processing, and the cloud server stores the obtained data for further analytics. Here, the fog gateway also minimizes the data to be stored on the cloud server as well as the bandwidth needed to send the clinically vital data to the cloud. It is interesting to observe that the {\em Fog Data} architecture proposed by the authors has a generic implication which is also suitable for different other types of sensors and applications. The fog nodes (gateways) basically work a middleware between the BSN and the cloud to do intermediate data processing. 

In the paper, the authors have validated their proposed SOA with two use cases -- speech monitoring of the patients with Parkinson's disease (PD) and Electrocardiogram (ECG) monitoring. They have been able to reduce the data by feature extraction, pattern mining and compression at the fog gateways. The fog computer extracted the QRS complex from ECG signals using real time processing on Intel Edison. They have also shown that intermediate data compression at the fog devices can reduce ECG data by more than $98\%$ in most of the cases. This work can be extended further to get speech features like shimmer, jitter etc. at the intermediate fog nodes for speech disorder recognition. {\em Fog Data} architecture can also be used to validate the Echo Wear, a smart-watch technology for voice and speech treatments, which has been discussed in a follow-up work~\cite{dubey2015echowear} by the same authors. 

\subsubsection{Human activity monitoring based on service orientation} 
\begin{figure}[!t]
	\centering
	\includegraphics[width=0.5\textwidth]{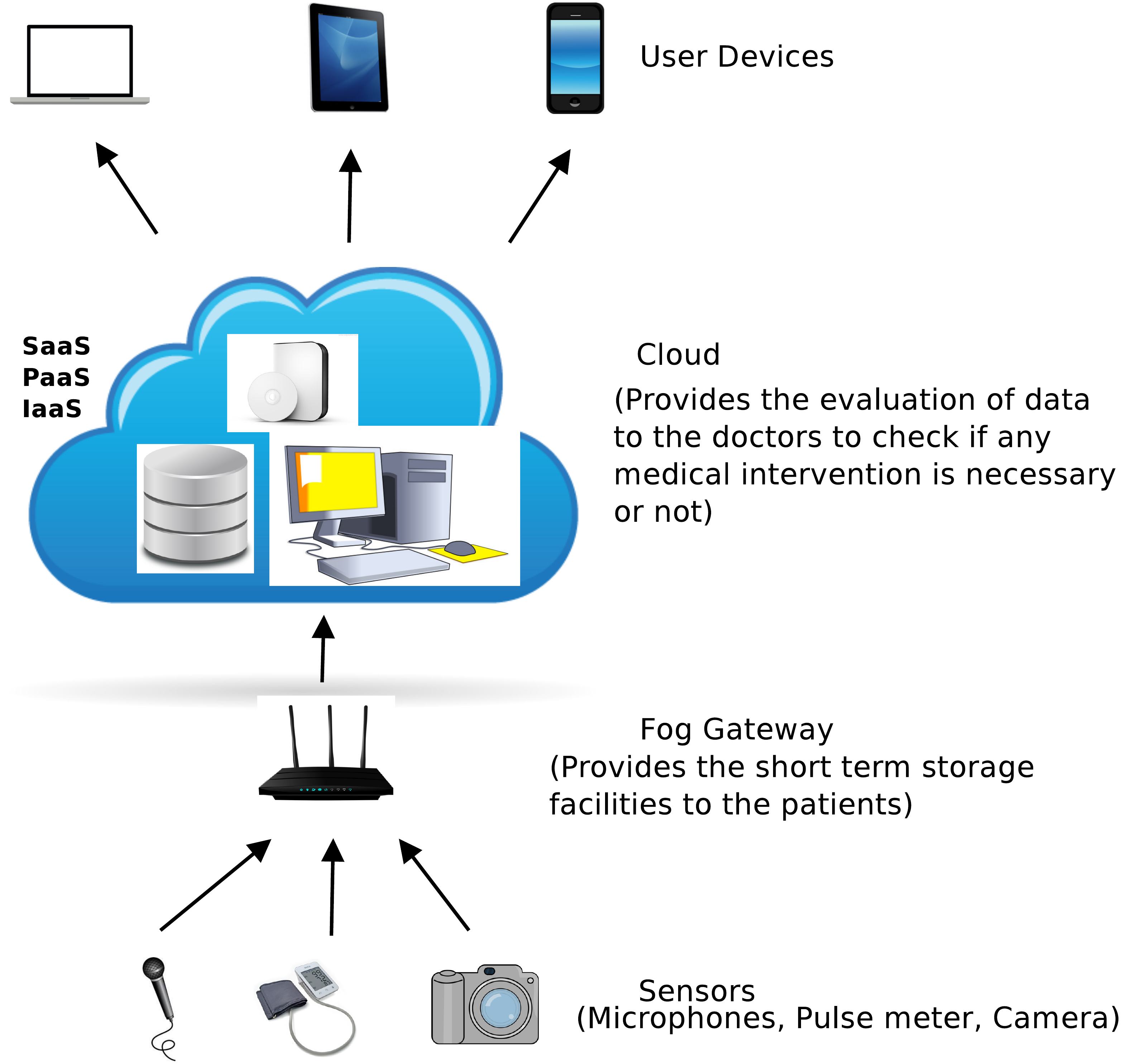}
	\caption{Combined Fog-Cloud Architecture for Activity and Health Monitoring\cite{stantchev2015smart} 
	}
	\label{fig:fog-cloud}
\end{figure}

In the existing literature, the fog based service oriented architecture has been studied to monitor human activity and health. Now, we discuss about these works in the following. In \cite{stantchev2015smart}, the authors have developed a fog-cloud integrated SOA offering SaaS, PaaS and IaaS for human activity and health monitoring, which is based on their previous work on developing a service oriented system for activity and health monitoring~\cite{stantchev2006architectural}. The router or the gateway collects different sensor data and sends it to the cloud. The end users get software, server and also hardware as services for monitoring, processing and reporting the information generated from the sensed data. The overall architectural design of this combined fog-cloud architecture has been shown in fig.~\ref{fig:fog-cloud}. Different sensors, such as microphone, pulse meter, camera etc. send data to the gateway. Then, the gateway performs the short term storage of the data and it is connected to the cloud server. The fog level gateway provides short-term storage facilities to the patients. The cloud server provides SaaS, PaaS and IaaS facilities to the users. The work of the cloud layer is to facilitate the doctors to check the data in order to verify if any medical intervention is necessary or not. This work has shown a nice integration of cloud computing with fog computing technologies in order to provide an architecture suitable for activity and health monitoring applications. The nodes can interact by wired as well as wireless channel. The authors have given a process oriented view, where the sensors remain active, notice changes in the human activities and also monitor their health conditions. This integration has been able to provide better QoS as well as security of the services. 

Table~\ref{table:fogarchihealth} gives a summary of the works that utilize fog based SOA for health-care applications. In the table, we have shown various micro-services that can be executed sequentially or in parallel to achieve the end goals. Additionally, the table indicates the services and architectural components at the fog devices and at the cloud.

\begin{table*}[!ht]
	\centering
	\caption{Comparison of different fog computing based Service Oriented Architecture for Healthcare}
	\begin{tabular}{|>{\raggedright}p{2cm} ||>{\raggedright}p{2.5cm} |>{\raggedright}p{2.5cm} |>{\raggedright}p{2.5cm} |>{\raggedright}p{2.5cm} |>{\raggedright\arraybackslash}p{2.5cm} |} 
		\hline
		{\bf Literature} & {\bf Application Area} & {\bf End Service} & {\bf Micro-services} & {\bf Services at the Fog} & {\bf Services at the Cloud} \\ \hline \hline 		
		Cao \etal \cite{cao2015distributed} & Healthcare  & Fall detection & RSS computation, ADL
		filtering for fall like event detection, Data preprocessing, Classification module 
		& RSS computation, ADL filtering & Data preprocessing, Classification module \\ \hline
		Dubey \etal \cite{dubey2015fog} & Healthcare  & Clinical data processing & Feature extraction, Pattern mining, Compression, Signal processing pipeline, Onsite database, Communication control & Feature extraction and Compression & Pattern mining, Online database and communication control\\ \hline
		Stantchev \etal \cite{stantchev2015smart} & Healthcare  & Activity and health monitoring & Data storage, Data evaluation to check if any medical help is needed & Data storage & Data evaluation to check if any medical help is needed \\ \hline
		Zamfir \etal~\cite{zamfir2016towards} & Healthcare & IoT based pervasive health monitoring & Sensor data collection and fusion, data analytics & Data fusion and local processing & Large scale data analytics \\ \hline
		Gia \etal~\cite{gia2015fog} & Healthcare & ECG feature extraction & Embedded data mining, distributed storage, notification service, real-time processing, visualization, and diagnosis & Feature extraction including heart rate, P wave and T wave via a flexible template, based on a lightweight wavelet transform mechanism & Storage and future data analytics \\ \hline
		Zao \etal~\cite{zao2014pervasive} & Healthcase & Pervasive brain monitoring & Brain computer data pre-processing and compression, Massive parallel data processing etc. & Brain computer data pre-processing and compression, real time data processing, data caching, computation offloading & Massive parallel data processing \\ 
		\hline 
		Ahmad \etal~\cite{ahmad2016health} & Healthcare & Personalized health recommendations & Data curing services, intermediate data generator, data analyze, cloud access security broker (Health data sharing guidelines, data governance -- access policies, analytics on data visibility, data encryption)  & Data curing, intermediate data generator and analyze & Cloud access security broker \\ 
		\hline 
	\end{tabular}	
	\label{table:fogarchihealth}
\end{table*}

\subsubsection{Service orientation in fog for emergency control} 
\begin{figure}[!t]
	\centering
	\includegraphics[width=0.5\textwidth]{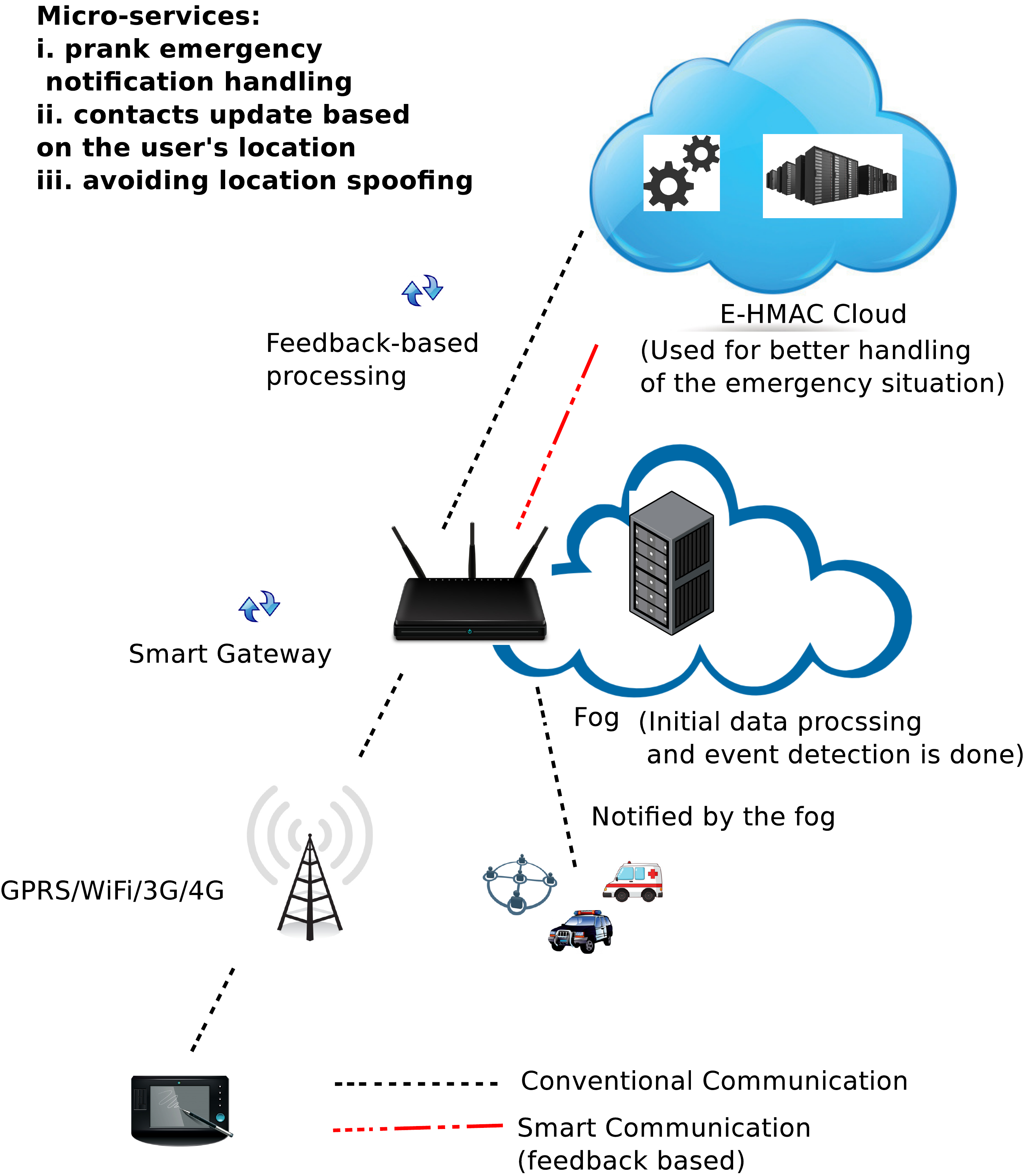}
	\caption{Smart Gateway based Fog Architecture for Emergency Control\cite{aazam2015hamc}}
	\label{fig:emergency}
\end{figure}

Fog computing can be leveraged to tackle emergency situation that requires real-time response in order to take proper action quickly. In this direction, the following works have been done in the context of service oriented fog architecture. In \cite{aazam2015hamc}, the authors have used a smart gateway, called a {\em micro data center} (MDC), which brings the cloud storage and processing services closer to the customers. The smart gateway acts as a fog node here. This architecture helps to notify others about an emergency situation very quickly. Fig.~\ref{fig:emergency} shows the architectural components of this system. The smart devices are connected to the cellular network, WiFi access point etc. These access points are in turn connected to the smart gateway that acts as a fog computing layer. Fog computing edge devices communicate with the cloud for further data processing.

In \cite{aazam2015hamc}, the authors have tackled emergency situations. The emergency situations require to contact the proper disaster management department in order to make the situation normal. It can be noted here that there have been some previous works on emergency management based on mobile cloud computing based platforms, like~\cite{greer2012personal,hasegawa2008emergency,ribault2012simulation}. However, no prior work has developed a proper emergency notification mechanism that would be fast as well as efficient at the same time~\cite{zhou2011emergency,thakur2010shield}. In the paper \cite{aazam2015hamc}, a service architecture for emergency alert and management has been proposed using a combination of fog and cloud computing. The mobile application {\em Emergency Help Alert Mobile Cloud} (E-HAMC) can be activated by pressing only a single button. 
The gathered data is sent to the fog nodes i.e. the smart gateway for initial processing and event detection. After that, the data is pre-processed and sent to the cloud for better handling of an emergency situation in future. More specifically, the micro services are as follows.
\begin{enumerate}
	\item[(i)] Prank emergency notification handling: The victim sends the picture of the emergency situation which is sent to the fog by the application. If the victim is not in any situation to send the picture, any passer-by can do that.
	\item[(ii)] Contacts update based on the user's location: In the event of the change of city or country of the user, the application contacts the cloud server to synchronize the user's contact list with the available departments dealing with different disasters.
	\item[(iii)] Avoiding location spoofing: When the victim sends emergency motification, the location details are taken from the connected global positioning system (GPS) or base tranceiver station (BTS) of the user's mobile device. The location spoofing is avoided by the application in this way.
\end{enumerate}
The authors have observed that the latency was around $6$ times less by using fog than when data is to be directly sent to the cloud server. The application is able to contact the concerned department about the emergency situation. E-HMAC also has the feature of sending message to already stored contact numbers. Their prime aim is to avoid the complexity and the latency in case of an emergency situation. 

\subsubsection{Summary of works on fog based SOAs} 
Apart from the above works, there are multiple other domains where fog computing based SOAs have been used for system and application development. For example, fog based service orientation has been explored over various applications like cloud based robotics~\cite{begum2016towards}, where the micro-services are executed at the robots; whereas data processing and intelligence extraction is done over the cloud, development of self-organization controls~\cite{prazeres2016soft}, cyber-manufacturing systems~\cite{wu2016fog}, smart home applications~\cite{zhang2016home} to bring back local services at the home gateways from the cloud server, infrastructure management over optical network devices through micro data-centers~\cite{zhang2016infrastructure}, and so on. The service orientation is essential in the context of providing real-time latency sensitive services to the end-users in fog computing domain. Based on the above discussion, we can say that the fog nodes are implemented to have the services deployed in them. Also, some of the complex services are deployed in the cloud nodes. The edge devices do the proper inter-play with the cloud servers to execute these complex micro services. 
So, the challenge is to separate the micro services so that they can be deployed in the proper devices i.e. in fog or in cloud device. Table~\ref{table:fogarchiothers} summarizes various other works that uses fog based SOA in different domains of computing.

\begin{table*}[!t]
	\centering
	\caption{Comparison of different fog computing based Service Oriented Architecture for Various Applications}
	\begin{tabular}{|>{\raggedright}p{2cm} ||>{\raggedright}p{2.5cm} |>{\raggedright}p{2.5cm} |>{\raggedright}p{2.5cm} |>{\raggedright}p{2.5cm} |>{\raggedright\arraybackslash}p{2.5cm} |} 
		\hline
		{\bf Literature} & {\bf Application Area} & {\bf End Service} & {\bf Micro-services} & {\bf Fog Devices and Services} & {\bf Services at the Cloud} \\ \hline \hline 	
		Aazam \etal \cite{aazam2015hamc} & Emergency control  & Emergency notification  & Prank emergency notification handling, Contacts update based on the user’s location, Avoiding location spoofing & Initial data processing and event detection & Future handling of the emergency situation\\ \hline
		Begum \etal \cite{begum2016towards} & Real time Robotics & 3D environment service and motion control service & Intelligent sensor data fusion, recognition of environment features, sensing of the 3D environment, collision free motion of robotic arms & Sensor data fusion and recognition of environment features, motion control of robotic arms & 3D environment construction from the sensed data and collaborative planning of robotic arm movement \\ \hline 
		Prazeres \etal~\cite{prazeres2016soft} & Self-organizing Control & Message service oriented middleware for fog computing & Self-organizing monitoring (localization, discovery, composition, security etc.), failure recovery, gateway deployment, management and balancing of profiles, & Monitoring, failure recovery, gateway deployment & Management and balancing of profiles \\ \hline 
		Wu \etal~\cite{wu2016fog} & Cyber-manufacturing System & High performance computing and real-time stream analytics & Automatic fault and failure detection, self-diagnosis, and preventative maintenance scheduling & Local sensing and self-diagnosis, local failure detection & Large scale data analysis for root cause analytics of a failure \\ \hline 
		Zhang \etal~\cite{zhang2016home} & Smart Home & Home automation & Data collection and summarization from home automation and personal wearable devices, incoming data processing (ventilation control, lighting control etc.), reporting service, entertainment service, back-up and recovery, analytics and proxy & Low level services and control like ventilation control, lighting control etc. & Analytics and reporting service, backup and recovery, proxy services \\ \hline 
		Zhang \etal \cite{zhang2016infrastructure} & Optical Networks & Infrastructure management & Distributed micro data-center (MicroDC) to support delay-sensitive bandwidth-intensive residential, enterprise, and wireless backhaul services -- an optimization framework is developed to handle this & MicroDC services - an optimization framework for balancing  deployment cost, power awareness and optical link degradation factors & Network backhaul management \\ \hline 
	\end{tabular}	
	\label{table:fogarchiothers}
\end{table*}

\subsection{Application Specific Architectures}

Apart from the service oriented view of fog computing architecture, application specific fog infrastructure developments have been also evolved over the past couple of years. There architectures primarily focus on the application requirements, and decide which part of the computation can be executed over the edge or near-edge devices, and which part need to be executed over the cloud. The fog-cloud orchestration has been explored in these works to improve application programmability, to reduce computation delay and overhead, and to improve energy efficiency. Just like SOA based fog paradigm, these works also extend towards various application domains, starting from health-care to intelligent transportation system, smart cities and other smart environments, security monitoring and so on. Here we give a summary of these fog architectures with a focus on how the edge devices and cloud interconnect themselves through the fog middleware.  

\subsubsection{Fog Frameworks for Pervasive Health-care} Pervasive health-care monitoring applications are motivated by data. The huge amount of data generated in the context of IoT devices have created enormous opportunities by creating smart applications, which in turn reduces costs. In the field of IoT based health-care, there is a huge amount of big data that is continuously been generated from various body mounted and environment monitoring sensors. The global size of big data in health-care sector was roughly $200$ Exabytes in $2012$~\cite{chen2014data}. Motivated by this, a number of fog computing infrastructures have been focused on efficiently managing health-care data through the fog computing environments~\cite{dubey2015fog,stantchev2015smart,shi2015fog,cao2015distributed,monteiro2016fit,rahmani2015smart,nandyala2016cloud,gu2015cost,moosavi2016end,lo2016mobile,fratu2015fog,andriopoulou2017integrating,rahmani2017exploiting,chakraborty2016fog}. In~\cite{nandyala2016cloud}, the authors discuss a basic IoT based health-care monitoring system for homes and hospitals, where the fog nodes run IoT enabled real time monitoring and forward the data to IoT gateways. The IoT gateways summarize the data and forward the data summary to the cloud for further analytics. In contrary to the SOA, here the fog devices (IoT gateways) do not execute any specific services, but run a basic data summarization module to reduce the network overhead, transportation latency and duplicate data delivery. Gu {\textit et al.}~\cite{gu2015cost} have developed a medical cyber-physical system (MCPS) based on the fog computing framework, where a optimization framework is used to decide and minimize the cost for offloading data summarization activities to the fog devices, while to maximize computation efficiency. In~\cite{lo2016mobile}, the authors have developed a big-data analytics platform for medical data analytics over the fog computing framework. They developed a cloudlet-based mobile cloud-computing infrastructure for health-care big data applications, where the decision problem is to decide the amount of computation to be offloaded to the cloudlets. Andriopoulou \textit{et al.}~\cite{andriopoulou2017integrating} have designed a IoT based fog computing framework where part of the computation intelligence has transferred to the edge devices from the cloud. A recent work by Chakraborty \textit{et al.}~\cite{chakraborty2016fog} has developed an experimental framework to illustrate the utility of fog computing for  time-sensitive medical applications. In their architecture, the fog nodes maintain a list of the critical threshold values for various medical sensor data. If the incoming value crosses that threshold, then the data is offloaded to the cloud. Further in their architecture, the data is also intermediary stored at the fog devices for few minutes, in case the doctor wants for a quick short-term report. In a nutshell, fog computing has become an enabler for IoT based medical data processing, where both the SOA based intelligence computation as well as application specific medical data summarization tasks are performed with higher data accuracy and data consistency compared to the traditional cloud based framework. 

In table~\ref{table:fogarchipervasive}, we have summarized the fog computing frameworks for pervasive health-care systems. In the table, we have compared different works based on the architecture, services, communication between fog devices and the role of fog nodes in pervasive healthcare domain.    

\begin{table*}[!t]
	\centering
	\caption{Comparison of Different Fog Architectures for Pervasive Health-care Systems}
	\begin{tabular}{|>{\raggedright}p{2cm}||>{\raggedright}p{2cm} |>{\raggedright}p{2cm}|>{\raggedright}p{2cm}|>{\raggedright}p{2cm}|>{\raggedright}p{2cm}|>{\raggedright\arraybackslash}p{2cm}|} 
		\hline
		{\bf Research work} & {\bf Architecture} & {\bf Services} & {\bf Fog Devices}  & {\bf Communication between Fog Devices} & {\bf Application} & {\bf Advantages of Using Fog}\\
		\hline \hline 
		Nandyala \etal ~\cite{nandyala2016cloud} & Four tiers: Smart Devices or Things Network, Gateway(Fog),  Core and Cloud & Healthcare monitoring & Gateway  & Wired as well as wireless & Patient data sensing, data transmission and notifying medical staff and family members  & Low latency, improved scalability, reliability, flexible processing \\ \hline
		Gu  \etal ~\cite{gu2015cost}  &  System is modeled as an undirected graph, an MINLP formulation on the minimum cost problem is done & Cyber-Physical System for healthcare & Base stations  & Wireless & Patient monitoring  & Leveraging fog computing, a cost-effective system is designed \\ \hline
		Lo’ai A \etal ~\cite{lo2016mobile} & A cloudlet based mobile cloud computing infrastructure  & Healthcare services  & Gateways  & Wireless & Big data analysis for healthcare applications & By using fog real-time analysis of patient records was done\\ \hline
		Chakraborty \textit{et al.}~\cite{chakraborty2016fog} & Four layers: Client nodes, Data generator node, Fog node and Cloud node  & Patient monitoring &  Access points & Wireless & Time critical medical data analysis & Low and predictable latency \\ \hline
		Cao \textit{et al.}~\cite{cao2015distributed} & Three modules: Mobile device, Communication module, Cloud server  & Real-time fall detection & Mobile devices & Wireless & Threshold based fall detection, Detecting false alarms &  Less Response time, less energy consumption and fall detection accuracy \\ \hline
	\end{tabular}	
	\label{table:fogarchipervasive}
\end{table*}

\subsubsection{Fog Computing Architecture for Intelligent Transportation Systems and Vehicular Technologies} Another important area, where the researches have concentrated to building up interesting fog computing architectures and frameworks, is the domain of intelligent transportation system (ITS) and vehicular technologies~\cite{hou2016vehicular,bitam2015vanet,chen2017exploring,park2017network,malandrino2016price,kim2015shared,zhang2016social,salonikias2015access}. 
As a consequence, the concept of vehicular fog computing~\cite{hou2016vehicular} has been emerged recently. Vehicular fog computing has additional challenges over the basic fog computing framework, as follows. 
\begin{enumerate}
	\item[(i)] The edge nodes are not only mobile, but sometimes the mobility can be highly dynamic with high speed. Therefore, the underlying network is also dynamic, and sometime intermittent -- it may not be possible to always set up a connection to the remote cloud server. 
	\item[(ii)] The computation requirements are based on vehicular control engines, and therefore accuracy and safety criticality need to be ensured. 
	\item[(iii)] Because of the possibility of having intermittent connectivity with the remote cloud server, access control becomes an issue for vehicular fog computing environments~\cite{salonikias2015access}. However the delay in access control decisions should not affect the safety criticality of the local computation and decisions.
	\item[(iv)] In a vehicular environment, failure or sporadic behaviors of a few sensor nodes may affect the control decisions taken over a fog. Although such anomalous behaviors of sensors are easy to find out over the cloud, they may not be so easy over a fog computing environment. Therefore, ensuring correctness of the local computation is a challenge which needs to be ensured for intelligent or autonomous vehicles.  
\end{enumerate}

In \cite{salonikias2015access}, a ITS based fog architecture has been proposed, which has four areas -- {\em Core information and communication technologies} (CI), {\em Road Side} (RS), {\em Vehicles and Humans} (VH), and {\em Sensors and Actuators} (SA). Here CI acts as a cloud service provider while RS and VH are working as the fog nodes. SA can be wireless sensor network, smart signs, smart traffic lights etc. 


CI follows the SaaS model of cloud computing and is responsible for giving application services. Content and infrastructure management along with data processing resources and data warehousing are part of the CI level. Consumers are present in the VH area, and they are basically vehicles having humans carrying smart devices (like smartphones, tablets etc.) connected to the ITS network. Fog computing is utilized in this work in order to provide low latency and specific services like location aware services. By incorporating fog computing in this scenario, we can have several advantages. As an example in the case of a service like providing alternative routes, the route data would be uploaded to the nearby fog servers. This would make the cloud server less congested. Contextual information like traffic conditions are known to the fog nodes by this operation. So, the total time to upload, process and download rerouting instructions to vehicles would be very less.

\begin{figure}[!t]
	\centering
	\includegraphics[width=0.5\textwidth]{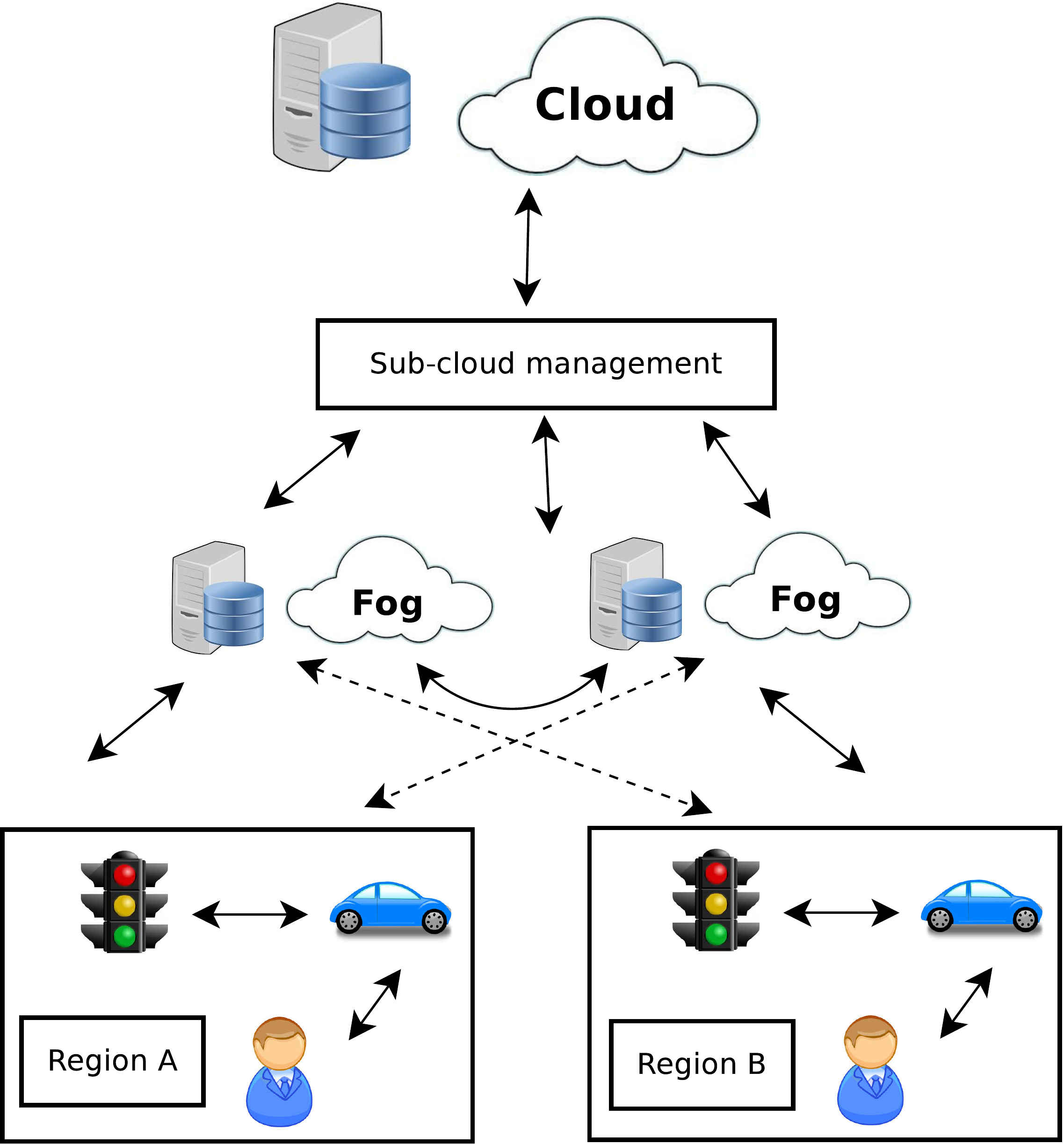}
	\caption{Usage of Fog Computing in Social Vehicle Swarms \cite{zhang2016social}}
	\label{fig:socialvehicleswarms}
\end{figure}

In another work, Zhang \etal \cite{zhang2016social} have discussed about social vehicle swarms in order to study and analyze a socially aware vehicular network. An agent based model has been used to find the hidden patterns. The authors have also introduced supportive technology and methods, deep reinforcement learning, data mining and fog computing i.e. sub-cloud computing to improve living conditions and quality of experience. Due to the advantage of providing rapid response, fog computing is more flexible in social vehicle swarms than cloud computing. In this scenario, the fog computing is done close to all the vehicular agents. Fig.~\ref{fig:socialvehicleswarms} depicts the proposed model. Fog computing allocates data to collectors according to their preferences, increasing the chance of real-time communication. The sub-cloud management layer does the communications between the fog and the cloud. The fog nodes store and process data related only to an event with low importance, whereas the difficult problems are transmitted to the cloud. 

Table~\ref{table:fogarchivehicle} summarizes the fog computing architectures for enabling low response time computation, local data storage and caching, security and privacy over intelligent vehicles and vehicular communications. In the table, we give a brief summary of various works in this direction with a highlight on how fog computing is utilized for providing various services over different interesting applications. 

\begin{table*}[!t]
	\centering
	\caption{Comparison of Different Fog Architectures for Intelligent Transportation Systems}
	\begin{tabular}{|>{\raggedright}p{2cm}||>{\raggedright}p{2cm} |>{\raggedright}p{2cm}|>{\raggedright}p{2cm}|>{\raggedright}p{2cm}|>{\raggedright}p{2cm}|>{\raggedright\arraybackslash}p{2cm}|} 
		\hline
		{\bf Research work} & {\bf Architecture} & {\bf Services} & {\bf Fog Devices}  & {\bf Communication between Fog Devices} & {\bf Application} & {\bf Advantages of Using Fog}\\
		\hline \hline 
		Salonikias \etal \cite{salonikias2015access} & Three layer architecture -- core information and communication technologies (Cloud), vehicles and humans (Fog) and sensors-actuators  & Attribute based access control & Vehicles, Personal devices like smart-phones and tablets, Roadside units  & Wireless at the edge, wired between roadside units & Authentication and access control for vehicular clients & Low latency, less congestion at cloud, local data storage at roadside units\\ \hline
		Hou \etal~\cite{hou2016vehicular} & Vehicular ad hoc network and vehicle to infrastructure  & Parking behavior analysis & vehicles, roadside units, base stations & Wireless & Capacity planning of parking areas & Explores the impact of fog-cloud interaction on the vehicles \\ \hline
		Bitam \etal~\cite{bitam2015vanet} & Vehicular ad hoc network & Vehicle management & Vehicles  & Service access points (wireless) & Safety applications, vehicular software, web services, business applications & A common cloud architecture spanned over vehicular nodes \\ \hline
		Chen \etal~\cite{chen2017exploring} & Three layer vehicular cloud network -- Central cloud layer, road-side cloud layer and vehicular cloud layer & Vehicular data scheduling & Vehicles and roadside units & Wireless at edge and wired at backbone & Vehicular communication & Reduction in network load, less response time \\ \hline   
		Park \etal~\cite{park2017network} & Connected vehicles - software defined network (controller) connects the vehicles &  Intelligence in vehicular network for handling connection failures & Vehicles & Wireless & Reliable communication & Explore mobility for handling link failures and management of network state information\\ \hline  
		Malandrino \etal~\cite{malandrino2016price} & Vehicle to infrastructure (V2I) networks & Data caching at fog nodes & Base stations and aggregation switches & Wired backbone & Mobile-edge caching & Reduction in response time for vehicular services \\ \hline  
		Zhang \etal~\cite{zhang2016social} & Vehicle swarms -- vehicles which are socially connected (say, the vehicles of a hotel) & Privacy preserving data mining over socially connected vehicular swarm & Vehicles & Wireless & Information extraction and quality of experience (QoE) assurance for socially connected vehicular swarms & Ensures security and privacy - computation is confined within the local fog nodes (vehicles) \\ \hline  
	\end{tabular}	
	\label{table:fogarchivehicle}
\end{table*}

\subsubsection{Context Aware Computing at the Fog}
Context aware computing is an essential feature of a system. Context can be defined as any information that can be used to characterize the situation of an entity or environment. A system is context aware if it uses a context to provide relevant services to the user \cite{abowd1999towards}. In case of fog based ITS, personal information (i.e. current location) is transferred, processed and stored in the fog and the cloud for context-aware services. This creates the privacy concerns to the users, which fog computing can take care of by ensuring not to reveal the person's identity to the cloud.

\begin{figure}[!t]
	\centering
	\includegraphics[width=0.5\textwidth]{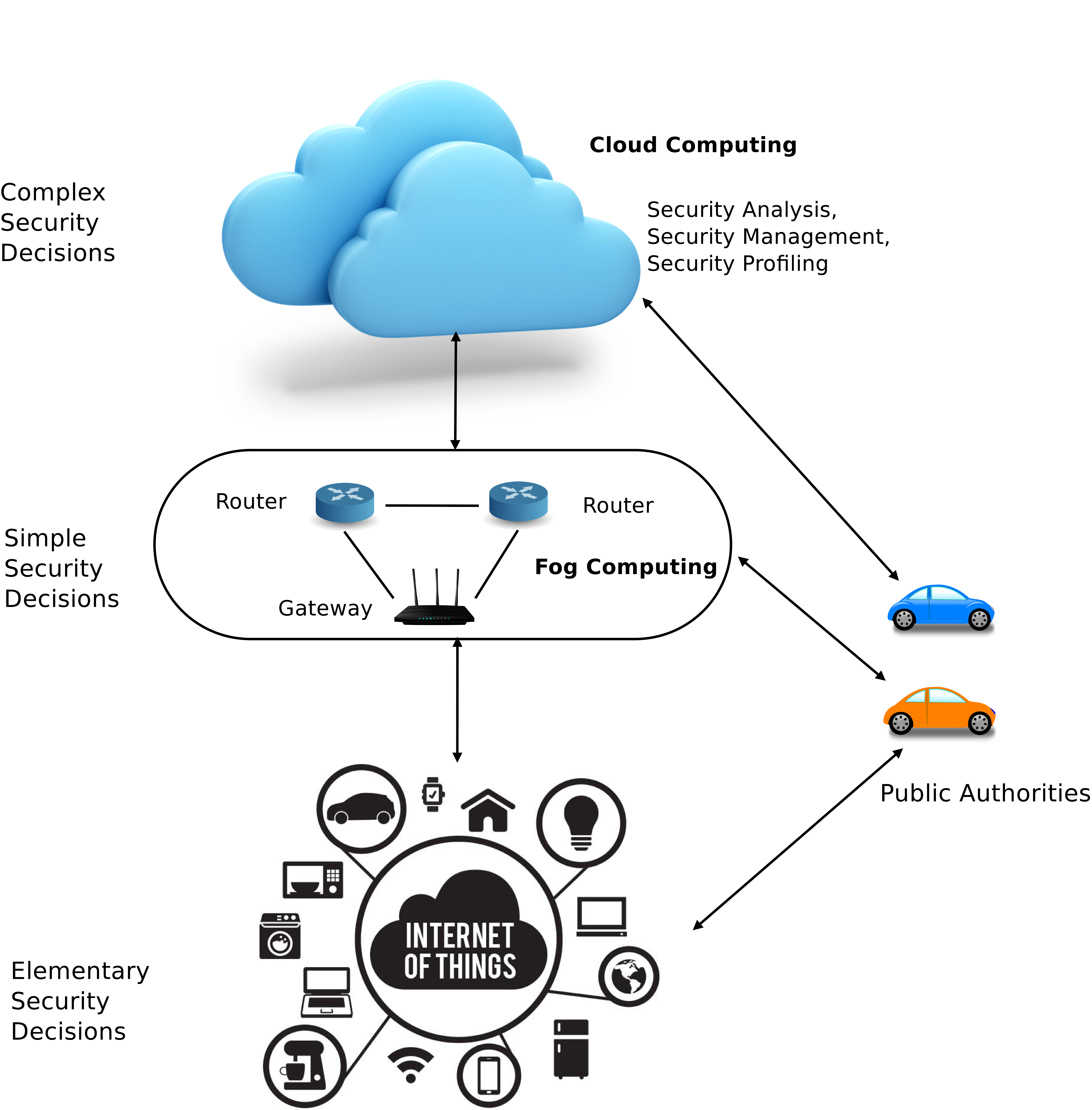}
	\caption{A Security based Architecture using Fog Computing\cite{sehgal2015smart}}
	\label{fig:8}
\end{figure}

\textbf{Security framework based on fog architecture:} Due to population explosion, it has become challenging for the civil authorities to provide security cover to the citizens. The authors in \cite{sehgal2015smart} have proposed a security based architecture in order to overcome this issue. This proposition is based on the concept of IoT devices, fog or edge computing and cloud computing. It consists of three layers, (i) IoT layer, (ii) fog computing layer, and (iii) cloud computing layer. Fig.~\ref{fig:8} shows a generic view of the proposed architecture. The elementary security decisions are taken in the IoT layer. The IoT layer is connected to the fog computing layer, where the simple security decisions are taken. The complex security decisions are taken in the cloud computing layer, which is connected to the fog computing layer. All the three layers i.e. the IoT layer, the fog computing layer and the cloud computing layer are connected to the public authorities. The fog computing layer consists of different routers and gateways. The cloud computing layer does the security analysis, security management as well as security profiling.

Inter-layer entities are connected using various communication technologies including wireless as well as wired communication. If any physical threats are detected, then public authorities are immediately alerted. The lowest layer of the architecture is IoT layer where the IoT devices communicate and cooperate with each other using a wireless ad hoc network. This IoT layer gathers knowledge about the physical surroundings. The middle layer is the fog computing layer where the routers, gateways, bridges etc. are extending the cloud computing features. The real-time and latency sensitive applications are performed at the fog computing layer. Based on the gathered data, the security related decisions for a woman may be taken in this layer. The top layer is the cloud computing layer which is a computing and storage layer. This is offered as a service for the IoT devices. The cloud collects data from the IoT infrastructure in order to provide security related decisions for a user.

\begin{figure}[!t]
	\centering
	\includegraphics[width=0.5\textwidth]{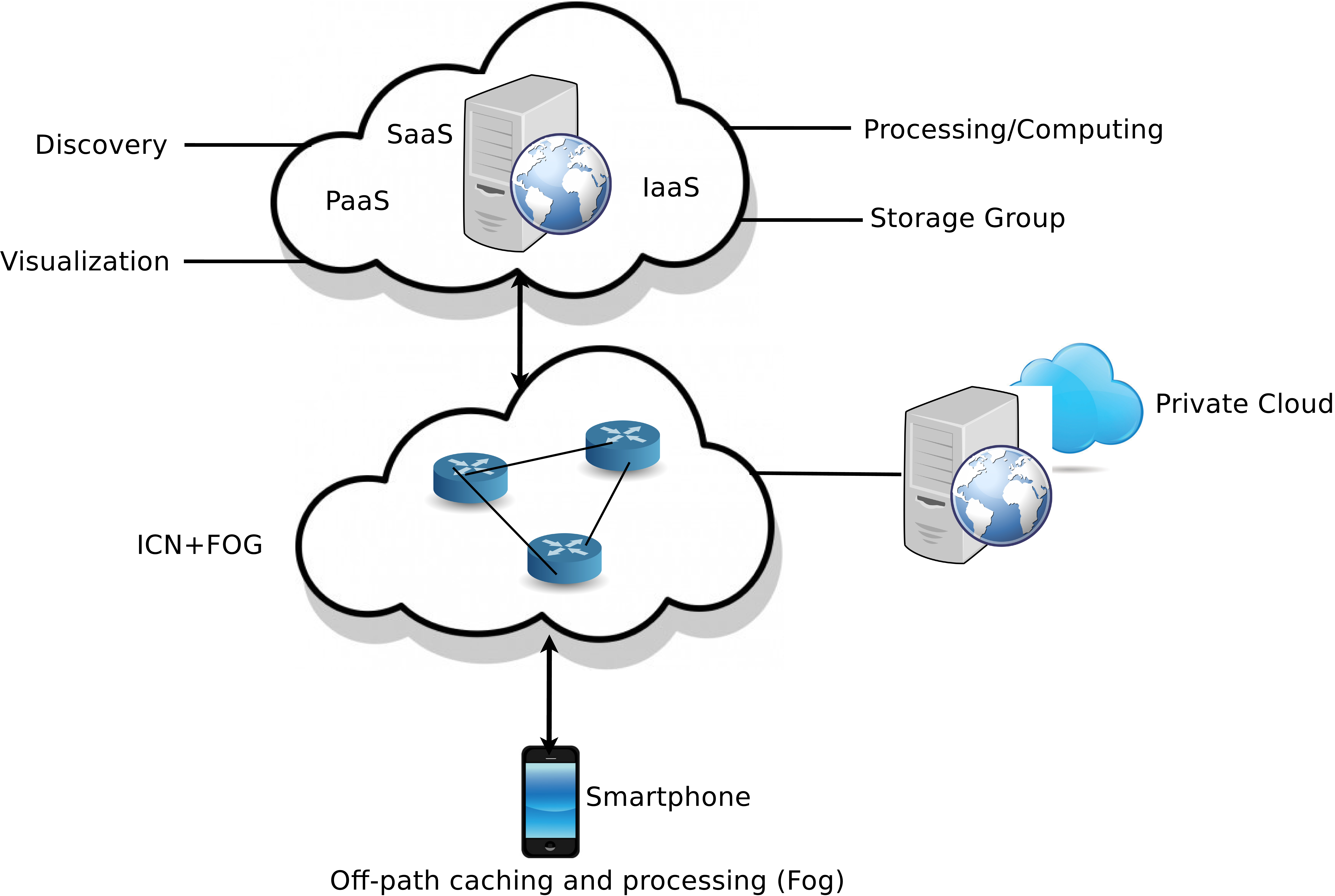}
	\caption{A Fog Computing based Architecture for the Information Centric Networking\cite{abdullahi2015ubiquitous}}
	\label{fig:9}
\end{figure}

\textbf{Fog over Information Centric Networking:} In \cite{abdullahi2015ubiquitous}, a fog computing architecture for the Information Centric Networking (ICN) is proposed for the off-path caching to the IoT. Fig.~\ref{fig:9} depicts the architecture. The smartphone communicates with the information centric networking and fog computing layer. The ICN and the fog computing layer are connected to the private cloud. Cloud computing facilitates the discovery of services, visualization, processing as well as storage. This architecture uses fog computing in order to provide a ubiquitous computation framework.

\begin{figure}[!t]
	\centering
	\includegraphics[width=0.5\textwidth]{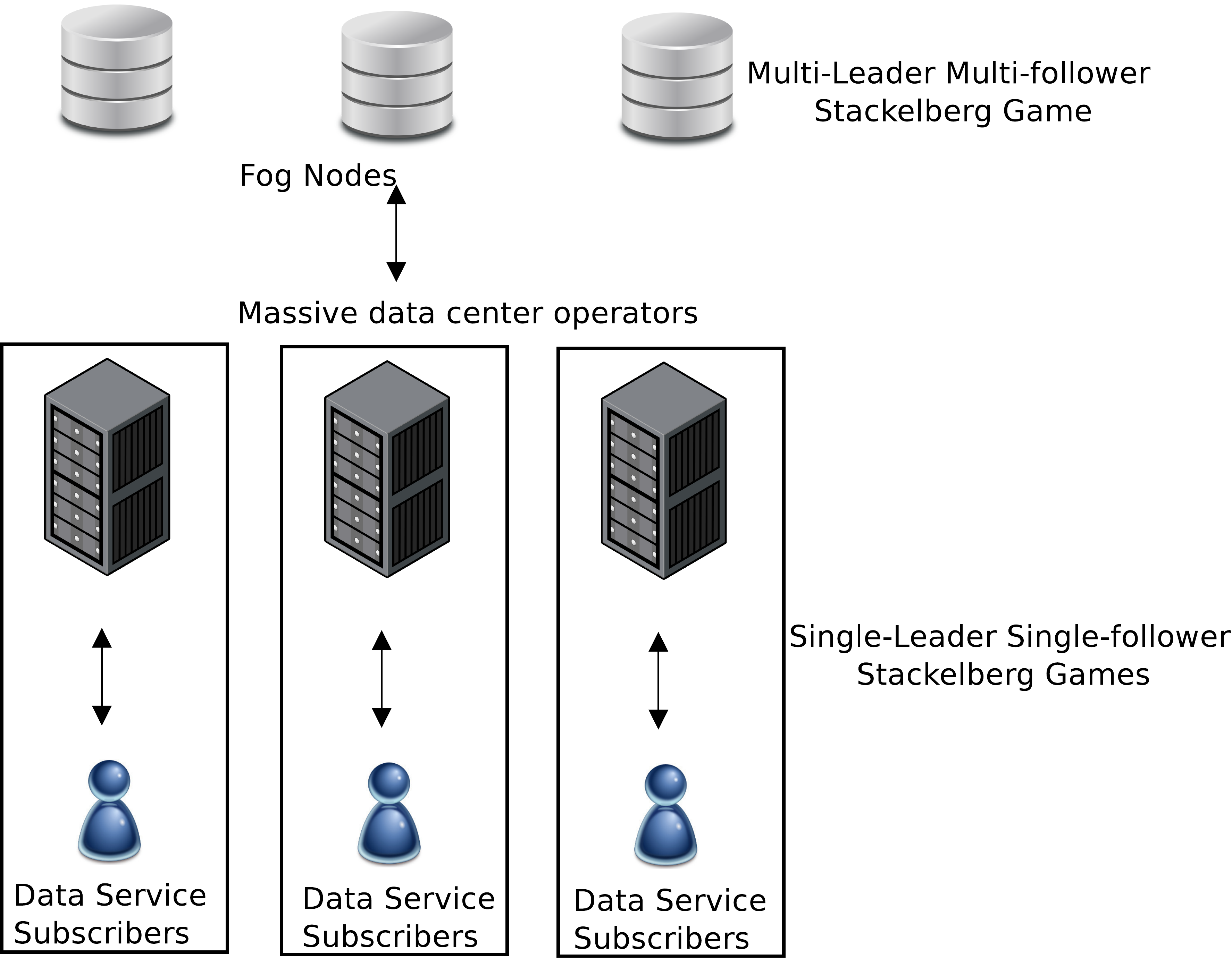}
	\caption{A Hierarchical Game based Model for Fog Computing\cite{zhang2016fog}}
	\label{fig:10}
\end{figure}

\textbf{Hierarchical game based model for fog computing:} The surge of data services motivates the introduction of data center networks to serve the clients. Again, the data center subscribers have created the need to virtualize the resources so that resource allocation among all data service subscribers can be provided. So, the authors in \cite{zhang2016fog} have proposed a hierarchical game based model for fog computing. The resource management problem is solved by this proposed architecture. The system architecture of this model is shown in fig.~\ref{fig:10}. The fog nodes are connected to the massive data center operators which is again connected to the data service subscribers. The fog nodes and the data centers run two different variants of Stackelberg game to optimize the resource allocation based on the demands from the users. The interaction between the data center operators and the fog nodes are modeled as multi-leader multi-follower Stackelberg game, whereas the interaction between data center operators and data service subscribers is modeled as a single-leader single-follower Stackelberg game, as shown in fig.~\ref{fig:10}. In case of the multi-leader multi-follower Stackelberg game, the leaders are the fog nodes, and the data center operators work as the followers. At the next level, where the system is modeled as a single-leader single-follower Stackelberg game, the data center operators work as the leaders, and data service subscribers are the followers. The interesting formulation of the problem comes from the fact that the interaction between the data center operators and the data service subscribers are independent of each other, and therefore the problem can be modeled as a single-leader single-follower Stackelberg game. Here the fog nodes help in optimizing the resource allocation in such kind of hierarchical architecture. The authors have shown simulation results to indicate that the resource management performance can be improved based on the game model; and all the fog nodes, massive data center operators and data service subscribers are able to achieve satisfying utilities.

\begin{figure}[!t]
	\centering
	\includegraphics[width=0.5\textwidth]{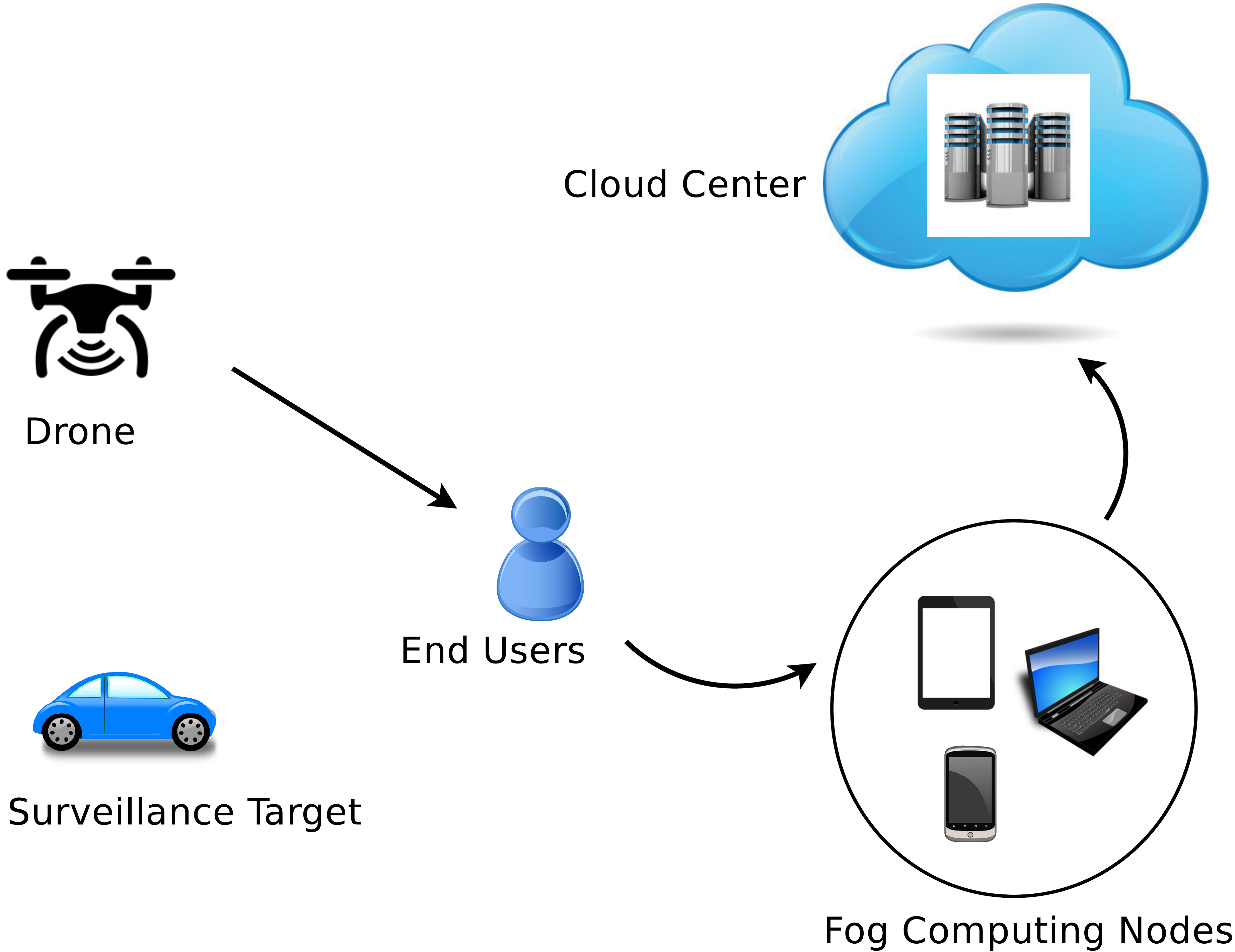}
	\caption{A Smart Surveillance Architecture\cite{chen2016dynamic}}
	\label{fig:11}
\end{figure}

\textbf{Surveillance architecture based on fog computing:} In \cite{chen2016dynamic}, the authors have proposed a three layer surveillance architecture. These three layers are shown in fig.~\ref{fig:11}. The {\em user layer} or {\em surveillance application layer} collects data through smart devices. The {\em fog computing layer} does the real-time data processing and storage. Several devices can be used for the fog computing purposes such as smart tablets, smartphones, laptops in a police car, on-board computing devices on the drone etc. Effective video processing is the key issue for a surveillance system because of the need for real-time data processing. In order to get this, the output frame rate from the video processing system should be equal to or higher than the input frame rate. In the proposed surveillance system, a drone is used as a sensor to monitor the area of interest. The objective is to identify the movement of suspicious vehicles from the video data. Instead of sending the whole video frame, a sub-area including the suspicious vehicle is extracted from the original frame and sent to the fog computing unit. Again, the fog computing layer prevents the local significant data from being sent to the cloud node. In order to reduce the workload, only the relevant data is extracted and sent to the cloud by the fog layer. So, the latency of transmitting data from the surveillance area to the cloud is reduced. The cloud computing layer is used for future data analytics. The requirements of real-time surveillance task has been meet by this approach. The proposed system has the capacity to handle multiple targets without using multi-target tracking algorithm. The drone captures the data of the surveillance target and sends it to the end users. The end users send the data to the fog computing nodes which does the initial processing at the edge devices. The complex tasks are sent to the cloud data center.

Table \ref{table:fogarchi} summarizes various context dependent architectures developed over the fog computing framework. In summary, we can say that the context sensitive architectures have similarity in the architectural components. The works have been done in many application areas i.e. transportation systems, IoT, surveillance systems etc. Some works have considered security as an important aspect of the architecture.

\begin{table*}[h!]
	\centering
	\caption{Comparison of Different Context Dependent/Sensitive Architectures of Fog Computing}
	\begin{tabular}{|c ||c |L |L |} 
		\hline
		{\bf Research work} & {\bf Architectural Components} & {\bf Security Aspects Consideration} & {\bf Application Area}\\ \hline \hline
		Salonikias et al. \cite{salonikias2015access}  & Cloud, Fog and Sensors  & Yes & Transportation Systems\\ \hline
		Sehgal et al. \cite{sehgal2015smart}  & Cloud, Fog and IoT  & Yes & IoT\\ \hline
		Abdullahi et al. \cite{abdullahi2015ubiquitous} & Cloud, Fog and smart devices  & Yes & Information Systems\\ \hline
		Zhang et al. \cite{zhang2016fog}  & Hierarchical game based model  & No & Information Services\\ \hline
		Chen et al. \cite{chen2016dynamic} & Cloud, Fog and user devices  & Yes & Surveillance systems\\ \hline
		Liu et al. \cite{liu2016paradrop} & Cloud and Wi-Fi AP  & Yes & General latency-sensitive systems\\ \hline
		Yoon et al. \cite{yoon2016low} & Raspberry Pi and Wi-Fi AP  & No & Video transcoding systems\\ \hline
		Aazam et al. \cite{aazam2014fog} & Cloud, Fog and IoT  & Yes & IoT\\ \hline
		Gupta et al. \cite{gupta2016ifogsim} & Fog Simulator  & Yes & General latency-sensitive systems\\ \hline
	\end{tabular}	
	\label{table:fogarchi}
\end{table*}


\section{Technologies for Enabling Fog}
\label{technology}
The fog computing framework has been developed and implemented over various technologies, ranging from conventional networking infrastructure to employment of special low cost servers at the edge. Broadly, the researchers and developers have explored two pathways to deploy fog infrastructure -- (a) by exploiting the capability of edge devices available at the conventional communication infrastructures, such as base stations at cellular communications, routers or gateways near the edge, access points for wireless local area networks, etc.; and (b) through the deployment of special low cost fog servers near the edge network. In this section, we broadly discuss such various technologies that have been explored for setting up the fog infrastructure. 
%

\subsection{Fog Computing Framework over 5G Networking System}
The basic advantage of utilizing cellular infrastructure for fog computing is that it uses a hierarchical architecture. A number of networking components in cellular infrastructure, like the base station, the switching centers, the serving and packet gateways etc. have partial computing capability, and this capability can be utilized to develop efficient fog applications on top of the cellular architecture. Therefore, a large number of research works, such as~\cite{peng2016fog,peng2015recent,yan2016user,bastug2014living,checko2015cloud,tandon2016cloud,intharawijitr2016analysis,militano2015device, tran2016collaborative, hung2015architecture} and the references therein, have explored cellular infrastructure, particularly the fifth generation (5G) networks that is {\em Long Term Evolution Advanced} (LTE-A), for developing fog computing applications. Some of these applications try to improve the network capabilities and functionalities for LTE-A network, whereas some others utilize the various edge components of LTE-A for developing fog based third-party applications.  The various works, where fog computing has been applied on 5G cellular networks, are as follows. 

\begin{figure*}[!t]
		\centering
		\includegraphics[width=1\textwidth]{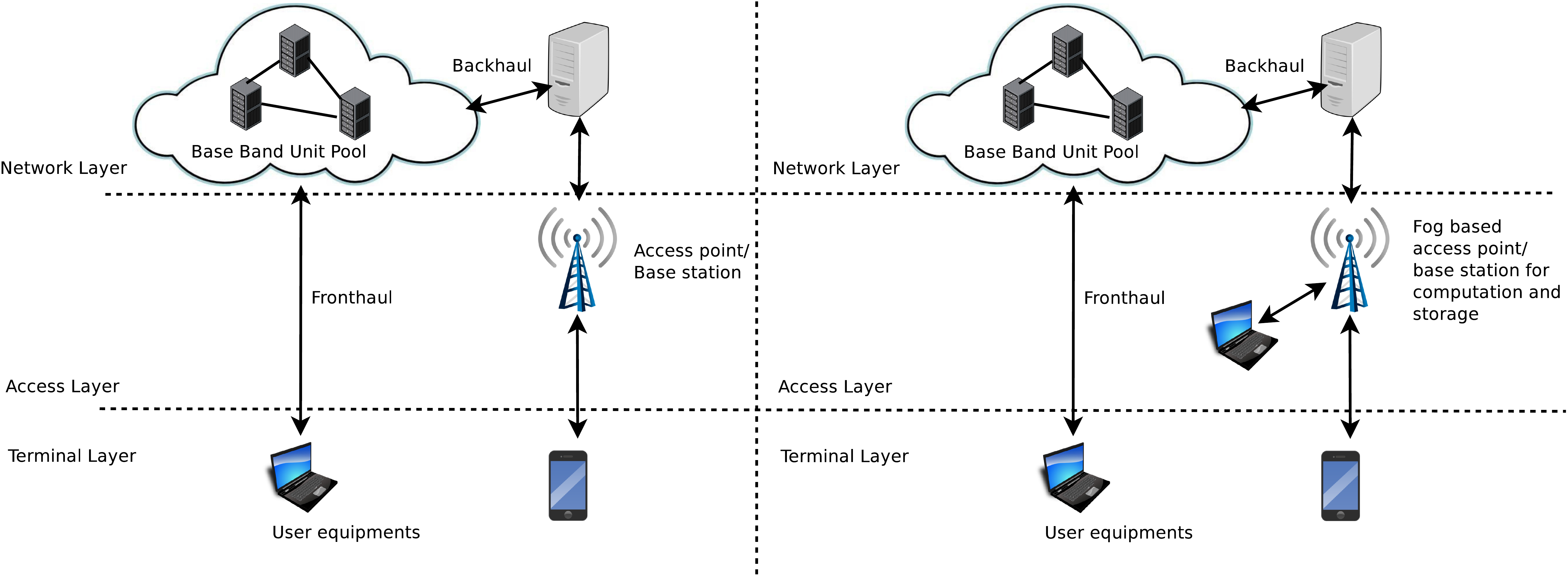}
		\caption{Cloud RAN and Fog RAN\cite{peng2016fog}}
		\label{fig:cloudandfogran}
	\end{figure*}

\textbf{Fog based radio access networks:} The LTE-A network requires various level of computations for the effective high speed data communication purpose, the primary being the signal processing activities. In LTE-A, the signal processing is done over a component, called {\em Radio Access Networks} (RAN). The fog computing framework is primarily being explored for improving the capability of RAN in LTE-A networks, with an objective towards increasing the spectral efficiency along with energy efficiency for cellular networks. In their seminal work, Peng \etal~\cite{peng2016fog} have proposed a RAN architecture for 5G systems based on fog computing, which is an effective extension of cloud based RAN (CRAN) as being explored in various recent literatures~\cite{checko2015cloud,wubben2014benefits,pompili2016elastic,agiwal2016next}. The concept of cloud RAN based LTE-A networks is as follows. A LTE-A base station, called {\em Evolved Node B} (eNB), has broadly two components - the radio frequency (RF) module that takes care of the electromagnetic signal transmission and reception, and the baseband processing unit (BBU) that takes care of the signal processing activities, like modulation and demodulation, signal encoding and decoding, noise cancellation etc. In Cloud RAN architecture, the BBUs from multiple LTE-A eNBs are hosted on a single cloud, called the BBU pool, whereas the eNBs now only contain the RF modules. This improves energy efficiency, cost effectiveness, high availability and a easy management solution for the RAN infrastructure. However, the major shortcoming for this infrastructure is that the delay between the RF module and the BBU module hosted over a remote cloud affect the signal processing accuracy and also introduces significant traffic overload to the network fronthaul (the link from the RF module to the cloud hosting the BBU). To reduce the fronthaul load and delay, Peng \etal~\cite{peng2016fog} have proposed the fog based RAN architecture (Fog-RAN) that can reduce the delay and load at the fronthaul, whereas can provide the advantages of having virtualized BBUs. Fig.~\ref{fig:cloudandfogran} depicts the cloud RAN and fog RAN architectures. The major difference between the two architectures is as follows. In fog based RAN, an additional computing resource has been placed at the edge devices, which takes care of the partial signal processing activities and thus, reduces the load at the cloud side. 
	
There are multiple advantages of having fog based RAN as discussed in~\cite{peng2016fog}. This would provide high spectral benefits with significantly improved energy efficiency. Fog computing based local RAN helps to avoid large-scale radio signal processing in the centralized unit. 5G system is envisioned to provide energy efficiency growth by a factor of at least $10$ and system capacity growth by a factor of at least $1000$ compared to 4G systems \cite{peng2015recent}. In order to achieve this goal, CRAN network have been proposed to be an integral part of the LTE-A networks~\cite{peng2015system}. However, CRAN is unable to meet the capacity and time-delay constraints due to high fronthaul load, as we mentioned earlier.  In order to overcome the issues of CRAN based systems, the authors in~\cite{peng2016fog} have incorporated fog computing in RAN architecture where the storage, communication, control, measurement, management etc. are offloaded at the edge of the network. The Fog-RAN architecture is very advantageous in the sense that it can be easily scalable, which makes it adaptive to dynamic traffic and radio environments. 
	
A few successive works have considered the fog-RAN architecture, and have addressed various technical and research issues associated with it. The authors in \cite{yan2016user} have worked with fog based RAN in order to have high spectral and energy efficiency in the fifth generation wireless communication system. The work has derived the coverage probability and ergodic rate for both fog based access point users and device-to-device users. Various other works have been done to improve the 5G network efficiency with the help of fog computing paradigm. The edge processing and virtualization are the most efficient aspects in the context of 5G networks \cite{bastug2014living,checko2015cloud}. The fog based caching at the edge devices in a radio access network has been done by the works of Tandon \etal \cite{tandon2016cloud}. The work focused on the identification of the optimal caching along with fronthaul and edge transmission policies. The authors in \cite{xiang2015joint} have proposed the reduction of burden on the fronthaul by edge devices' caches. The proposed fog based RAN (F-RAN) architecture is used to study the cache incorporation into cloud-RAN. The authors have studied joint mode selection and resource allocation problem in F-RANs supported device to device network. They have seen that proposed architecture maximizes the system energy efficiency and cache incorporation improves the system performance. 
	
\textbf{Delay sensitive applications over 5G:} 5G systems need more latency-sensitivity than the 4G systems. Though cloud computing services are enabling the functionality of the 5G systems, but the long communication distance becomes an issue for delay sensitive applications. Fog computing is being explored in 5G systems to minimize this delay. The communication delay and the computing delay are both important for delay-sensitive applications. In this respect, the authors in \cite{intharawijitr2016analysis} have proposed a mathematical model for fog computing. The authors have considered a group of users in the same area as one source node and a fog server as one fog node. The relation of fog nodes and source nodes is shown as a complete bipartite graph. A source node is connected to a fog node. Each source node produces a workload with a rate of poisson process. Each fog node holds some workloads at a time. There is a transmission latency between source node and fog node. The transmission latency represents the round-trip time excluding the computing time of a fog node. 

%
	
In their work, the authors have given an objective function satisfying some constraints. The objective is to minimize latency and to allow workload up to a certain level. They have considered a group of users in the same area as one source node and a fog server as one fog node. The latency functions have been formulated which estimates the processing time of all current workloads held in a fog node. Here, the total latency is equal to the sum of the computing latency and the communication latency. The objective function defined in this way tries to find the minimum of the blocking probability, represented as the ratio between a number of rejected workloads and the total number of all workloads in the entire system. Before meeting the objective function, the system has to satisfy all the constraints. The total latency of an accepted workload must not exceed the accepted threshold which is the maximum tolerant latency of a service. The simulation evaluation of the proposed system shows that the lowest latency policy always provides the least blocking probability when there are fewer source nodes than fog nodes. Comparing the random policy and the maximum available capacity policy, both policies provide a nearly equal blocking probability in most of the cases, even though the number of nodes is dynamic. The authors have also explored the optimal value for the latency threshold in terms of the blocking probability. 
	
\textbf{Machine to machine communication in fog computing:} Fog computing plays a crucial role in IoT applications. Fog has the advantage of providing location awareness and low-latency interactions to Machine-to-Machine (M2M) applications. Vallati \etal \cite{vallati2015exploiting} have analyzed the configuration of LTE network to support the future M2M fog computing applications. The work showed that it is able to provide low latency interactions between devices. The works of \cite{oueis2015fog} have discussed about the issue of load balancing in fog computing based 5G systems. The local computation clusters process all the requests of computation offloading by multiple clients. The authors have proposed an efficient algorithm for fog clustering. The algorithm is able to meet higher user satisfaction ratio with a proper resource management.

It can be noted here that the 5G based cellular system and the fog computing framework are very related to each other in terms of compatibility. 5G system promises to provide the users sub-millisecond latency and more than $1$ Gbps transmission speed. As fog computing has the potential of serving the end users within this small latency value, fog is a perfect fit for 5G systems in comparison with the cloud computing. Various recent works~\cite{militano2015device,tran2016collaborative,hung2015architecture} have discussed about the applications of fog computing in the technological point of view of 5G systems.
Table~\ref{table:fog5gtable} summarizes and compares different works on fog computing for 5G systems.  

\begin{table*}[!t]
	\centering
	\caption{Comparison of Different Fog Computing Paradigms for 5G Systems}
	\begin{tabular}{|>{\raggedright}p{2cm}||>{\raggedright}p{2cm} |>{\raggedright}p{3cm}|>{\raggedright}p{2cm}|>{\raggedright}p{3cm}|>{\raggedright\arraybackslash}p{3cm}|} 
		\hline
		{\bf Research work} & {\bf Architecture} & {\bf Services} & {\bf Fog Devices}  & {\bf Application} & {\bf Advantages of Using Fog}\\
		\hline \hline 
		Intharawijitr \etal ~\cite{intharawijitr2016analysis} & A mathematical model regarding the fog architecture & Low latency services & Servers connected at the edge & Various 5G applications  & Improves service latency for end users  \\ \hline
		Peng \etal ~\cite{peng2016fog} & Cloud computing network layer, Terminal layer and Access layer & Delay sensitive services &  Cellular base station  & Various 5G applications  & Providing latency sensitive services in 5G  \\ \hline
		Yan \etal ~\cite{yan2016user} & Fog based radio access network & High spectral and energy efficient service provisioning & Fog servers near the edge & Various 5G applications  & Virtualization of RAN near the edge -- this reduces fronthaul delay significantly \\ \hline
		Oueis \etal ~\cite{oueis2015fog} & Fog based 5G system & Multiple users requiring computation offloading & Small cell base stations  & Load balancing in fog computing & High users' satisfaction percentage of a minimum of $90\%$ for up to $4$ users per small cell \\ \hline	
		Xiang \etal ~\cite{xiang2015joint} & Fog computing based radio access network  & Energy efficient delay sensitive services &  Fog servers near the edge  & Joint mode selection and resource allocation problem in Fog-RANs supported device to device communication & Energy efficient low latency services   \\ \hline				
		Hung \etal ~\cite{hung2015architecture} & Cloud radio access networks and fog networks  & Real-time services in 5G  & Fog servers near the cellular edge & 5G applications & Low-latency services have been obtained by architecture harmonization between CRAN and Fog-RAN \\ \hline			
		Tandon \etal ~\cite{tandon2016cloud} & Fog radio access networks  & Caching in edge nodes  & Cellular base stations and fog servers near the edge & Minimizing content delivery delay & Interplay between cloud and fog nodes helps in minimizing delivery latency \\ \hline	
		Vallati \etal ~\cite{vallati2015exploiting} & Machine to machine applications  & Providing low latency interactions between the devices  & Cellular gateway & Latency sensitive communications between devices & Fog devices provide fast response \\ \hline				
	\end{tabular}	
	\label{table:fog5gtable}
\end{table*}

\subsection{Fog Computing at the Wireless Edge}

Various works have used wireless edge devices like Wi-Fi access points (AP) to host the fog computing platform. 

\textbf{Edge computing platform using wireless edge devices:} Liu \etal~\cite{liu2016paradrop} have proposed a specific edge computing platform known as {\em Paradrop} which provides computing and storage resources at the extreme edges of the network, which are the Wi-Fi APs or the wireless gateways. The advantage of using a Wi-Fi AP as a fog node is that it has a unique contextual knowledge of the end-devices. Their proposed platform focuses on specific design issues about the architecture, a programming interface and orchestration framework. The {\em Paradrop} platform consists of three components --  (i) a hosting substrate in the Wi-Fi APs, which supports multi-tenancy, (ii) a cloud server working as a back-end and (iii) an application program interface (API) through which third-party developers can deploy the services across such different APs. This gives a unique advantage of fog computing, where popular edge devices like the Wi-Fi APs are utilized for local computation by exploiting its knowledge about the end-devices. 
	
\textbf{Low-cost video transcoding solution at the wireless edge:} The work of Yoon \etal~\cite{yoon2016low} has proposed a low-cost video transcoding solution running at the wireless edge, i.e. the wireless APs. In this approach, they have used a low-cost hardware, Raspberry-Pi which provides real-time video transcoding solution. The proposed approach can provide adaptive solution to sudden network dynamics and it is able to cover client's feedback quickly. The deployment is broad and quick as it is transparent, low-cost as well as scalable. The proposed solution can provide higher video bit-rates without causing intermediate video stalls.
	
\textbf{Smart indoor localization mechanism:} Researches have developed interesting applications by utilizing the capability of fog computing at Wi-Fi access points. In~\cite{sciarrone2016smart}, Sciarrone \etal have developed a smart indoor localization mechanism by applying fog computing paradigm at Wi-Fi APs. The limited computation capability at Wi-Fi APs has been utilized to develop a radio map based on received signal strength (RSS). This radio map, called the RSS fingerprint, is then used to develop an indoor positioning mechanism. The authors show that by utilizing computation offloading at Wi-Fi APs, the system can achieve good localization accuracy with significant power saving, which is as high as $80\%$ in many cases. 
	
\textbf{Caching and policy enforcement at the access point:} In~\cite{iotti2017improving}, the authors have developed a methodology for improving the quality of experience based on fog computing at wireless edge. They have developed a model, where proactive caching and policy enforcement can be applied at the access points of a wireless hotspot, that can significantly boost up the service quality. Zhu \etal~\cite{zhu2013improving} have shown that fog based proactive content caching at the hotspots (access points and proxies) can significantly improve Internet access performance. 

Such types of smart gateways have been explored in various other works~\cite{aazam2014fog,aazam2014smart,rahmani2017exploiting,yannuzzi2017new,wang2017survey}. We observe that wireless gateways work as a key enabler for deploying fog services because of their computational capabilities to execute additional services. 

\subsection{Fog Computing over Other Network Components -- Routers, Switches and Gateways}
The fog computing paradigm has been explored to provide services for the basic network protocols as well as for various other network applications over network components of a conventional local area networks, such as the routers, the gateways, the network switches and so on. In~\cite{slabicki2016performance}, the authors have compared the performance of three network protocols -- {\em constraint application protocol} (CoAP), {\em simple network management protocol} (SNMP) and {\em network configuration protocol} (NETCONF), over the fog computing framework. The authors have analyzed the impact of the selection of the communication architecture (cloud or fog) on the performance of data exchange for the above three network application/management protocols. The performance of the above three protocols have been compared for three different communication architectures -- (a) direct synchronization between devices, (b) synchronization through the local gateway (fog node), and (c) synchronization via the cloud servers. The observations from the experiments are as follows. The authors have shown that the synchronization by the cloud requires up to the three times longer time than the data exchange through a local gateway that acts as the fog server, and up to six times longer than the direct communication between the network nodes. They have further shown that the transmission through SNMP and CoAP results in similar delay, which is very lower than the delay introduced through the NETCONF protocol. This work shows that fog computing can even help in improved performance for various network application and management protocols.    

In~\cite{vilalta2016end}, the authors have developed an integrated SDN/NFV based fog computing paradigm for optical networks, where virtual network functions are dynamically deployed at the fog nodes to handle network data processing and to ensure QoS over network service provisioning. Mayer \etal~\cite{mayer2017fog} have shown that fog computing can enable social sensing under limited network connectivity, where the edge devices can store and process intermediate data from a social sensing framework. The data from various user devices, like smartphones and wearables, can be collected at the network devices, like routers and gateways, where intermediate data processing can be done to provide services, under the limited connectivity scenario when cloud connectivity is not available. A number of works, such as~\cite{ku20175g,tao2017foud} and the references therein, have explored fog based architecture to support radio access technologies over 5G cellular network. We have discussed these radio access cellular technologies earlier in this survey. 

\section{Exploring the Fog Computing Features for Developing Services}
\label{features}
This section discusses several features which are utilized and extended in fog based systems. These services range from data analytics and mining, IoT based services, security and privacy, network services and so on. In \cite{luan2015fog}, the authors have discussed about several fog computing features that can be summarized as follows.
\begin{enumerate}
	\item[(i)] \textit{Target User}: In fog computing framework, the main target users are the edge users, and more particularly the mobile users. The popularity of mobile devices has created enormous opportunities for mobile edge computing that can be extended to the fog paradigm.
	\item[(ii)] \textit{Service Type}: The service type is localized and limited. The focus of fog computing is basically on the deployment location. Services are provided with respect to the context of them. Fog nodes are situated in the vicinity of the sensing devices.
	\item[(iii)] \textit{Hardware}: Under fog computing frameworks, the hardware in general has limited storage, computation power and wireless interface. These devices are gateways, routers, base stations, access points as well as various other edge devices.
	\item[(iv)] \textit{Distance to Users}: The fog computing devices are in the physical proximity of the users, and the devices communicate through single-hop wireless connection in most of the times. This has created the advantage of getting real-time responses for the users.
	\item[(v)] \textit{Working Environment}: The working environment is not constraint to indoor only. Several fog devices participate in the formation of edge network. The placement of edge devices is a critical issue in the context of IoT.
	\item[(vi)] \textit{Deployment}: The deployment of fog nodes can be centralized as well as distributed by local businesses.
\end{enumerate}
Next, we discuss various service, networking and application platforms that can be utilized to develop fog based systems.

\subsection{Software Defined Networking (SDN) and Network Function Virtualization (NFV)}
The number of devices connected to the Internet have increased at a rapid pace. This increase is driven by two sources: user devices and sensors/actuators. The impressive growth may soon be get suppressed by the enormous number of sensing/actuating IoT devices placed virtually everywhere. The management of network of billions of heterogeneous devices is very challenging and complex. In \cite{vaquero2014finding}, the authors have discussed about the device ubiquity as the opportunity for fog computing. There is an issue of configuration and maintenance of different types of services running on billions of heterogeneous devices. The fog computing paradigm needs the heterogeneous devices and their running services to be handled in a more homogeneous manner. This requires a software for proper orchestration of the devices. For this kind of requirement, Vaquero \etal~\cite{vaquero2014finding} have mentioned about different enabling technologies, such as software defined networking (SDN) and network function virtualization (NFV), for fog computing. NFV~\cite{hawilo2014nfv} is the most preferred technology for this kind of management of the services. NFV provides the on-demand network services (such as a firewall, a router or a wide area network (WAN) accelerator, a virtual private network (VPN) etc.) or user-services (such as a database) whenever it is needed. SDN~\cite{lantz2010network} is one of the enabling technology which is needed for deploying virtual network functions (VNF) in a NFV environment, in order to have some network services by software only. 

Bhardwaj \etal~\cite{bhardwaj2016fast} have explored the current technologies that can be leveraged in the design of edge function platforms. The back-end driven offloading of tasks to the edge is a way to address bandwidth use and latency issues. The authors have also developed a solution which has the feature of providing the security measures of the edge functions. Also, they have proposed and evaluated a platform known as AirBox for fast, scalable and secure offloading of edge functions. A number of research works, such as~\cite{xu2016towards,huang2016software,kahvazadeh2017securing,sahoo2017sdn,chen2017integrated,fawcett2017siren} and the references therein, have explored SDN and NFV based frameworks for providing fog based application services. 

\subsection{Data Mining and Live Data Analytics at the Edge}
In paper \cite{dubey2015fog}, a dynamic time wrapping (DTW) algorithm is proposed for mining patterns in time-series data. This has been used for various applications such as business, finance, single word recognition, analysis of ECG signals etc. Euclidean distance fails to detect similarity between similar and out-of-phase series. Whereas, the DTW can detect similarity between two series regardless of different length, and phase difference. Also, they have used clinical speech processing chain (CLIP) which is a series of filtering operation applied to speech data for computing the relevant metrics. In~\cite{sharma2017live}, the authors have developed a live and online data analytics platform based on fog computing. In an IoT environment, the major challenge is to develop an effective mechanism to extract important features from the massive amount of heterogeneous data generated from various IoT devices. Such online data analytics platform over IoT network can provide real-time information and feedback to the end-users. Another challenge in IoT environment is how to utilize the data-aware intelligence to enhance the performance of data analytics. Accordingly, the authors have proposed a framework for live data analytics through coordinated processing between the edge devices  and the cloud computing framework by integrating advantages from both the platforms. The proposed framework has the capability to exploit the network-wide knowledge as well as the historical information available at the cloud to process edge data analytics while satisfying various performance requirements of heterogeneous IoT networks. 

Data intensive analysis is one of the major challenges in a smart city environment, because of the ubiquitous deployment of various different types of sensors and actuators devices. Such application requires location awareness and latency sensitive monitoring as well as intelligent control based on geo-distribution of the sensors and actuators over the smart city environment. In~\cite{tang2017incorporating} the authors have developed a big data analytics platform based on fog computing framework. To develop a data analytics use case in smart city environment, the authors have developed a prototype of a smart pipeline monitoring system based on fiber optic sensors and sequential learning algorithms,  to detect events that can threaten pipeline safety. The working prototype is used to experimentally evaluate the event detection performance of the recognition of $12$ distinct events based on edge data analytics at the fog. \cite{garcia2015edge} provides a good summary of various edge data analytics techniques that have been explored in the existing literature. 

%
\subsection{Fog based IoT Data Processing and IoT Services}
IoT is a pertinent use case of the fog computing paradigm. With the proliferation of IoT devices, it is expected that in future the number of connected devices will exceed the number of computers connected to the Internet today \cite{mietz2013semantic,lu2017lightweight}. It is also estimated that the global mobile traffic would increase from $2.6$ to $15.8$ Exabyte by $2018$ \cite{abdelwahab2016network}. This scalability is a need of the hour which the cloud computing cannot solve alone. IoT has the following requirements.
\begin{enumerate}
	\item[(i)] \textit{Heterogeneity}: IoT devices are heterogeneous in the sense that they have come from different vendors and they have different communication protocols. In this heterogeneous environment, we need to abstract the physical devices into high-level entities.
	\item[(ii)] \textit{Scalability}: Geographically distributed IoT devices generate huge amount of data. These highly distributed devices need proper orchestration for any complex work.
	\item[(iii)] \textit{Location aware computing}: There is a need for content locality of services in IoT scenarios \cite{li2015decentralized}. The locality aware computing makes the system more secure as the data remains within a particular administrative domain. However, in case of cloud computing, the storage and computation activities are performed at a remote location making it more prone to security risks.
	\item[(iv)] \textit{Service discovery}: The IoT network is in general very large, and it deals with many entities which have different unknown services. We need to perfectly orchestrate the IoT devices for the service discovery.
\end{enumerate}

In \cite{bonomi2012fog}, the authors have discussed some of the use cases of fog computing based IoT, such as the scenarios like connected vehicles, smart grid, wireless sensor and actuator network etc. Most of the applications in IoT scenarios cannot be handled by the current compute and storage models bounded to data centers. IoT nodes are distributed in a large scale in different geographical locations. These applications seek real-time decisions based on data analytics. Consequently, these applications require high throughput within short time periods. The requirements of computation and storage resources are of much importance in the context of IoT. However, the question is where should these resources be placed. Mobile cloud computing have several issues in this context. The authors in \cite{lei2013challenges} have discussed about these challenges. The main issue is the mobility of the IoT devices. In \cite{yannuzzi2014key}, the authors have discussed the usefulness of fog computing in the presence of IoT, where the IoT nodes act as fog nodes. The fog computing enables high throughput on demand for short time periods while ensuring the mobility of the IoT devices. Sometimes, there might not be any connectivity with the cloud. In such cases, fog nodes provide the service to the end-users within less amount of time. 

An IoT platform needs to have the following six domains -- (1) the domain of the ``things" most of which are mainly mobile to mobile (M2M), host to mobile (H2M) or M2M gateways; (2) the network domain, covering the edge, the aggregation and the core; (3) the cloud domain; (4) the service and application domains; (5) the users domain and (6) the fog nodes. These fog devices provide compute, storage and network capabilities to the ``things". These fog devices are scattered from the end devices up to the cloud. The mobility of the IoT devices has been taken into the consideration by the fog computing paradigm. Fog scenarios requiring the reliability and/or minimum latency need to locate the intelligence where it is needed in the network. Therefore, the computation and storage facilities should be placed very close to the monitoring and actuation points. The advantages of fog computing make it a perfect fit for the IoT scenarios. These advantages are large-scale geographical distribution, data aggregation at the edge, provision of getting minimum latency etc. In order to build a credible platform for IoT applications, the combination of fog and cloud computing makes the perfect fit for it. The authors in \cite{nastic2016middleware} have proposed a middleware which supports the multi-level provisioning of IoT cloud systems. The middleware has a generic light-weight resource abstraction mechanism, a support for automated provisioning of edge resources, and a flexible provisioning model that enable self-service, on-demand usage of edge resources.

In \cite{aazam2014fog}, the authors have presented integration of cloud and IoT. However, this integration demands the necessity of data trimming, so that the core network and cloud data centers do not get burdened. In this context, fog computing plays a great role. A smart gateway based on fog computing is proposed for data preprocessing and data trimming here. As ``anything" is getting connected to the Internet now a days, there might be a possibility that at some stage, it is no longer necessary to upload data to the cloud. Sometimes, some data may not be required in the cloud end as well. In these cases, either the device must not generate the data or the gateway device must decide when it is needed to stop uploading the data. This ensures that unnecessary resource consumptions do not happen at the network or at the cloud. The power consumption issues is also get resolved by this approach. For this to take place, the gateway devices connecting IoT to the cloud should be having some extra capabilities to do a little processing before sending it to the Internet and then to the cloud. This kind of gateway is known as ``Smart Gateway"~\cite{aazam2014cloud,aazam2014smart}. The network and cloud resources are better utilized by this approach. The Fog computing in this context helps to provide high quality streaming to mobile nodes. The applications which can be optimized by fog computing are video streaming, gaming, augmented reality, etc. This applications require low latency requirements. Since Fog nodes are localized, it has the potential to give low latency with more context awareness. Thus, the IoT and cloud computing can be integrated in order to provide better and quick service provisioning, data trimming and data pre-processing.

Due to the low-latency service provisioning with improved QoS feature of fog computing, it has become a key enabler for consumer centric IoT applications and services which need real-time responses. The authors in \cite{datta2015fog} have proposed a fog computing based architecture for connected vehicles. The authors have discussed about the M2M data processing with semantics, discovery and management of connected vehicles in the context of fog computing. The Road Side Units (RSUs) and machine-to-machine (M2M) gateways are acting as fog nodes. 

In the IoT domain, the application development can be challenging due to the presence of heterogeneous resources, widely distributed devices and processing etc. In order to overcome these challenges, the authors in \cite{giang2015developing} have proposed a distributed dataflow (DDF) programming model for IoT which utilizes computing infrastructures across the fog and the cloud. The work showed that the approach helps in the development process of fog based IoT applications.

The authors in \cite{hong2013mobile} have proposed Mobile Fog which is a high level programming model for future IoT applications which are geographically distributed, latency-sensitive as well as large-scale. The authors have analyzed the camera network and connected vehicles for the programming model in order to show the efficiency of the proposed model.

Sarkar \etal \cite{sarkar2015assessment} have analyzed the fog computing and cloud computing platform by mathematically formulating the parameters of fog in the context of IoT applications. The parameters which are considered here are power consumption, service latency, $CO_2$ emission and cost. The results showed that in IoT environment, fog computing outperforms cloud computing with the increase of latency-sensitive applications. But, it was observed that in case of less number of low-latency services, fog computing is an overhead compared to the traditional cloud computing.

Human security has become an important aspect. In the presence of IoT devices, the omnipresent cloud computing as well as fog computing, it has become possible to provide physical security to people. The authors in \cite{sehgal2015smart} have provided a security framework incorporating pervasive computing, IoT, cloud and fog computing to provide safety to the individuals. The security decisions are taken in different layers based on the complexity and urgency. The IoT layer is used for gathering knowledge of the physical surroundings and taking elementary security decisions. The cloud layer take more complex decisions and the fog layer takes real-time security decisions.

In Table~\ref{table:tabIoT}, we have summarized and compared different IoT based works on fog computing. From the table, it can be inferred that most of the works has IoT devices, fog and cloud as the preffered architecture. The IoT based services are mostly real-time services.

\begin{table*}[!t]
	\centering
	\caption{Comparison of Different Technologies for IoT}
	\begin{tabular}{|>{\raggedright}p{2cm}||>{\raggedright}p{2cm} |>{\raggedright}p{2cm}|>{\raggedright}p{2cm}|>{\raggedright}p{2cm}|>{\raggedright}p{2cm}|>{\raggedright\arraybackslash}p{2cm}|} 
		\hline
		{\bf Research work} & {\bf Architecture} & {\bf Services} & {\bf Fog Devices}  & {\bf Communication between Fog Devices} & {\bf Application} & {\bf Advantages of Using Fog}\\
		\hline \hline 
		Yannuzzi \etal ~\cite{yannuzzi2014key} & User domain, Cloud domain and Fog domain & Real-time services to the users & Scattered from core to edge of the network   & Wireless &  IoT based latency sensitive applications & Mobility of the devices is ensured  \\ \hline
		Bonomi \etal ~\cite{bonomi2012fog} & Smart things network, field area network, Core network, Cloud & Real-time services to the users & Gateways, Routers  & Wireless &  IoT based latency sensitive applications & Mobility as well as scalability is ensured  \\ \hline		
		Aazam \etal ~\cite{aazam2015fog} & IoT, Smart gatweway acting as fog, Cloud server & Dynamic resource estimation and pricing & Smart gateways  & Wireless &  Emergency, Healthcare and any latency sensitive applications & Temporary storage, preprocessing, data security and privacy etc.  \\ \hline
		Hong \etal ~\cite{hong2013mobile} & IoT, Gatweway acting as fog, Cloud server & A high-level programming model for simplification of the development on a large number of heterogeneous devices distributed over a wide area  & Gateways  & Wireless &  Large scale situation awareness applications & Large scale deployment and latency-sensitive services  \\ \hline	
		Datta \etal ~\cite{datta2015fog} & Vehicular sensors, Access points and Cloud system & Consumer centric IoT services  & Access points  & Wireless &  Connected vehicles, Smart road intersection management, Smart grid & Real-time service provisioning  \\ \hline	
		Sarkar \etal ~\cite{sarkar2015assessment} & Mobile terminal nodes, Access points and Cloud gateway & Latency sensitive services  & Access points  & Wireless &  Any IoT based system & Fog is applicable in the context of IoT for delay-sensitive applications\\ \hline	
		Sehgal \etal ~\cite{sehgal2015smart} & Three layers: IoT, Fog and Cloud & Security services  & Routers and Gateways  & Wireless &  Security providing system & Fog works as a distributed expert system for latency-sensitive security applications\\ \hline	
		Giang \etal ~\cite{giang2015developing} &  IoT, Fog node and Cloud server & Distributed data flow programming model to provide IoT services   & Routers and Gateways  & Wireless &  IoT based applications & Scalability, Mobility \\ \hline	
	\end{tabular}	
	\label{table:tabIoT}
\end{table*}

\section{Security and Privacy in Fog Computing}
\label{securityprivacy}

\begin{figure}[!t]
	\centering
	\includegraphics[width=0.5\textwidth]{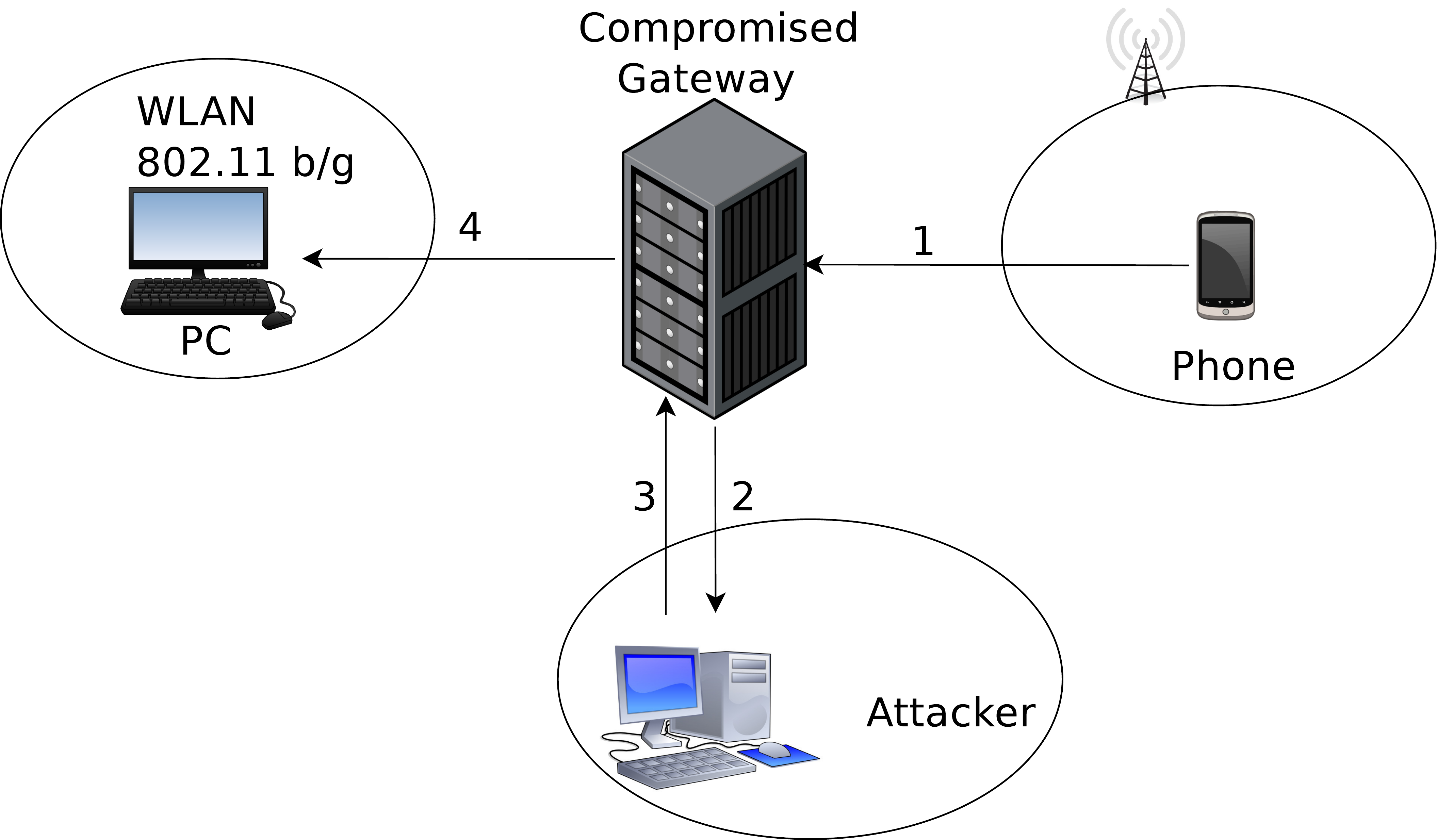}
	\caption{A man-in-the-middle attack in the Fog Computing. Steps: 1. Data is sent from the sender's device to the gateway. 2. Attacker fetches the data from the gateway and does some modification. 3. The modified wrong data is sent to the gateway. 4. Receiver gets the wrong data from the compromised gateway.}
	\label{fig:12}
\end{figure}

Fog devices are deployed in places out of rigorous surveillance and protection. Eavesdropping, data hijack etc. are the traditional attacks which can compromise the fog computing system. There are some existing intrusion detection techniques in the literature. These can be applied in the context of fog computing also \cite{modi2013survey}.  There are some signature-based methods in which the patterns of behavior are observed and checked against an already existing database of possible misbehaviors. Again, some anomaly-based method can be used in order to detect an intrusion. The observed behavior is checked with the expected behavior. If there is any deviation, then the intrusion is detected. The works of \cite{valenzuela2013real} have developed a process that detects anomalies in the input values that could have been modified by attacks. In~\cite{stojmenovic2015overview}, the authors have discussed about system security issues. The man-in-the-middle attack is one such attack which has the potential to attack the fog computing system. In this kind of man-in-the-middle attack, gateways serving as fog devices may be compromised or replaced by fake ones \cite{zhang2010man}. Private communication of the users gets hijacked once the attackers take control of the gateways. Fig.~\ref{fig:12} depicts the man-in-the-middle attack in the context of fog computing. In this attack scenario, the smartphone users send data to the gateways. However, the gateway can be compromised as the attacker gets the data from the gateway. After that, the attacker does some modification of the data and sends it to the gateway. Consequently, the gateway sends the wrong or outdated data to the client. Thus, the fog enabled gateway becomes vulnerable.

\subsection{Authentication and Authorization in Fog Computing}
There are not much studies on the authentication and authorization issues in the context of fog computing. Though there are many security solutions for cloud computing, these are not suitable for fog computing because the fog devices work at the edge of the network. The fog computing environment faces many challenges which are not present in the cloud computing environment. Fog devices generally have connectivity with the remote server. This connectivity is used to distribute authentication information and collect audit logs. However, this connectivity is very slow in certain fog computing environments. The dependency on the remote cloud servers for the authentication services is not needed sometimes as the devices might get authentication services locally when the distant cloud server is down. There are some privacy related researches in the context of fog computing. Fog computing services are provided to large number of end users by the fog nodes. So, it creates a requirement for ensuring the authenticity of a user in this environment. The authors in \cite{stojmenovic2014fog} have considered authentication at different levels of fog nodes as the main security issue of fog computing. As fog nodes are scalable, the traditional public key infrastructure (PKI)-based authentication is not efficient. The biometric authentication is prevalent in mobile and cloud computing. So, the popular biometric authentication methods such as fingerprint authentication, face authentication, touch-based authentication etc. can be applied in fog computing also.

\subsection{Privacy Assurance in Fog Environment}
The existing literature have discussed about different privacy issues in the context of IoT and how fog computing can be utilized to overcome these issues. These various issues are discussed next.

\subsubsection{Trust and Authentication} 
In fog computing scenarios the fog service providers can be different parties, such as i) Internet service providers or wireless carriers, ii) Cloud service providers who want to expand their cloud services to the edge of the network, or iii) End users who have a local private cloud and want to turn their local private cloud into fog. This ensures the trust factor in the fog computing environment \cite{yi2015security}. Ensuring trust between IoT devices plays an important role to create a secure environment in IoT systems. Reputation based trust models have been successfully deployed in many scenarios like online social networks. The reputation based trust model can be designed for IoT \cite{yi2015security}. However, this is challenging as it requires ensuring service reliability, preventing accidental failures etc \cite{alrawais2017fog}.

The security of IoT devices requires authentication. But, the problem with the IoT devices is that it don't have enough memory and CPU power to execute the cryptographic operations required for any authentication. The authors in \cite{alrawais2017fog} have mentioned that the fog devices can be used to execute authentication protocols.  

\subsubsection{Network Security} 
Wireless network security is a  big concern to fog computing. The typical attacks over a wireless environment are jamming attacks, sniffer attacks etc. In a typical network, we have to trust the configurations manually generated by a network administrator. Also, we need to separate the network management traffic from the normal data traffic. Tsugawa \etal \cite{tsugawa2014cloud} have discussed about this in their work. However, the fog nodes are deployed at the edge of the network, which creates a heavy burden to the network management. The incorporation of SDN can ease the implementation and management in the context of fog computing. This in turn reduces cost while ensuring the scalability of the fog devices.

In the context of IoT, there is a rise of face identification and resolution applications which are needed in order to have identity consistency of humans. In this direction, Hu \etal ~\cite{hu2017security} proposed an fog computing based framework which can identify a face. Face identification and resolution framework consists of client device layer, fog layer and cloud layer. The work helps in improving the processing capacity and saves the bandwidth. The work also proposed a scheme to solve the security and privacy issues. In order to solve these issues, the authors have considered the authentication and session key agreement scheme, data encryption scheme and data integrity checking scheme	to solve the confidentiality, integrity as well as availability issues.

The existing traffic light control system faces some challenges such as avoiding heavy roadside sensors, dealing with malicious vehicles and avoiding single point of failure. To overcome these issues, Liu \etal ~\cite{liu2017secure} have proposed a secure intelligent traffic light control system for vehicular adhoc network (VANET). In the architecture, the traffic lights are acting as fog nodes. The authors have proposed two secure schemes of traffic light controlling using fog computing. The hardness of the computational Diffie-Hellman puzzle and the hash collision puzzle are used in the security of these schemes.

\subsubsection{Secure Data Storage} 
In fog computing, the user's control over data is handed over to fog node, which introduces same security threats as it is present in the cloud computing environment. To address the security threats, audible data storage services are there in the context of cloud computing. Integrity, confidentiality and verifiability are provided in cloud computing scenarios to allow a client to check if its data is stored on untrusted servers. The work of Wang \etal~\cite{wang2010privacy} talks about a privacy-preserving public auditing for data stored in cloud. However, in the fog computing domain, some new challenges are present which needs to be looked into in order to have an efficient user experience. The works of ~\cite{hu2017security,basudan2017privacy,liu2017secure,koo2017privacy} have implemented secure data storage in fog systems. 

\subsubsection{Secure and Private Data Computation} 
In the fog computing scenarios, some of the computation takes place in the edge devices. The end users should verify the correctness of the computations which have been offloaded to the fog node. The privacy related challenges arises in the fog computing scenario when the sensitive user data is sent to the fog node for further processing. So, the sensitive user data has to be encrypted before sending to the fog devices. But, the lack of capability of IoT devices to encrypt as well as decrypt makes the data integrity a challenge \cite{alrawais2017fog}.

\subsubsection{Data Privacy at the End Devices} 
There is a possibility of the leakage of private information of the users, when end users are using the services of cloud computing, IoT etc. These privacy related issues are more in fog computing environment than the cloud computing scenario as the fog devices are placed in the vicinity of end users. These fog devices collect more sensitive information than the remote cloud servers. The privacy-preserving techniques have been there in different application scenarios like cloud \cite{cao2014privacy}, smart grid \cite{rial2011privacy}, wireless network \cite{qin2014preserving} etc. There are different privacy related considerations, such as data privacy, usage privacy as well as location privacy. Many privacy preserving IoT applications are present in the existing literature \cite{wei2012mobishare}. But, the resource constrained IoT devices are not well fitted to deliver efficient privacy-preserving systems.

The works of ~\cite{basudan2017privacy} discusses about the privacy-preserving protocol for enhancing security in fog based road surface condition monitoring system. The proposed system uses vehicular crowdsensing. The system consists of smart devices, rodeside units (RSU) and cloud. In this architecture, the RSUs and base stations are acting as fog devices. The authors have designed a data transmission protocol considering security aspects such as data confidentiality and integrity, mutual authentication, integrity, privacy as well as anonymity.

Wang \etal ~\cite{wang2017secure} proposed a secure and privacy-preserving real-time navigation service for Vehicular Adhoc Network (VANET). The fog nodes collect real-time road conditions by generating spatial crowdsourcing task. After that, fog nodes get traffic information from the vehicles in its coverage to find the optimal route to the destination. Vehicles get the continuous optimal route from the fog nodes until it reaches its destination. The scheme provides some security feaures such as authentication, confidentiality along with privacy-preservation. No one can link up a vehicle's navigation query and its identity. But, the trusted authority (TA) can trace the identity of the driver who provides false traffic information.

In fog computing scenario, there are privacy issues regarding the outsourced data due to the complexity of the system. Koo \etal ~\cite{koo2017privacy} have proposed a privacy-preserving deduplication protocol which is able to manage ownership in fog computing systems. The protocol is able to perform access control by user-level key management and update mechanisms. The proposed scheme is efficient in terms of communication and key management in fog where the ownership changes very often.

\subsubsection{Access Control} 
Access control is an important aspect for the consideration of the security and privacy of a user. The traditional access control is implemented in the same trust domain. Nevertheless, cloud computing uses the outsourced data and implements the access control cryptographically. In the fog computing domain, there is a need to support secure collaboration and interoperability between the heterogeneous resources. In this direction, the authors in \cite{dsouza2014policy} have proposed a policy-based resource access control in fog scenarios. However, in the fog computing environment designing of access control and at the same time meeting the resource constraints is a challenging work. 

\subsubsection{Intrusion Detection} 
Intrusion detection techniques are widely deployed in the cloud systems to mitigate attacks such as insider attack, flooding attack, port scanning etc. \cite{modi2013survey}. These techniques can also be used in smart grid system to monitor power meter measurements and to detect abnormal measurements that could have been compromised by attackers \cite{valenzuela2013real,qin2013defending}. In fog computing environment, intrusion can be detected by monitoring and analyzing log files, access control policies and user login information. But the intrusion detection creates many challenges in the fog computing systems in order to meet the low-latency.

\subsubsection{Rogue Node Detection} 
Some IoT node may pretend to be legitimate in order to exchange and collect the data generated by other IoT devices for malicious purposes. Ma \etal have proposed a framework which is able to detect the presence of rough access points in Wi-Fi networks \cite{ma2008hybrid}. In the IoT systems, addressing this issue is difficult due to the complexity in trust management in various scenarios \cite{alrawais2017fog}.

In Table~\ref{table:tabSecurity}, we have compared different security and privacy issues which arise in fog computing. Form the table ~\ref{table:tabSecurity}, we can infer that most of the works have considered authentication. Again, very less amount of work have considered data privacy at end devices.

\begin{table*}[!t]
	\centering
	\caption{Comparison of Different Security and Privacy issues in Fog Computing}
	\begin{tabular}{|>{\raggedright}p{1.8cm}||>{\raggedright}p{1.8cm} |>{\raggedright}p{1.8cm}|>{\raggedright}p{1.8cm}|>{\raggedright}p{1.8cm}|>{\raggedright}p{1.8cm}|>{\raggedright}p{1.8cm}|>{\raggedright\arraybackslash}p{1.8cm}|} 
		\hline
		{\bf Research work} & {\bf Application} & {\bf Fog nodes} & {\bf Authentication}  & {\bf Secure data storage} & {\bf Secure and private data computation} & {\bf Data privacy at end devices} & {\bf Access control}   \\
		\hline \hline 
		Hu \etal ~\cite{hu2017security} &  Security and privacy preservation scheme of face identification and resolution  & Router, gateway, dedicated server & Yes   & Yes & Yes  &  No &   Yes   \\ \hline
		Basudan \etal ~\cite{basudan2017privacy} & Secure road surface condition monitoring & RSU, Base stations & Yes  & Yes   & Yes & Yes  &  No     \\ \hline
		Wang \etal ~\cite{wang2017secure} &  Secure and privacy-preserving real-time navigation service & Road Side Units (RSU)  & Yes   & No & Yes  & Yes  & Yes     \\ \hline
		Liu \etal ~\cite{liu2017secure} &  Security based intelligent traffic light contol & Traffic light   & Yes   & Yes & Yes  & No  & No     \\ \hline
		Alrawais \etal ~\cite{alrawais2017fog} &  Improving the distribution of certificate revocation information among IoT devices for better security & Smart devices   &  Yes  &  No & Yes  & No  & Yes      \\ \hline
		Koo \etal ~\cite{koo2017privacy} &  Deduplication of encrypted data in fog storage & Routers, Gateways  &  Yes  & Yes & Yes  & Yes  &  Yes    \\ \hline
		Stojmenovic \etal \cite{stojmenovic2014fog} & Secure Machine to	machine	networks &Smart gateway & Yes   & No & No  & No  & Yes    \\ \hline
		Dsouza \etal~\cite{dsouza2014policy}& Secure collaboration between different user-requested resources & Smart devices &  Yes  & No & No  & No  & Yes     \\ \hline
	\end{tabular}	
	\label{table:tabSecurity}
\end{table*}

%% file: 5_qos_parameters.tex
\section{QoS Parameters in Fog Computing and Communication}
\label{qossection}
The end users always need better experience of the services obtained from the providers. In order to improve the quality of service of the end users, fog computing plays a potential role. We have pointed out about the different quality of improvement that can be brought about by the application of fog systems.

\subsection{Reliability}

The system should perform and give correct results in the context of real-time applications. We can leverage fog computing in order to achieve this. This is known as reliable fog based system. In the following, we discuss about the reliable fog systems.

In~\cite{madsen2013reliability}, the authors have pointed out various reliability issues in wireless networks. The reliability issues can be summarized as radiated electromagnetic interference, end-to-end packet reliability etc. Fog computing should address these issues minimizing the overall delay and with high reliability. In the context of IoT, we have to take into consideration the failure of individual sensors, the lack of coverage from access network in some region, the failure of whole network, the failure of the system platform, the failure of the user's interface connected to the system etc. Stojmenovic~\cite{stojmenovic2014fog} has discussed a SDN based reliable fog architecture with use cases like vehicular network and demand response management, where the SDN like architecture is utilized to manage communication over fog devices while ensuring end to end reliability.  In~\cite{huang2016reliable}, the authors have developed a real time reliable streaming mechanism by utilizing fog based systems. The key aspect of their proposal is that the streaming content is preallocated from fogs and clouds that can sustain the quality of service for real time streaming. The mobile devices at the edge reserve the content for streaming, which is based on three aspects -- (i) a stochastic prediction of its locations, (ii) the amount of content required for streaming, and (iii) the required parameters used for reservation. Therefore, the reliability of the real time streaming applications can be effectively maintained. The system can further ensure the performance of real time streaming when the devices change their mobility behavior, by tuning the streaming parameters at the fog devices. 

Application specific reliability can also be ensured using fog computing framework. A major utility of fog computing is that the edge devices can take care of reliable communication as well as reliable computation near the end devices. In~\cite{craciunescu2015implementation}, the authors have ensured the reliability of e-health applications using fog computing platform. The work addresses the technical challenge of having an impedance mismatch between the characteristics of today's cloud infrastructure and the requirements of smart connected object applications within the sensing environment. The authors have investigated the possibility to offload cloud tasks, such as storage and data signal processing, that improves the reliability of the cloud infrastructure.  A reliable communication platform for fast moving rail has been explored in~\cite{wang2017reliable}, where the authors have utilized fog computing platforms where a fog layer is placed in between the 3G infrastructure and the end users. This new fog layer introduces a series of mutually chained network gateways, which are located in different rail compartments to provide reliability. The authors have shown that such architecture can significantly improve the reliability and performance guarantee of the wireless communication for high speed railways. Reliable fog layers have also been introduced in the operating systems utilized at the smart devices.

Although smartphones can act as a promising fog device for computation at the edge because of their computation capability, however to support a wide range of applications as well as to adhere to the resource constraints, the software stack of the smartphones needs to be reliable and adaptable. The reliability over Android based fog computing environment has been explored in~\cite{dantu2017raina}. The authors have discussed the required modifications at the Android protocol stack to ensure computation reliability. 

\subsection{Energy Consumption}

In the fog computing environment, some of the requests are redirected to the cloud computing core for further data processing. For this to take place, there is a cost which is incurred in order to do the data uploading. In \cite{sarkar2016theoretical}, the authors have modeled the fog computing architecture theoretically and analyzed its performance in the context of IoT scenarios. The authors have shown that in the fog computing domain, the overall upload cost increases with the increase of the percentage of requests that are required to be redirected to the cloud core. The paper also mentions that the energy consumption due to transmission for the fog computing structure is lower than that for cloud computing. One interesting point that has been observed in this paper is as follows. If the low-latency IoT applications are very less, the energy consumption for fog computing will be higher. As the number of requests referred to the cloud increases, the processing energy increases almost linearly. Also, they have found that in the context of IoT applications, with approximately one-fourth of the requests need real-time services and the fog computing structure improves the mean energy consumption by $40.48\%$.

Energy management is of high importance for micro grids, homes and buildings. In \cite{vatanparvar2015energy,al2016energy}, the authors have proposed a platform which is energy efficient by using fog computing. The authors have implemented energy management as a service over fog computing domain. The real-time requirements needed for energy management are provided by the proposed system over fog computing domain. The open source software as well as hardware and the ability to be customized provide the user to get the control as a service in the point of view of energy management. As a result of this, the implementation cost and time-to-market gets decreased significantly. The system also has the following features,
\begin{enumerate}
	\item[(i)] low-power and low-cost devices for computation, storage and communication,
	\item[(ii)] scalability, and
	\item[(iii)] service oriented architecture for abstracting the communication and hardware heterogeneity.
\end{enumerate}

One of the advantages of cloud computing was that it was highly energy efficient as it could consolidate resources on physical machines. When we distribute machines in fog computing, that cannot be achieved. So, we need a perfect balance between falling back to cloud - which is poor for QoS, and fully utilizing the edge, which is energy wastage.

\subsection{Delay Sensitive Services over the Fog}

Fog computing can help in realizing the throughput performance of real-time delay sensitive applications. The authors in \cite{dastjerdi2016fog} have discussed about the applicability of fog computing in IoT environments. The IoT is the new paradigm where consumer electronic items, home appliances etc. get connected in order to make smart city, smart infrastructure \cite{buyya2016internet}. Though cloud computing could help by providing on-demand and scalable storage, processing of data etc in IoT domain, there are some latency-constrained applications for which cloud computing is not a good option. We can bring fog computing into the picture so that the real-time services can be provided. This will also make the perfect utilization of network bandwidth. The IoT applications compete for the limited resources. Fog computing seamlessly interplays with the cloud resources and the edge devices in order to provide the resources in IoT. This way the overall delay sensitive services are provided in fog computing paradigm.

In order to decrease the delay involved in website rendering, the fog computing approach can be implemented. The website rendering performance can be improved by the fog nodes and this delivers better results than the approaches present in the web servers. Zhu \etal \cite{zhu2013improving} have proposed their work in this direction. The work utilized the knowledge which is available in the network edge to make the system more adaptable to dynamic network conditions.

The low latency computing and communication services are not effectively provided by the existing mobile and telecommunication systems. The fog based radio access networks are able to provide the delay requirements of these systems\cite{shih2017enabling} seamlessly. Fog based radio access network brings the computing capability of the cloud to the edge of the network. The idea is to utilize the  computing resources of the existing radio access equipments on small or micro cells. F-RAN is able to provide the IoT applications these resources to minimize the end-to-end delay. The authors have also discussed about the trade-offs among performance, the computing cost and the communication cost.

The latency-sensitive applications needs a better infrastructure than the existing cloud based systems in order to improve response time as well as energy consumption. The computation offloading to the edge of the network can help in providing this aspect in the fog computing domain. The authors in~\cite{hu2016quantifying} talk about this. The work shows that the highly interactive mobile applications perform better by the application of fog computing in Wi-Fi as well as 4G LTE networks. The work also showed that offloading the services to a distant cloud consumes more energy than running in a local mobile node. The fog computing plays as a key enabler for the mobile augmented reality as well as cognitive assistance applications.

Hao \etal \cite{hao2017challenges} have discussed about the challenges which are present in the cloud computing scenario. The delay element is one of the research challenges which is present in the cloud domain. The authors have proposed a software architecture which is based on fog computing. The results have showed that real-time communications can be improved by this architecture in terms of delay. The work focused on synchronization policies, locking policies and migration policies.

The authors in \cite{mubeen2017delay} have worked towards the interplay of cloud, fog and IoT in control applications of the automation industry. The work does the proper management of offloading the controller to the cloud or the fog based on the delay requirement of the applications. The work also performed mitigation of delays caused by the network when the controller is offloaded to the fog or cloud. The usage of IoT devices have helped to perform local computation in local fog nodes for delay sensitive applications.

\subsection{Quality of Experience for the End Users}

Fog based systems have the potential to improve the quality of experience to the end users. The works of \cite{prazeres2016soft} have applied fog computation in the context of IoT in order to facilitate the interoperability of IoT ecosystems through the delivery of services over a virtual infrastructure. The authors have proposed a fog based IoT platform known as {\em SOFT-IoT} (Self-Organizing Fog of Things). The local users can access data and devices by the fog node and the remote users can use the cloud to access. This way a better quality of experience is provided to the end users. A gateway have been enabled to work as a smart device based on fog computing. This gateway is known as Fog of Things gateway in the proposed platform.

The future wireless access networks require an improved service provisioning in order to response better. In this direction, the works of \cite{iotti2017improving} have proposed a model of Internet network access based on fog computing. The model uses the virtual machines in order to dynamically move cloud or web contents to the edge nodes. The system performs proactive caching and implements traffic policies based on the interaction between access infrastructure and applications. The results showed that the system is able to optimize bandwidth usage with reduced latency. Thus, the system is able to provide better quality of experience with ideal resource management to the end-users of the wireless access networks.

The authors in \cite{osanaiye2017cloud} have described the fog computing architecture. The work also analyzed the different services and applications of fog computing in comparison to cloud computing. The fog computing systems are required in order to build the smart city, smart home etc. These are essential for better quality of experience for the users. The work also presented a smart pre-copy live virtual machine migration approach. This minimizes the downtime as well as the migration time to provide resource and service availability to the end users for better quality of experience.

The works of \cite{aazam2016mefore} have estimated resources using fog computing in order to enhance quality of experience in IoT systems. The IoT nodes have fluctuating behavior if they are mobile. Fog helps in estimating resources based on the behavior of the IoT nodes. As the IoT devices are heterogeneous, it is difficult to predict how much resource will be consumed and also the full utilization of the resources is unpredictable. The work which is based on fog helps in overcoming these issues. The proposed methodology is known as {\em Media Fog Resource Estimation} (MeFoRE) which provides resource estimation according to the behavior and historical data of the customer as well as enhanced QoS.

Social aware device to device communication is one of the most critical part in the context of fog computing. The authors in \cite{flores2016social} have developed a software sensor called {\em Detector} which is used to sense the infrastructure in the proximity of a mobile user. These discovered devices can be used to support the processing of other devices in the fog environment. The authors in \cite{farris2015social} have proposed a {\em social IoT} (SIoT) platform based on fog computing. This platform is able to detect the need to change the geographical location of the virtual object and to manage the inter-cloud mobility of processes and data. The proposed system is providing better quality of experience to the users than a cloud based SIoT platform.

\subsection{Network Caching in Fog for Providing QoS at the End Devices}

The cellular networks get burdened due to the consumption of large amount of data by the users. This is also true in the presence of vehicular network as the users consume entertainment data and navigation data. The network will be less congested if proper caching is in place. Malandrino \etal \cite{malandrino2016price} have defined a metric called price-of-fog metric which is the additional caching to deploy when moving from traditional, centralized caching to a fog based caching. In the fog based caching, the caches are closer to the network edge. This approach reduces the service time for vehicular applications.

The network caching has also been incorporated in {\em Information Centric Networking} (ICN). The fog computing nodes are used for edge caching in the network. The authors in \cite{abdullahi2015ubiquitous} have proposed a framework of applying Information Centric Networking (ICN) as API to ubiquitous computing. ICN cache is done at the edge nodes through fog computing by referring object with names instead of IP addresses. It helps in accessing information residing on the cloud nearer to the user. This framework helps in realizing the benefits of getting faster response in IoT. The works of \cite{hao2017challenges} have proposed about a fog computing based architecture which is using caching in fog nodes. This caching helps in getting faster response than the response obtained from cloud. The software architecture has cache manager component which is responsible for caching in fog nodes.

In \cite{paschos2016wireless}, the authors have discussed about how fog can be used for caching in future wireless networks. The base stations and the user devices can be used for caching purpose. The overall end to end quality of service gets improved by applying this edge side caching. There is a need to share resources in the widely deployed fog clusters. In this direction, the authors in \cite{jingtao2015steiner} have proposed a new caching scheme known as Steiner tree based caching scheme. The fog nodes produce a Steiner tree to minimize the total path cost which in succession minimize the cost of resource caching. The proposed caching scheme has been compared with the shortest path one. The proposed caching scheme is performing better in terms of reduced cost of data sharing.

Though fog computing performs better in comparison with cloud alone, these edge nodes have limited resources. So, there is a need to conserve energy. Wireless communication at a short distance is efficient to reduce energy consumption of these mobile devices in fog. Again, the utilization of edge nodes is to be improved and also, we need to guarantee that mobile nodes can access the edge nodes rapidly. In this direction, the authors in \cite{wang2016cachinmobile} have proposed a novel paradigm mobile caching network. This caching network has energy-efficient edge nodes and the paradigm is known as {\em CachinMobile}. The paradigm leverages social networking and device-to-device (D2D) communication.

The fog computing paradigm can help in processing end user requests in the fog nodes. The authors in \cite{fan2016web} have implemented web resource caching in edge devices to serve as a caching proxy server. The end devices can also be used for resource storage along with the edge devices. The proposed architecture has better downloading latency in comparison with single caching proxy approach. Also, the resource caching in fog network reduces the cost of data transmission through core network. Though this approach improves the latency requirements, it has not considered the security aspects when caching the resources at the end devices.


In Table~\ref{table:tabQoS}, we have compared different quality of services aspects of fog computing. From the table, we can infer that very less amount of works have considered quality of experience as well as in-network caching.

\begin{table*}[!t]
	\centering
	\caption{Comparison of Different QoS aspects of Fog Computing}
	\begin{tabular}{|>{\raggedright}p{1.8cm}||>{\raggedright}p{1.8cm} |>{\raggedright}p{1.8cm}|>{\raggedright}p{1.8cm}|>{\raggedright}p{1.8cm}|>{\raggedright}p{1.8cm}|>{\raggedright}p{1.8cm}|>{\raggedright\arraybackslash}p{1.8cm}|} 
		\hline
		{\bf Research work} & {\bf Application} & {\bf Fog nodes} & {\bf Reliability}  & {\bf Energy Efficiency} & {\bf Delay Sensitiveness} & {\bf Quality of Experience} & {\bf In-network Caching}\\
		\hline \hline 
		Madsen \etal ~\cite{madsen2013reliability} & Utility computing based applications & Smart devices  & Yes   & No & Yes  & No  & No \\ \hline
		Stojmenovic \etal ~\cite{stojmenovic2014fog} & Machine to machine networks & Smart gateway  & Yes   & No & Yes  & No  & No \\ \hline
		Huang \etal ~\cite{huang2016reliable} & Vehicular Networks & Smart devices  & Yes   & No & Yes  & No  & No \\ \hline
		Craciunescu \etal ~\cite{craciunescu2015implementation} & E-health applications & Smart devices  & Yes   & No & Yes  & No  & No \\ \hline
		Wang \etal ~\cite{wang2017reliable} & Rail applications & Smart gateways & Yes   & No & Yes  & No  & No \\ \hline
		Dantu \etal ~\cite{dantu2017raina} & Smartphone based applications & Smartphone  & Yes   & No & Yes  & No  & No \\ \hline
		Sarkar \etal ~\cite{sarkar2016theoretical} & IoT based applications & Smart devices  & No   & Yes & Yes  & No  & No \\ \hline
		Al \etal ~\cite{al2016energy} & IoT based applications & Smart devices  & No   & Yes & Yes  & No  & No \\ \hline
		Dastjerdi \etal ~\cite{dastjerdi2016fog} & IoT based applications & Gateways and private clouds  & Yes   & Yes & Yes  & No  & No \\ \hline
		Zhu \etal ~\cite{zhu2013improving} & Website rendering & Gateways  & No   & No & Yes  & No  & Yes \\ \hline
		Hu \etal ~\cite{hu2016quantifying} & Mobile Applications & Base station or WiFi access points  & No & Yes & Yes  & No  & No \\ \hline
		Hao \etal ~\cite{hao2017challenges} & Ubiquitous computing & Smart devices  & No & Yes & Yes  & No  & Yes \\ \hline		
		Mubeen \etal ~\cite{mubeen2017delay} & Automation Applications & Smart devices  & No & No & Yes  & No  & No \\ \hline	
		Shih \etal ~\cite{shih2017enabling} & Radio Access Networks & Base stations  & No & Yes & Yes  & Yes  & No \\ \hline	 	
		Prazeres \etal ~\cite{prazeres2016soft} & IoT based applications  & Gateways  & No & Yes & Yes  & Yes  & No \\ \hline	 	
		Farris \etal ~\cite{farris2015social} & Social IoT based applications  & Smart devices  & No & No & Yes  & Yes  & No \\ \hline			
		Flores \etal ~\cite{flores2016social} & Social aware device to device communication  & Smart devices  & No & No & Yes  & Yes  & No \\ \hline	
		Fan \etal ~\cite{fan2016web} & Web based applications  & Edge devices  & No & No & Yes  & No  & Yes \\ \hline	
		Wang \etal ~\cite{wang2016cachinmobile} & Device-to-device communications  & Edge devices & No & Yes & Yes & Yes  & Yes  \\ \hline
		Jingtao \etal ~\cite{jingtao2015steiner} &  Latency sensitive applications & Edge devices  & No & Yes & Yes  & No  & Yes \\ \hline	
	\end{tabular}	
	\label{table:tabQoS}
\end{table*}

%% file: 6_applications.tex
\section{Applications over Fog Computing Framework}
\label{application}
In this section, we discuss about the potential use-cases of fog computing as explored in various recent literatures, such as radio access network, e-health, VANET, smart city, augmented reality and real-time video analytics, content delivery and caching as well as mobile big data analytics. These application areas need bandwidth as well as latency-sensitivity. Fog computing can be used in these contexts to solve the challenges present in these domains. Here, we discuss how various requirements of such applications can be fulfilled with the help of fog computing framework. 

\subsection{Developments of Fog based Radio Access Networks}

In the development of 5G systems, the radio access points perform a key role in order to provide the proper deployment of the system. The radio access points help to properly deal with this increasing data traffic. Moreover, the radio access points can be used as the local computing and storage devices which can minimize the burden of the cloud server. These radio access points thus act as a fog node. In~\cite{oueis2015small}, the authors have proposed an approach for the problem of radio access points clustering for fog computing. This clustering is needed in order to provide adaptive sizing and resource management of computation clusters. Again, this approach guarantees a higher quality of experience (QoE) by proper usage of available resources. The solution gives higher user's satisfaction ratio while minimizing the communication power consumption.

Earlier, we have discussed the utilities of cloud based radio access networks (could-RAN or CRAN)~\cite{checko2015cloud}, where the concept of device virtualization can be utilized to support energy efficiency as well as spectral efficiency by segregating radio equipment from the baseband processing units (BBUs), and hosting the BBU over a remote cloud. However, the delay and traffic load at the network front-haul that interconnects the radio equipment and the cloud hosting the BBU pool impacts the performance of the CRAN architecture. Consequently, recent research works have explored fog based radio access networks, as pointed out in~\cite{hung2015architecture,peng2016fog,xiang2015joint,yan2016user,park2016joint} and the references therein. In fog based radio access networks, the advantage is that the BBU pool can be kept near the radio equipment, while utilizing the advantages of virtualized network functions. 

\subsection{Fog Computing for e-Health}

E-health monitoring is a much talked about application of wireless sensor network. It has the potential of ubiquitous monitoring of the patients \cite{schwiebert2001research,schwiebert2002biomedical}. A report from World Health Organization (WHO) states that the majority of the patients forget to take their medicines in accurate time \cite{sabate2003adherence}. In these situations, the smart sensors are capable of sensing the human data and send the relevant actuation based on the sensed data. However, these sensors have limited battery power and also they are alone not efficient for data processing as well as data storage. Health monitoring systems are perfect for context-aware computing, wherein the outcome of an application is related to the context sensed by the end devices. Fog computing acts as a key enabler in these situations. Fog nodes can effectively store and pre-process the sensed data. They forward the relevant features from the sensed data to the remote cloud server for further analysis. Based on the gathered data from the fog devices, the cloud server can compare the patient's medical history and can provide the required prescription for further treatment. Cloud computing alone cannot provide these services in e-health situations, as many of the computation are local and latency sensitive. The work of Shnayder \etal~\cite{shnayder2005sensor} talks about some issues that the cloud server face. The large volume of data generated by the heterogeneous and geographically distributed IoT devices have created a large volume of data~\cite{vera2013business}, which needs extra attention for data storage as well as processing. Fog computing is best suited in these situations because of its ability to provide context aware, latency-sensitive and highly scalable solutions. Cloud and fog together can revolutionize the healthcare sector like any other sectors. The research and development of smart e-health technologies have the potential of providing remote treatment to the patients reducing the requirement of visiting the medical practitioner~\cite{fong2013mobile}.

\begin{figure}[!t]
	\centering
	\includegraphics[width=0.5\textwidth]{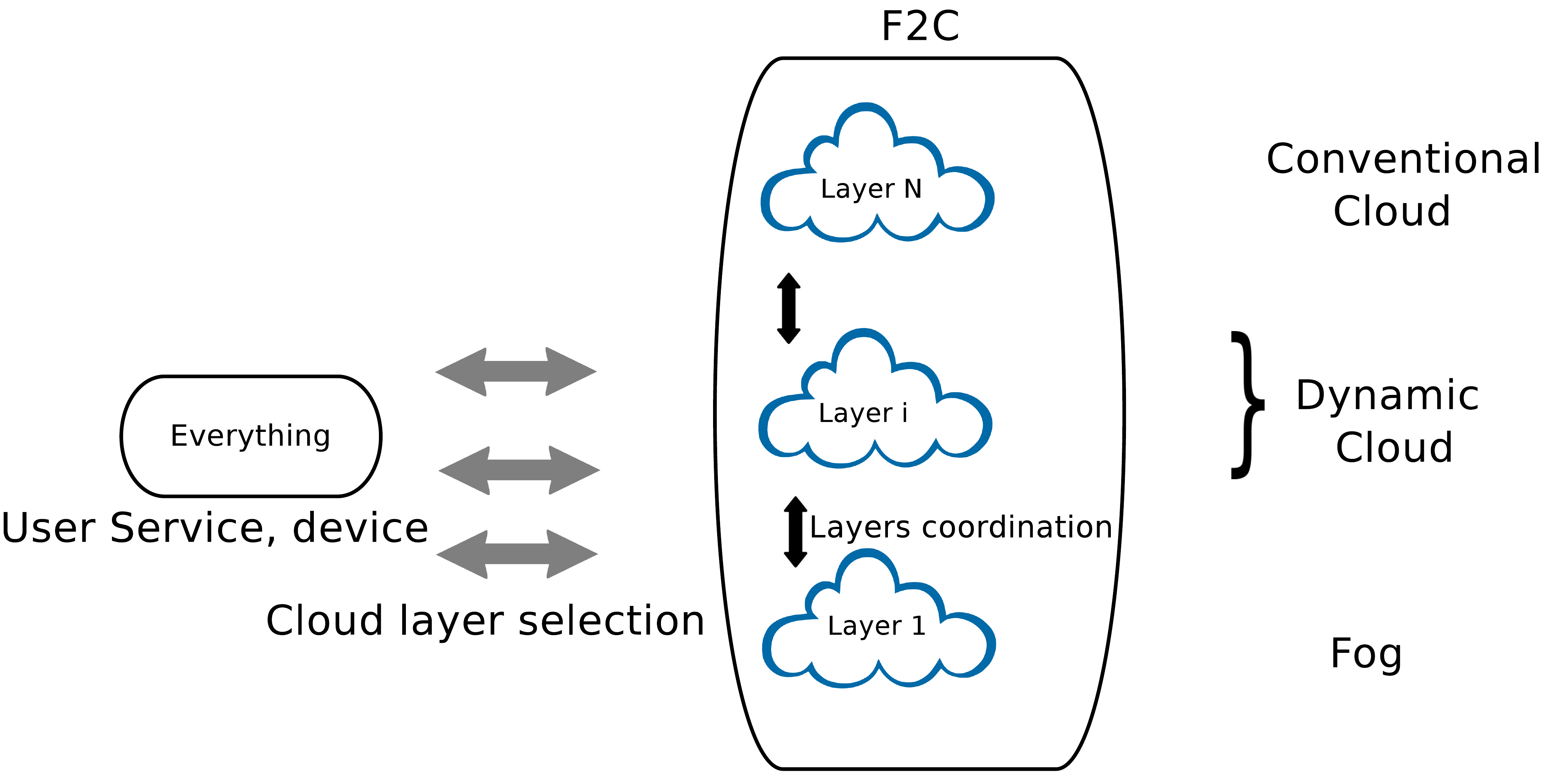}
	\caption{A Fog Computing based Healthcare Application~\cite{masip2016fog}}
	\label{fig:foghealthcare}
\end{figure}

In~\cite{masip2016fog}, the authors have proposed an architecture namely fog-to-cloud. This system has many capabilities as follows, 
\begin{enumerate}
	\item[(i)] real-time monitoring of the patient's oxygen doses,
	\item[(ii)] real-time estimation of the patient's effort,
	\item[(iii)] patient’s therapy based on activity,
	\item[(iv)] context information collection and processing, and
	\item[(v)] patient's therapy tuned to context information.
\end{enumerate}
Fig.~\ref{fig:foghealthcare} shows the proposed architecture for fog-to-cloud system. The fog-to-cloud system has the capacity to utilize the most effective set of resources (i.e. infrastructure as well as data) to run a service. This services can be executed in a cloud or in a fog device or even in a combination of them. The proposed fog-to-cloud system enables a context-aware and patient-tailored tuning of the oxygen volume.  The user services and devices select the required layer for getting the service. These layers can be fog layer, dynamic cloud as well as conventional cloud layers.

In~\cite{craciunescu2015implementation}, the authors have implemented fog computing in order to do real-time monitoring of the patients and to notify users in case of a gas leak, abnormal range of patient’s pulse or oxygen level, falling of patients. The authors have studied the possibility to offload some services from the cloud servers to the edge of the network. This decreases the delay in performing these tasks within the cloud. The real-time processing is performed at the home personal computer and the extracted metadata is sent to the cloud for further analytics.

\subsection{Fog Computing for Vehicular Environments}

The integration of IoT devices, cloud computing and SDN has generated a smart vehicular network known as smart transportation system. Again, the requirement of location aware services near to the sensing devices have created the demand for data processing and storage near the edge devices through the fog computing framework. Fog computing helps to realize the delay-sensitivity of the smart transportation system. In these scenarios, the time required for data transfer and decision making process is very less in order to avoid vehicle collision. There are several security threats to smart transportation system. In order to overcome these issues, there is a need for a system which would cater the data availability, confidentiality and integrity. System authentication is also another security aspect to have a consideration.

\begin{figure}[!t]
	\centering
	\includegraphics[width=0.5\textwidth]{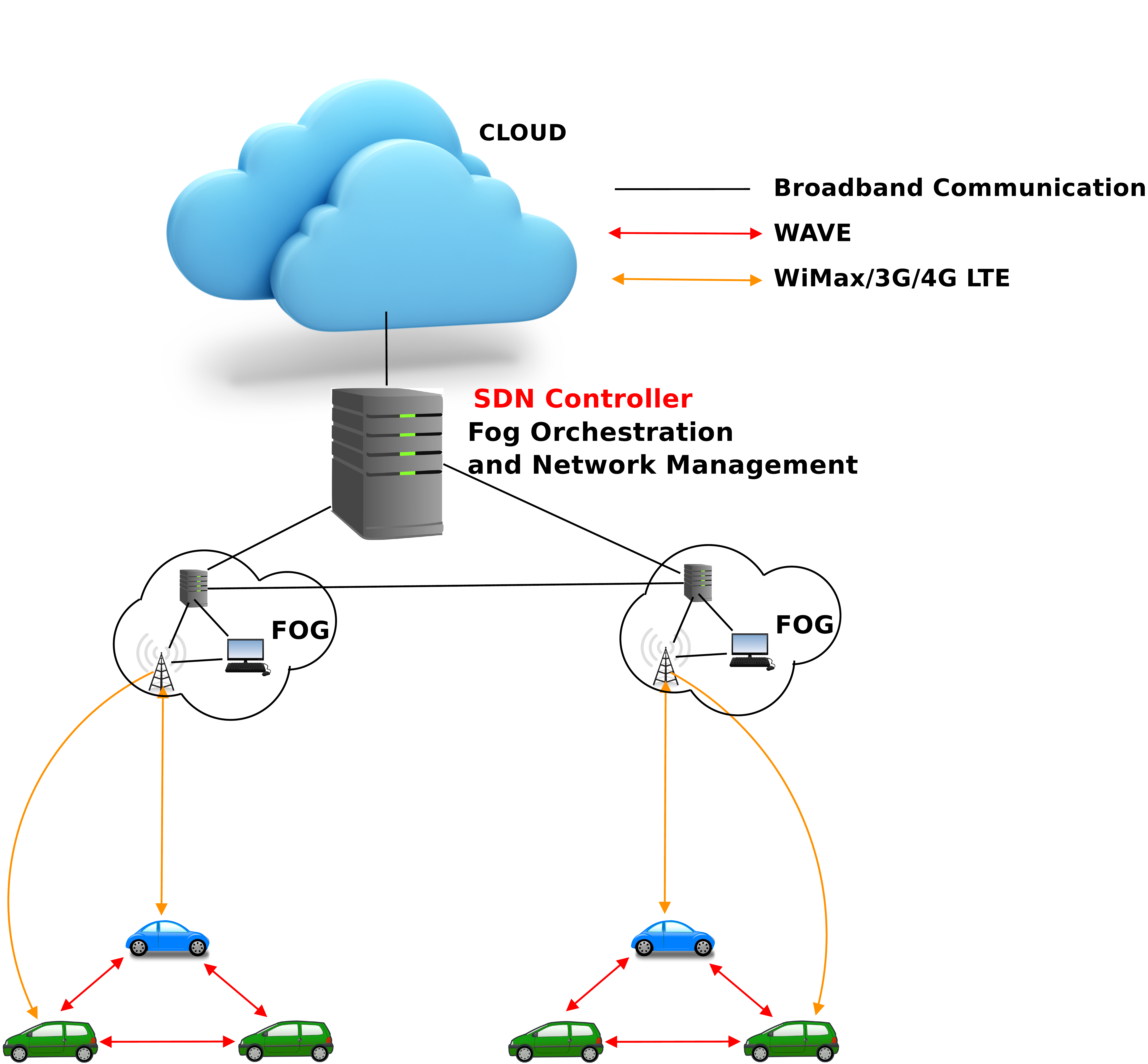}
	\caption{Fog and SDN based Architecture for Vehicular Adhoc Networks~\cite{truong2015software}}
	\label{fig:fogsdnvanet}
\end{figure}

SDN has the potential to provide efficient management and deployment of network services. The SDN based system provides flexibility, scalability, programmability as well as global knowledge. Whereas the fog computing paradigm provides latency sensitive and context-aware services. In~\cite{truong2015software}, the authors have proposed an fog and SDN based architecture (FSDN) for vehicular adhoc networks (VANETs). Fig.~\ref{fig:fogsdnvanet} shows an instance of this architecture. The network of vehicles are connected to the fog layer by the cellular networks. The fog network is connected with the SDN controller. SDN layer does the fog orchestration and network management services. The SDN controller is connected to the cloud computing layer.

The FSDN VANET architecture can optimally configure service deployments, dynamically reconfigure itself for better quality of service. The various layers for FSDN architecture are as follows. 
\begin{enumerate}
	\item[(i)] \textit{SDN Controller}: It has the global knowledge which helps to control all the network behavior of the system. Fog orchestration and resource management is also done here.
	\item[(ii)] \textit{SDN Wireless Nodes}: The vehicles work as the end-users as well as forwarding element.
	\item[(iii)] \textit{SDN Road-Side-Unit (RSU)}: It is a fog device. It runs OpenFlow and it is controlled by the SDN controller.
	\item[(iv)] \textit{SDN Road-Side-Unit Controller (RSUC)}: A group of RSUs are connected to a RSUC through broadband connection before accessing to the SDN controller. It also has OpenFlow and it is also controlled by SDN controller. These are also fog devices.
	\item[(v)] \textit{Cellular Base Station (BS)}: It can also provide fog services and is controlled by SDN controller.
\end{enumerate}

\begin{figure}[!t]
	\centering
	\includegraphics[width=0.5\textwidth]{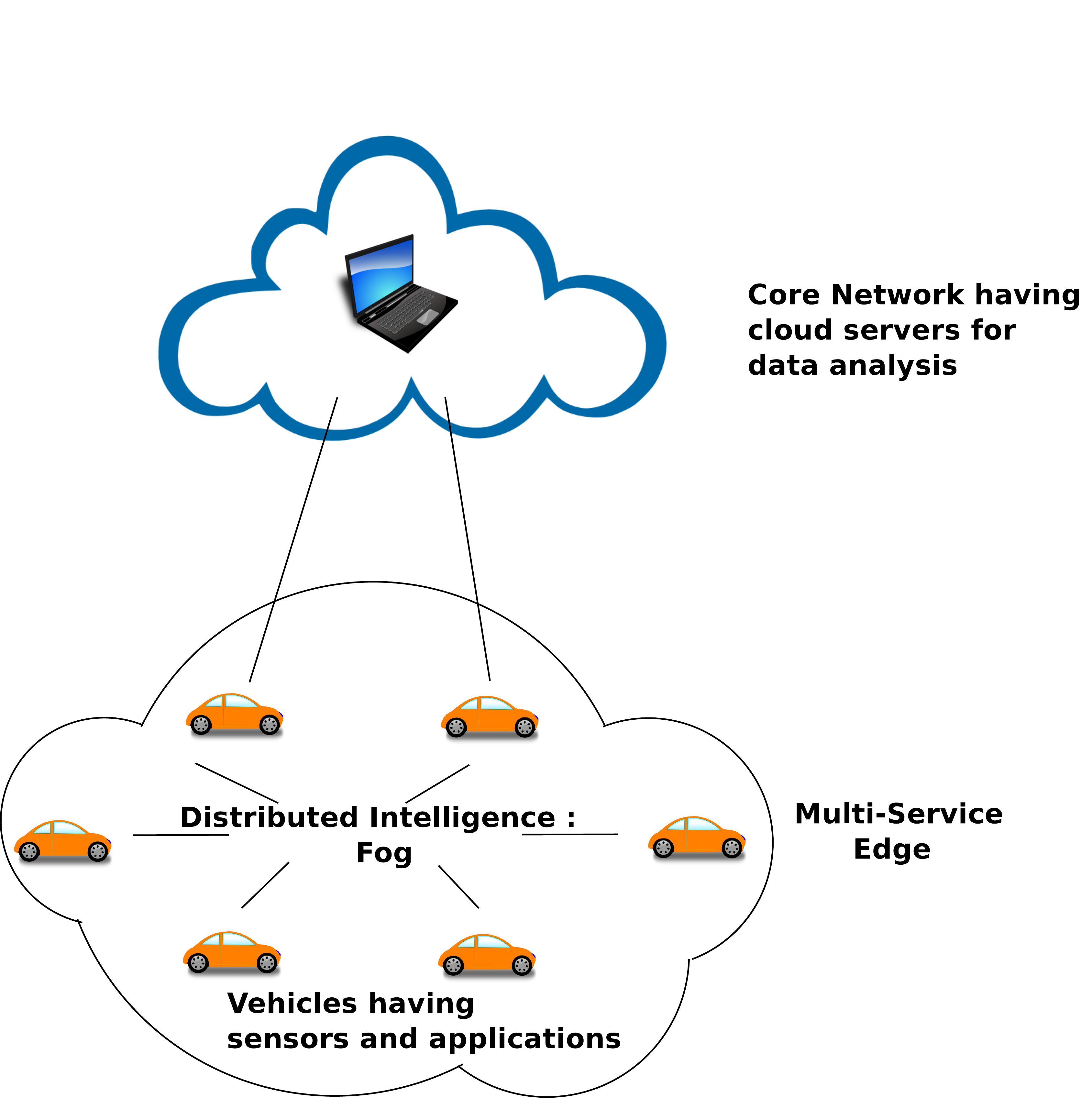}
	\caption{An Architecture for Vehicular Network based on Fog Computing~\cite{houvehicular}}
	\label{fig:vehicularnet}
\end{figure}

In~\cite{houvehicular}, an architecture for vehicular network based on fog computing i.e. vehicular fog computing (VFC) is proposed. This is an overview of vehicles as the infrastructures for communication and computation, which is a new paradigm referred to as vehicular fog computing. Fig.~\ref{fig:vehicularnet} gives an idea about the usage of fog computing for vehicular networks. The vehicles having sensors and applications have created a fog network which is also known as the multi-service edge. Fog layer provides the distributed intelligence. The fog layer is connected to the core network having cloud servers for data analysis.

By utilizing the computation of the edge devices, fog based vehicular environment can give better QoS to the end users. The proposed VFC has the advantage of providing more reliable communication with higher capacity. The computational performance gets improved due to the usage of currently underutilized computational resources of individual vehicles. 
The authors in \cite{gosain2016enabling} have proposed an architecture which has the capability of sensing and controlling applications running on cars. This helps in collecting the datasets for public safety surveillance. The connected vehicle applications can be optimized in terms of the latency and bandwidth by using the edge computing cloud infrastructure.

\subsection{Fog Computing for Smart City Applications}

Fog computing deals with the shifting of the computation logic at the edge of the network where the data needs to be quickly processed and the required actions are needed to take place very fast. The whole application logic should not be offloaded to the cloud server as this make the network congested. This is also applicable in the context of smart city based IoT applications where thousands of smart objects, vehicles etc. interact in order to provide effective services. In~\cite{bruneo2016stack4things}, the authors have proposed {\em Stack4Things}, an Open Stack based framework which is capable of managing the IoT infrastructure. This framework involves the Infrastructure-as-a-Service and Platform-as-a-Service layers. This system can decide with efficient analysis about which specific tasks to delegate to the cloud infrastructure. The proposed Stack4Things is using fog computing paradigm for smart city applications. They have provided the framework which is relevant in smart mobility scenario where vehicles interact with smart objects to provide highly responsive geo-localized services. The authors have developed a distributed application, which can facilitate the end user with services deriving from the interaction of its smart car with different smart city objects i.e. smart traffic lights, smart streetlights, smart advertising billboards etc. They have handled the mobility of the nodes by allowing the developers to dynamically build or destroy cloud based virtual networks. 

Aazam \etal \cite{aazam2014fog} have worked towards building a smart gateway based communication along with fog computing for smart city based applications. The trimming and pre-processing of data before sending to cloud is very important in order to lessen the burden on the cloud server. This helps in quick service provisioning in the context of smart cities. The real-time delay sensitive applications can be responded in quick time as the system reduces commnication overhead of the core network.

Sehgal \etal \cite{sehgal2015smart} have proposed a fog based security framework to provide safety to the individuals. The delay sensitive security decisions are taken by fog layer. The authors have mentioned that the framework can be applied to other domains like smart health monitoring, smart cities etc.

With the advent of fog and cloud computing, there is an increased research interest for distributed data analytics. There is a need for scalable energy-efficient platforms to enable distributed data analytics. In this direction, the authors in~\cite{jayaraman2014cardap} have proposed a platform named Context Aware Real-time Data Analytics Platform (CARDAP). The complex distributed mobile analytics applications like sensing activity of citizens in smart cities etc. allow the deployment of the proposed platform. The proposed platform helps in real-time mobile data mining for data reduction. The experiments show that the platform is energy efficient.

Vaquero \etal \cite{vaquero2014finding} have given a comprehensive definition of fog computing. The fog is referred to as the convergence of a set of technologies which were present previously. The ubiquity of devices require fog based services in order to provide wide-spread resources, heterogeneity and real-time service provisioning. In this context, fog computing helps the smart city based ubiquitous applications also.

Table~\ref{table:tabsmartcity} summarizes various works done in smart city based fog applications. 

\begin{table*}[!t]
	\centering
	\caption{Comparison of Different Fog based Smart City Applications}
	\begin{tabular}{|>{\raggedright}p{2cm}||>{\raggedright}p{2cm} |>{\raggedright}p{2cm}|>{\raggedright}p{2cm}|>{\raggedright}p{2cm}|>{\raggedright}p{2cm}|>{\raggedright\arraybackslash}p{2cm}|} 
		\hline
		{\bf Research work} & {\bf Architecture} & {\bf Services} & {\bf Fog Devices}  & {\bf Communication between Fog Devices} & {\bf Application} & {\bf Advantages of Using Fog}\\
		\hline \hline 
		Tang \etal ~\cite{tang2015hierarchical} & Four layers: Data centers, Immediate computing nodes, Edge computing nodes, Sensing networks on critical infrastructure  & Low latency services & Edge devices  & Wireless & High computing and intelligence services for smart cities  & Minimizes the time to service  \\ \hline
		Yannuzzi \etal ~\cite{yannuzzi2014key} & User domain, Cloud domain and Fog domain & Realtime services to the users & Scattered from core to edge of the network   & Wireless &  IoT based latency sensitive smart city applications & Mobility of the devices is ensured  \\ \hline
		Bruneo \etal ~\cite{bruneo2016stack4things} & Open Stack-based framework & IaaS, PaaS & Edge devices   & Wireless & IoT based smart city applications & More dynamic service provisioning  \\ \hline	
		Jayaraman \etal ~\cite{jayaraman2014cardap} & IoT, Gateway and cloud server & Sensing activities of citizens in smart cities & Gateways & Wireless & Context aware distributed mobile data analytics in smart cities& Energy efficient and scalable system  \\ \hline					
		Aazam \etal ~\cite{aazam2014fog} & IoT, Smart Gateway as fog device and cloud server & Data trimming and preprocessing & Smart gateways & Wireless & Data preprocessing in smart city applications & Low latency   \\ \hline									
		Vaquero \etal ~\cite{vaquero2014finding} & SDN based edge server and cloud server & 
		Different services in the context of smart city & SDN based Edge cloud & Wireless & Smart city based applications & Proper service management, data privacy  \\ \hline		
		Sehgal \etal ~\cite{sehgal2015smart} & Three layers: IoT, Fog and Cloud & Security services  & Routers and Gateways  & Wireless &  Security providing system, Smart city based applications & Fog works as a distributed expert system for latency-sensitive security applications\\ \hline	
		
	\end{tabular}	
	\label{table:tabsmartcity}
\end{table*}

\subsection{Augmented reality and Real-time Video Analytics based on Fog}

Augmented reality applications are favored on smartphone, smart glasses etc. This gives an augmented view of the objects with much more information which are relevant to the users. Google Glass, Sony SmartEyeglass are among the recent projects. These applications require high computation power for processing video streaming. Also, they demand high bandwidth for data transmission. Human are very sensitive to delays in a series of consecutive interaction with the smart devices. A processing delay of more than tens of milliseconds breaks down the user experience. Fog computing in this context improves the processing and transmission delay. It also improves the throughput required for a perfect user experience. The authors in~\cite{ha2014towards} designed and implemented a wearable cognitive assistance spanning on Google Glass and Cloudlet. This offers hints on social interactions by real-time scene analysis. The nearby cloudlets are used for offloading computation-intensive tasks. The smart city and smart connected vehicle demands real-time surveillance, traffic management etc. Fog computing has the potential of providing resources to store captured video frames, transcode and process video frames. This is needed for object recognition, object tracking and data mining etc. Privacy issues can also be taken care by applying some techniques in the fog devices. Thus, Fog computing helps to increase throughput and reduce latency for augmented reality and real-time videos~\cite{yi2015survey}.

\begin{table*}[!t]
	\centering
	\caption{Comparison of different application areas of fog computing}
	\begin{tabular}{|c ||c |c |p{1.8cm}|p{2.2cm} |L |} 
		\hline
		{\bf Research work} & {\bf Fog nodes} & {\bf Scalability} & {\bf Dependency on Cloud} & {\bf Use-case} & {\bf Technology/Design approach}\\ \hline \hline
		Oueis et al. \cite{oueis2015small} & Radio access Points & High & Low & Radio access network & Formulated the distributed clustering problem as a joint optimization of the computation and communication resources \\ \hline
		Craciunescu et al. \cite{craciunescu2015implementation} & Edge devices & Not known  & Medium & Healthcare & Two layers: Fog node and cloud system \\ \hline
		Truong et al. \cite{truong2015software} & Edge devices & High & High & VANET & Software Defined Network\\  \hline
		Hou et al. \cite{houvehicular} & Vehicles and edge devices & High  & High & VANET & Three layers: Sensors and applications, Multi-service edge, Core\\ \hline
		Bruneo et al. \cite{bruneo2016stack4things}  & Edge devices & Not known & High & Smart City & OpenStack\\ \hline
		Ha et al. \cite{ha2014towards} & Edge devices  & Not known & High & Augmented reality and real-time video analytics & Google glass and cloudlet\\ \hline
		Zhu et al. \cite{zhu2013improving} & Edge devices  & Not known & Low & Content Delivery and caching & Three layers: Embedded systems and sensors, Edge, Core, Cloud \\ \hline
	\end{tabular}	
	\label{table:fogapplication}
\end{table*}

\subsection{Fog based Content Delivery and Caching}

Some web content delivery technologies cannot adapt to the requests from user after there is an optimization at the server side. Some optimizations can be done in the client side in order to have an improved web performance. So, there is a requirement of integrating caching and content delivery by using fog computing. The authors in~\cite{zhu2013improving} have leveraged fog computing in order to have an web optimization. Fog node's presence near to the client side helps to know client side user experience in order to optimize the rendering of web page. Similarly, caching in fog nodes helps to save the bandwidth and reduce latency~\cite{yi2015survey}.

\subsection{Mobile Big Data Analytics over Fog}

Big data processing is currently one of the most popular research topic for big data architecture in cloud and mobile cloud~\cite{yang2013framework,qian2013timestream}. Fog computing provides elastic resources for large scale data processing without suffering from the issues of the cloud. In cloud computing paradigm, the event or data will be transferred to the data center and result will be sent back to end user after a series of processing. An integration of fog and cloud can handle the big data related operations effectively. This in turn reduces the computation power on data processing. Data processing in the fog is the key technique to perform analytics on large scale of data generated in the context of IoT~\cite{yi2015survey}.

Table~\ref{table:fogapplication} summarizes various application areas of fog computing as discussed in the current literature. From the table, we can infer that VANET, smart city and augmented reality based applications high dependency with cloud.

%% file: 9_fog_computing_associations.tex
\section{Fog Computing Associations}
\label{associations}
In the computing world, we have different collaborations among the industry and the academic institutions. The same concept is true for fog computing fraternity. The following efforts have been made in the community which are related to the fog or edge computing.

\subsection{OpenFog Consortium}

{OpenFog Consortium}\footnote{ \url{https://www.openfogconsortium.org/} (Last accessed: 1 July, 2017)} is a community of industry and academia formed to provide the research of fog computing in order to solve the bandwidth, latency and communication challenges in the context of IoT, artificial intelligence etc. It was founded by ARM, Cisco, Dell, Intel, Microsoft and Princeton University edge computing laboratory in 2015. Their work is related to developing a framework for efficient and reliable networks and intelligent endpoints combined with secure, privacy-friendly information flows between clouds, endpoints and services based on open standard technologies.

\begin{table*}[!t]
	\centering
	\caption{Comparison of Different Overheads Associated with Fog Computing}
	\begin{tabular}{|>{\raggedright}p{3cm}||>{\raggedright}p{2cm} |>{\raggedright}p{2cm}|>{\raggedright}p{2cm}|>{\raggedright\arraybackslash}p{2cm}|} 
		\hline
		{\bf Research work} & {\bf Computational Overhead} & {\bf Storage related Overhead} & {\bf Maintenance related Overhead}  & {\bf Application migration related Overhead} \\
		\hline \hline
		Madsen \etal ~\cite{madsen2013reliability} & Yes & Yes & Yes & No \\ \hline
		Stantchev \etal ~\cite{stantchev2015smart} & No & Yes & Yes & No \\ \hline
		Oueis \etal ~\cite{oueis2015fog} & Yes & No & Yes  & No \\ \hline
		Dubey \etal ~\cite{dubey2015fog} & Yes & No & Yes  & No \\ \hline
		Hao \etal ~\cite{hao2017challenges} & Yes & No & Yes & Yes \\ \hline
		Osanaiye \etal ~\cite{osanaiye2017cloud} & Yes & No & Yes  & Yes \\ \hline
		Bittencourt \etal ~\cite{bittencourt2015towards} & Yes & No & Yes & Yes  \\ \hline
		Gia \etal ~\cite{gia2015fog} & Yes & Yes & No & No \\ \hline
		Aazam \etal ~\cite{aazam2015hamc} & Yes & Yes & No & No  \\ \hline
		Liu \etal ~\cite{liu2016paradrop} &  Yes & Yes & Yes & Yes  \\ \hline
	\end{tabular}	
	\label{table:tabOverhead}
\end{table*}

\subsection{Central Office Re-architected as a Datacenter}

{OpenCORD}\footnote{ \url{http://opencord.org/} (Last accessed: 1 July, 2017)} initiative aims at building micro-datacenters at the network edge for network providers to run their virtual network functions on. Such virtualized infrastructure near the edge of the network can be a place where fog computing can come into being. The major service providers like AT\&T, SK Telecom, Verizon, China Unicom and NTT communications are already supporting CORD.

\subsection{ETSI Mobile Edge Computing}

{Mobile Edge Computing (MEC)} is a recent initiative from telecommunication providers aiming to provide compute and storage capabilities akin to cloud computing in the radio access network (RAN) close to mobile subscribers. Placing general purpose infrastructure services in the RAN enables application developers and service providers to leverage low latency and high bandwidth access to users, along with accurate information about network conditions (e.g. user location, radio link quality, etc.). Such characteristics enable the delivery of applications and services with properties like heavy data volume, real-time response and context-awareness - that were not possible with traditional cloud-based applications. A key aspect of MEC is multi-tenancy wherein the infrastructure, though owned and managed by the network provider, will be opened to third party service providers allowing rapid deployment of applications and services for subscribers and enterprises and enable vertical solutions. 

The European Telecommunication Standards Institute (ETSI) proposed a reference architecture for MEC in a whitepaper\footnote{\myurl ~(Last accessed: 1 July, 2017)}. A new Industry Specification Group (ISG) is proposed to be set up in ETSI to allow the creation of industry specifications for Mobile-edge Computing (MEC). Although the deployment of MEC may vary in terms of topological location (eNB, aggregation points or radio network controller site), ETSI has proposed a blueprint for the MEC platform which describes, on a high level, the necessary components. The major challenges identified include network integration, application portability, security, legal and regulatory considerations.

%% file: 10_overhead_of_using_fog_computing.tex
\section{Overheads Associated with Fog Computing Environments}
\label{overheadfog}

Fog computing has been incorporated in many applications in order to help the cloud for better response to user services. Though fog computing is highly distributed, loosely coupled as well as perfect fit for real-time latency sensitive applications, there are some overhead which we need to consider when applying fog computing. Various works have analyzed these trade-offs and developed solutions while considering such overhead for a fog based application. In this section, we discuss about these different overheads related to fog computing, as explored from various related literature.

\textbf{Computational overhead}: Fog computing helps to minimize the response time in presence of real-time latency sensitive applications. The computation involved in the fog devices creates some overhead in the system. The fog nodes need to be properly orchestrated in order to do the utilization of the fog environment. These controls of fog nodes generate more control packets in the system. That is why, there is an overhead for computation of the services. 

\textbf{Storage related overhead}: In the context of IoT based fog computing, there is a surge of data that is being generated. We need to properly store these data produced by the fog nodes. The extra fog nodes in the system contributes to much more storage devices than the cloud system alone.

\textbf{Maintenance related overhead}: Deployment of large number of fog nodes in the context of IoT requires maintenance of the nodes. The fog computing system is subjected to some maintenance related overhead in comparison with the cloud. The extra work of maintaining a fog node impacts the system operation. The fog nodes need to be controlled in order to have a consistency in the fog computing system.

\textbf{Application migration related overhead}: In order to properly utilize the system resources, there is a need of migrating services from one fog node to another. This migration of services and resource provisioning have generated an overhead in the fog computing environment.

In Table~\ref{table:tabOverhead}, we have compared different works based on the types of overheads associated with the proposed model and discussed in that literature. Such overheads are important factors behind the design of fog computing environments. Accordingly, we discuss various research objectives and open research areas on fog computing in the next section. 

%% file: 7_limitations_and_future_scope.tex
\section{Future Scopes and Open Research Areas in Fog Computing}
\label{futurescope}
In the presence of multiple distributed end devices, the generation of huge data and its processing need attention. Fog computing in this respect plays a great role of maintaining these devices. The existing literature discusses different aspects of fog computing. However, there are limitations in the exiting works as we mentioned time to time, and also there are several open issues that need to be addressed for designing a fully deployable system over fog framework by utilizing all of its advantages.  In this section, we summarize various research areas that need to be explored for developing next generation systems and applications over fog computing platform. 


\subsection{Cloud-Fog Orchestration}

A basic idea behind the development of fog computing concept was to bring back part of the computation to the edge devices so that processing and transmission overhead of data to the cloud can get reduced. However such orchestration between cloud and fog envisions various research challenges that need to be addressed. The various open research challenges for cloud-fog orchestration are summarized below. 

\textbf{Partitioning of tasks or services:} An important research area is partitioning of tasks or services between the fog nodes and the cloud nodes. This particular problem has several stages -- (a) estimation of resources at fog nodes, (b) task partitioning based on the resource availability at fog nodes and the expected response time for task completion, (c) estimation of overheads for task partitioning and migration, (d) estimation of overheads for result accumulation for various sub-tasks, (e) optimal placement of sub-tasks at various fog nodes and to the cloud, and so on. It can be noted here, that many a times solutions for these problems are application specific. However, based on the current survey, we observe that no generalized framework exists to develop a solution for such task partitioning problem in a fog environment. Although task partitioning and task allocation problems are well studied in the literature for multiple scenarios, for example, under distributed environments~\cite{ranganathan2002decoupling,foster1998grid,waldspurger1992spawn}, for parallel computing systems~\cite{rudolph1991simple,chu1987task,lee1990efficient}, and recently, for mobile and cloud computing environments~\cite{kosta2012thinkair,yang2013framework,gkatzikis2013migrate}, the scenario is different for a fog computing environment. The cloud plays a key role in the fog computing environment, and unlike the mobile cloud computing environments, the tasks are in general resource hungry. In a typical fog environment, a part of a complete task is offloaded to the fog devices from the cloud, and therefore the cost-benefit trade-off is important in this scenario. In such a scenario, the target is to offload the tasks from a {\em resource slack environment} (cloud) to a {\em resource constraint environment} (fog) with the objective of better response time and better support of privacy. Therefore, the task migration algorithms need to be very crucial and perfect with high degree of accuracy. This is the major challenge in developing a fog computing system, and thorough research needs to be done to develop the necessary frameworks.

\textbf{Enforcing semantics in fog computing:} The fog computing environment consists of many heterogeneous sensors, actuators, edge devices and cloud servers. Fog infrastructure is largely distributed in the context of IoT. In these largely distributed environment, the understanding of service oriented computing is a major issue. That is why, incorporating semantics for services is problematic. For example, multiple applications may use the same actuator service and may take conflicting actions. The actuation service for an application may be different from the actuation service of another application. There is a requirement of managing the work-flow of different services of an application. The work-flow of the application has to be properly managed with semantically correct services so that the end goal is met. The research challenge in this respect is to apply proper meaning to the actions in order to perform the application. There are previous works on enforcing semantics in distributed computing paradigm \cite{mietz2013semantic,el2013semantic}. However, these methodologies are not directly applicable for fog computing, because (a) fog is a partially distributed system, where the role of clouds need to be defined, (b) the role of cloud is application dependent for many cases, therefore, development of a generic framework is difficult, (c) the resource availabilities at individual fog nodes are very dynamic, and (d) for a SOA based architecture, the dependency among micro services may be complex with a mixture of dependable services. These concepts have to be looked into with respect to fog computing scenarios. 

\textbf{Multi-domain orchestration:} The fog infrastructure is a continuum of resources extending from the network edge to the cloud data centers, making different parts being managed by different entities. End-to-end service delivery would require coordination between multiple domains with potentially heterogeneous control policies. It is necessary to design standard interfaces between domains so that services can be provisioned, while satisfying end-to-end performance constraints. The research challenge is that different network domains have local knowledge of the network due to its own network protocol. There is a requirement of provisioning global knowledge of the network topology across multiple domains. There has been some existing works in cloud computing domain on multi-domain orchestration~\cite{sonkoly2015multi,mandal2011provisioning}. However, because of the distributed nature of fog nodes, maintaining resource allocation under multi-domain systems with heterogeneous policies is difficult and needs special attention. There is a scope to work on the orchestration of multiple domains in the fog based systems with heterogeneous policy constraints. 

\textbf{Interaction among fog devices:} one of the major objectives of fog computing is the requirement of real-time service provisioning. However, ensuring real time service provisioning in a distributed environment under heterogeneous system (with different resource availabilities under various policy domains) is difficult. In order to enable a fog node to generate response within a very less amount of time, the dependency on the other fog nodes should be taken care of. The edge devices interact among themselves for different service calling and data sending. Further, there can be complex dependencies among multiple services. For example, some services under an application can be executed in parallel, however, the other services may have dependencies on the outcome of previously executed services. Accordingly, we can represent the dependencies among multiple services as a dependency tree (or graph, based on the application). Parsing such dependency tree or dependency graph of services for an application under a distributed environment requires complex interaction among various fog nodes.  The research challenge here is to make these interactions very fast, so that the overall system generates the output within a predefined time threshold to ensure real time service execution guarantee. 


\subsection{Virtualization of Fog Devices}
The fog devices are resource-constrained, and that is why, there is a need to properly utilize the resources by running multiple operating environments and applications on a single fog device. Further, the resource allocation as well as various services need to be coordinated for proper orchestration of application services. In a typical fog environment, a single fog device may host tasks or services from two different applications or from two different users. Under such scenarios, the followings are the important requirements; 
\begin{enumerate}
	\item \textit{Service separation and encapsulation:} The services or tasks from two different applications or users may need to have separate environment. Therefore, service separation and encapsulation is important, when both the services or tasks run on a single fog device. 
	\item \textit{Application fairness:} To ensure application fairness, resource reservation and provisioning to the services over a single fog devices need to be monitored, and the management algorithms need to take care of the fairness aspect.
	\item \textit{Data privacy:} When services from different applications or users run on a single fog node, data privacy for individual application or user needs to be ensured. 
	\item \textit{Fault tolerance:} This is an important aspect. When a fog device fails, the services running on that device need to be migrated to a different device, while maintaining application and user transparency. Seamless migration of services is an essential requirement to ensure high availability of resources over fog computing environment. The environment should support various types of fault tolerance, like crash faults, network faults and byzantine faults. 
\end{enumerate}
The above requirements can ve ensured with virtualization technologies~\cite{singh2008server,nagarajan2007proactive}, where a single virtual machines (VM) can encapsulate services from a single application or user, and services from different applications or users will be executed under different VMs.  Virtualization is the technology that helps to run multiple operating environments with dynamic policies and applications in a single computer. Therefore, virtualization can be an important aspect for fog computing. Further, virtualization can support over provisioning of resources and on-demand resource allocation to improve resource availability at the fog layer. However, to the best of our knowledge, there exists few systems and methodologies for fog computing~\cite{bellavista2017feasibility,morabito2017virtualization,al2017cognitive,saurez2016incremental,bittencourt2015towards,hong2016dynamic}, which consider virtualization at the fog layer. Nevertheless, with the invent of storage virtualization~\cite{huber2010performance} and network virtualization~\cite{chowdhury2009network} technologies apart from server virtualization, this technology can be an important driving force behind the development of fog computing architecture. 

However, there are multiple challenges to support virtualization architecture at the fog layer. Here we discuss such challenges and the open research areas in this direction. 

\textbf{Containerization or encapsulation of services:} 
Encapsulation of services to a VM is an important aspect. However, traditional VM design and VM management softwares (called hypervisors) are primarily designed for data centers and they consume significant amount of system resource. Virtualization at the fog layer requires lightweight control on the encapsulated services, because the devices at the fog layer are in general resource constraint. One of the approach of implementing virtualization over resource constraint devices is to deploy the services in a container~\cite{soltesz2007container,celesti2016exploring}. Container software is an operating system level virtualization technique. The different services can be placed in different containers so that the larger application can be executed faster. A container image can hold a service for an application. These containers can be deployed in different IoT or fog devices. A central controller node can place the services in different containers. However, there is a requirement for designing a container management system for fog environment which can take care of the service dependency graph (or tree) for an application, and accordingly initiate the containers to encapsulate services. The encapsulation of services to a container also require proper estimation or monitoring of resources at various fog nodes. This resource monitoring also needs to be done with less overhead. Therefore, we have another challenge on container management, as we discuss next. 

\textbf{VM or container resource allocation for fog devices:} 
As mentioned, the traditional hypervisors are not suitable for fog devices, as the fog devices are resource constrained. So, there is a need for modifying the architecture of the hypervisor or container management middleware for fog devices. The hypervisor for fog devices should have the following characteristics -- (i) lightweight, that is less resource consuming, (ii) virtualization of the required applications based on service dependencies, (iii) resource estimation or automated resource monitoring capability for the fog layer, (iv) resource allocation for individual VMs and containers according to the capacity of the fog device, on which the VM or the container will be placed, and finally (v) aggregation of results from multiple services running on multiple VMs or containers over different fog devices. The research challenge is to address these problems under a distributed or partially distributed environment. It can be noted that the hypervisor or the container manager (or sometime called controller) need to properly executed these tasks in a time synchronized manner. 

\textbf{VM or container migration:} 
As we have already mentioned, VM or container migration is an important requirement to make the system fail-safe and fault tolerant. Further, IoT devices are mostly mobile, hence this mobility has to be addressed. Whenever there is a movement of a IoT device, there is a possibility of changing of access points. In these situations, the end users should be able to get the desired result within a delay-constraint situation. So, we need to do the service migration in order to carry on the computation task. Fog computing has to leverage its features in order to make migration of services easier. Again, the down-time of migration for the virtual machines or the containers should be minimized.  In~\cite{bittencourt2015towards}, a fog computing based architecture for virtual machine migration is proposed. However, the architecture has multiple constraints, and it does not consider service dependability while initiating the migration. Further dynamic workload during the mobility needs to be addressed while designing the migration strategies for fog devices. 

\textbf{Lightweight VM placement, migration, result aggregation and actuation:} 
A typical fog environment works in the principle of sensing-actuation, where the sensors sense the environment, then some decisions are taken based on the sensing data, and finally, the actuator services are triggered. These sequence of tasks need to be properly orchestrated in a fog environment. As the fog devices have minimal resources, the virtualization on these devices should be lightweight. The research challenge is to design a method so that the fog devices are given optimal number of tasks meeting the resource constraints. As we have already mentioned, containers are implemented by lightweight virtualization~\cite{morabito2015hypervisors,willis2014paradrop,pahl2015containers}, and the container technology can be a nice building block for implementing fog services at various fog nodes. The containers can be placed in fog devices, which would perform micro-services of a large application. Again, the containers can be migrated to a computationally less burdened fog node. This way, the resources of a fog network can be efficiently utilized. In~\cite{bellavista2017feasibility,morabito2017virtualization,al2017cognitive,saurez2016incremental}, the authors have analyzed the feasibility of using docker based container technology for a fog computing environment. They have shown that container can host fog services over fog devices. However, there are multiple research challenges that need to be addressed. An optimization framework needs to be developed to design the optimal placement of services based on container resource requirements, resource availability at the fog devices, response time for a specific placement, container initialization delay, service delay based on resource availability, and result aggregation and actuation delay. The associated overheads need to be analyzed, and the cloud-fog orchestration needs to be developed accordingly.  

\subsection{Application Development Platform over Fog}
Though there are many application development platform over fog computing in the existing literature, these are not generic rather application domain specific. So, there is a requirement of making an uniform application development platform in the fog based systems. The question comes, whether it is possible to develop a generic application deployment platform for fog based systems. We argue that with the invention of software defined and software controlled technologies, it should be possible to develop generic application build-up and application deployment platforms over fog based systems, at least for a class of applications with similar properties. The fog applications, being large scale, poses many challenges. There are two major requirements -- (1) a generic application development framework or application programming interface (API) needs to be provided, which can act as the baseline for extracting the services and the service dependency graphs from the applications, and (2) based on the service requirements and the service dependency graphs, an interpreter or a middleware needs to be developed that can distribute the services across the fog devices and accumulate the results for the actuation purpose. The various open research areas in this direction are as follows. 

\textbf{Large scale application development:} The applications sometimes need large scale distribution of fog nodes. Also, these fog nodes interact with the cloud data centers for data storage and processing. The challenges are (i) proper orchestration of the devices, (ii) scalability of the devices and so on. This generates the research scope in this fog based systems. The works of \cite{hong2013mobile}, talks about a programming model for latency sensitive large scale IoT applications. Saurez \etal \cite{saurez2016incremental} have proposed Foglets, which is a programming infrastructure for large scale deployment of latency sensitive geographically distributed applications. However, application classification and requirement analysis is necessary that can trigger the developments of generic fog architectures. 

\textbf{Fog programming infrastructure for application development:} 
There is a requirement of a programming language for fog-cloud orchestration. This would help in interacting with the underlying fog devices by a common middleware software, which would be instructed by this programming language. The features of this programming language can be (i) allowing a programmer to use a standard programming language in order to write their own algorithms, applications or services for their need, (ii) to provide an abstraction that the high level services are performed ensuring the QoS requirements while the middleware would execute the micro-services in distributed fog devices,  and (iii) to ensure scalability of fog nodes. Such programming API can work as the baseline for fog application development and deployment, and can also help in rapid prototyping of fog based systems.

\subsection{Resource Management at Fog Devices and Networks}
Fog devices are resource-constrained, as we have already pointed out. That is why, there is a need for proper distribution of resources i.e. CPU, memory, storage, workstations, network elements, sensors, actuators etc. in fog based systems. We need to also manage the logical resources as well. These logical resources can be operating system, energy, network throughput, protocols etc. The various research directions in this area are as follows. 

\textbf{Resource estimation at fog nodes:} 
Fog, being a distributed layer of computation, has a requirement of proper resource estimation at various fog nodes, so that resource provisioning can be done. The resource estimation needs to be done by the fog controller, which is a software middleware, and manages the service deployments and result accumulation for the fog based applications. However, the challenge is to do this resource estimation with minimal overhead and high accuracy. Further, the fog environment is dynamic, and the edge devices can support mobility. Due to this, the resource availability at various fog nodes is also dynamic in nature. This dynamic nature of the system needs to be handled while estimating the available resources at the fog nodes. 

\textbf{Resource allocation ensuring fairness and QoS:} 
The different resources should be properly allocated to different devices by applying the knowledge of resource estimation. Also, the resource allocation should be fair in the sense that it is required to meet the end-user's QoS requirement. For example, tasks like real-time video streaming should be given higher bandwidth in comparison with the hypertext transfer protocol (HTTP) web browsing traffic. Ensuring fairness with service differentiation and QoS is a difficult task, and so the problem is challenging for fog based systems. 

\textbf{Network resource management in fog:} 
As the network resources are distributed in fog based systems, we need to ensure the correct network connectivity among the resources. This is a challenging task in the context of fog computing.  The technologies like software defined networking (SDN) can be implemented in order to perform the control of network resources in the fog based systems~\cite{truong2015software}. This would perform like a middleware which can solve the problem of management of distributed network resources in the fog computing systems.

\subsection{Security and Privacy of Fog Nodes and Networks}
In a fog environment, the data is analyzed at the fog devices. Fog devices are typically less secure compared to the cloud, as they are physically closer to the attackers. Further, fog devices are resource constraint, and therefore typical security attacks, like denial of services (DoS), man in the middle (MITM), and session hijacking, are easy for the fog layer. Therefore, there is a requirement of meeting the security needs of the users while ensuring privacy. The data from one domain might not be shared with other domain in order to preserve privacy and security.

\textbf{Security issues:} The wide geo-distribution of fog infrastructure brings with it security challenges by increasing the defense perimeter. The easy physical access to fog or edge devices makes applications vulnerable to attacks by malicious system software~\cite{checkoway2013iago}. Another possible attack model is when a malicious fog instance behaves as a genuine one~\cite{stojmenovic2014fog}. Such a rogue fog could launch man in the middle attacks and compromise the privacy of users connected to it. 
The challenge here is to ensure that the authentication services are provide locally when the distant cloud server is down. Again as the fog nodes are scalable, the traditional public key infrastructure (PKI) based authentication is not efficient in this context. Therefore a proper authentication, authorization and accounting system needs to be developed for the fog environment. As the security is application specific many a times, the security features can be embedded within the fog API and fog middleware. The network security is also an important aspect that needs to be addressed while developing a fog computing platform. 

\textbf{Privacy issues in fog:} 
The privacy related issues in fog computing systems are due the fog node's presence in the vicinity of the end users. So, the fog devices have more user related private sensitive information. The challenge is to ensure trust between the end devices in order to create a secure environment in IoT based fog systems. The approach is to encrypt the sensitive user data before sending to the fog devices. But, the resource constraint IoT devices face challenge to encrypt as well as decrypt the user's private data. Further, a single fog device may handle data from multiple applications or different users. Under such scenarios, proper data encapsulation needs to be done at the fog API or fog middleware, such that data from one application is not visible to another application. 

%
%

%% file: 8_conclusion.tex
\section{Conclusion}
\label{conclusion}
With the increase of IoT devices, the traditional computing needs many changes. Though the distributed computing is present for many years, but it fails to provide the users the required services within the service level agreements. So, we need the support of the cloud servers to provide the end users virtualized resources. There are many applications of cloud computing in the context of IoT devices. However, the cloud alone cannot provide the required real-time responses with location aware services. Again there are several requirements such as wide spread geographical distribution of IoT devices, mobility, supporting very large number of nodes, omnipotent role of wireless access, device heterogeneity etc. which needs to be supported in these systems. That is why, we need computation and storage near the edge devices in order to provide the clouds the required handshaking for the QoS. This introduces the concept of fog computing, which has several benefits as discussed in this survey. Fog computing acts as an enabler for delay sensitive real-time data services. It gives better QoS for the users in several situations. By implementing the edge devices as the computing nodes, we are able to satisfy the need with reliability. In this paper, we have presented a thorough survey of the different systems, QoS parameters and applications of fog computing. We have explored the existing literature in order to find the latest developments as well as the different use cases of fog. In these works, the major issues regarding the security, privacy etc. are discussed. Nevertheless, there are several limitations and challenges of these systems, which are discussed elaborately in this survey. In a nutshell, this survey gives a thorough summarization of the various existing works on fog computing, analyses their pros and cons critically, and discusses the open directions of research in this emerging domain of computing.